\documentclass[conference]{IEEEtran}

\newif\iftechreport
\techreportfalse
\techreporttrue 

\ifCLASSINFOpdf
\else
\fi
\ifCLASSOPTIONcompsoc
  \usepackage[caption=false,font=normalsize,labelfont=sf,textfont=sf]{subfig}
\else
  \usepackage[caption=false,font=footnotesize]{subfig}
\fi
\usepackage{fixltx2e}

\usepackage[english]{babel}
\usepackage{cite}

\usepackage{listings}
\usepackage{mathpartir}
\usepackage[cmex10]{amsmath}
\usepackage{array}
\usepackage{mathrsfs}
\usepackage{amssymb}
\usepackage{amsthm}
\usepackage{thmtools}
\usepackage{thm-restate}
\usepackage[only,llbracket,rrbracket,llparenthesis,rrparenthesis,lightning,Lbag,Rbag]{stmaryrd}  
\usepackage{mathtools}
\usepackage{synttree}
\usepackage{fixltx2e}
\usepackage{acronym}
\usepackage{pbox}
\usepackage{microtype}
\usepackage{cancel}
\usepackage{mathrsfs}
\usepackage[textsize=small
]{todonotes}

\usepackage[title]{appendix}

\usepackage{xspace}
\usepackage{flushend}
\usepackage{galois}
\usepackage{mdframed}
\usepackage{tikz-cd}

\iftechreport
\usepackage{hyperref}
\else
\usepackage[bookmarks=false]{hyperref}
\fi

\usepackage{hyperref}
\usepackage{cleveref}
\crefname{mylem}{Lemma}{Lemmas} 
\crefname{appsec}{Appendix}{Appendices}

\newcommand{\skipcom}{\text{{\ttfamily\bf skip}}\xspace}
\newcommand{\ifcom}{\text{{\ttfamily\bf if}}\xspace}
\newcommand{\thencom}{\text{{\ttfamily\bf then}}\xspace}
\newcommand{\elsecom}{\text{{\ttfamily\bf else}}\xspace}
\newcommand{\whilecom}{\text{{\ttfamily\bf while}}\xspace}
\newcommand{\docom}{\text{{\ttfamily\bf do}}\xspace}
\newcommand{\assumecom}{\text{{\ttfamily\bf assume}}\xspace}
\newcommand{\assertcom}{\text{{\ttfamily\bf assert}}\xspace}

\newcommand{\err}{\lightning}

\newcommand{\triangleqom}{\triangleq_{\err}}

\newcommand{\agree}{\ensuremath{\mathbb{A}}}
\newcommand{\bagree}{\ensuremath{\mathbb{B}}}

\let\oldshortmid\shortmid
\renewcommand{\shortmid}{\mathord{\oldshortmid}}

\newcommand{\semins}[2]{ \llbracket #1  \rrbracket #2}  
\newcommand{\semexp}[2]{ \llbracket  #1 \rrbracket #2  } 
\newcommand{\st}{\sigma} 
\newcommand{\stm}{\tau} 
\newcommand{\St}{\Sigma} 

\newcommand{\Sterr}{\hat{\Sigma}} 
\newcommand{\States}{\operatorname{States}} 
\newcommand{\sterr}{\hat{\st}}
\newcommand{\rest}{}
\newcommand{\seminsc}[2]{\llbrace #1 \rrbrace #2} 

\newcommand{\seminsm}[3]{\llparenthesis #1 \rrparenthesis_{\rest #2} #3}

\newcommand{\seminsmabs}[3]{ \llparenthesis #1 \rrparenthesis_{\rest #2}^\sharp
  #3}
\newcommand{\seminscabs}[2]{ \llbrace #1 \rrbrace^\sharp #2}

\newcommand{\seminsmalt}[4]{\llparenthesis #1 \rrparenthesis_{\rest #2}^{\operatorname{ann}=#4} #3}
\newcommand{\seminscalt}[3]{\llbrace #1 \rrbrace^{\operatorname{ann}=#3} #2} 
\newcommand{\boolann}{\operatorname{a}}

\newcommand{\pset}[1]{\mathcal{P}(#1) }
\newcommand{\pstates}{\mathcal{P}(\States) }
\newcommand{\pstateserr}{\mathcal{P}_{\err}(\States) }
\newcommand{\relform}{\mathcal{L}}
\newcommand{\prelform}{\pset{\mathcal{L}}}
\newcommand{\prelformerr}{\mathcal{P}_{\err}(\mathcal{L})} 

\newcommand{\entail}{\mathbin{\mathord{\imp}^\sharp}}

\newcommand{\concleq}{\sqsubseteq}
\newcommand{\absleq}{\mathbin{\sqsubseteq^\sharp}}
\newcommand{\dotconcleq}{\mathbin{\dot{\concleq}}}
\newcommand{\dotabsleq}{\mathbin{\dot{\sqsubseteq}^\sharp}}
\newcommand{\concjoin}{\mathbin{\sqcup}}
\newcommand{\absjoin}{\mathbin{\sqcup^\sharp}}
\newcommand{\concmeet}{\sqcap}
\newcommand{\absmeet}{\mathbin{\sqcap^\sharp}}

\newcommand{\pured}{\mathcal{D}}
\newcommand{\hybmod}{\mathcal{M}}
\newcommand{\hybint}{\mathcal{I}}
\newcommand{\seminscabspured}[2]{ \llbrace #1 \rrbrace^\sharp #2}
\newcommand{\seminscabshybmod}[3]{ \llbrace #1 \rrbrace^\sharp_{#3} #2}
\newcommand{\seminscabshybint}[4]{ \llbrace #1 \rrbrace^\sharp_{#3,#4} #2}

\newcommand{\alphares}[1]{\alpha_{\rest #1}}
\newcommand{\alphast}{\alphares{\st}}
\newcommand{\gammares}[1]{\gamma_{\rest #1}}
\newcommand{\gammast}{\gammares{\st}}
\newcommand{\fv}[1]{\operatorname{fv}(#1)}
\newcommand{\guardop}{\operatorname{grd}}

\newcommand{\guard}[2]{\guardop^{#1}\left(#2\right)}
\newcommand{\guardm}[3]{\guardop_{#1}^{#2}#3}
\newcommand{\guardc}[2]{\guardop^{#1}#2}

\newcommand{\dotsubseteq}{\mathbin{\dot{\subseteq}}}
\newcommand{\dotpreccurlyeq}{\mathbin{\dot{\preccurlyeq}}}

\newcommand{\alphapair}{\alpha^{\rhd}}
\newcommand{\gammapair}{\gamma^{\rhd}}
\newcommand{\concdom}{\mathscr{C}}
\newcommand{\absdom}{\mathscr{A}}

\newcommand{\Mod}{\operatorname{Mod}}

\newcommand{\intervals}{\operatorname{Int}}
\newcommand{\statesint}{\States^{\intervals}}
\newcommand{\statesinterr}{\statesint_{\err}}

\newcommand{\leqintervals}{\leq^{\sharp,\intervals}}
\newcommand{\dotleqintervals}{\mathbin{\dot{\leq}^{\sharp,\intervals}}}

\newcommand{\capintervals}{\sqcap^{\sharp,\intervals}}

\newcommand{\alphaint}{\alpha^{\intervals}}
\newcommand{\gammaint}{\gamma^{\intervals}}
\newcommand{\seminscabsint}[2]{\llbrace #1
  \rrbrace^{\sharp,\intervals} #2}
\newcommand{\redtoint}{\operatorname{toint}}
\newcommand{\redtointst}{\redtoint_{\st}}

\newcommand{\redtoprelform}{\operatorname{tofor}}
\newcommand{\redtoprelformst}{\redtoprelform_{\st}}
\newcommand{\redtoprelformres}[1]{\redtoprelform_{#1}}
\newcommand{\gammastint}{\gammast^{\intervals}}

\newcommand{\alphastint}{\alphast^{\intervals}}

\newcommand{\guardint}[1]{\guardop_{#1}^{\sharp,\intervals}}
\newcommand{\appint}[2]{\operatorname{app}_{#1}^{#2}}
\newcommand{\blwint}[1]{\operatorname{blw}(#1)}
\newcommand{\abvint}[1]{\operatorname{abv}(#1)}

\newcommand{\dt}[1]{\emph{\textbf{#1}}} 

\newcommand{\imp}{\mathbin{\Rightarrow}} 

\newcommand{\hbra}{
  \hbox to \columnwidth{\vrule width0.3mm height 1.8mm depth-0.3mm
    \leaders\hrule height1.8mm depth-1.5mm\hfill
    \vrule width0.3mm height 1.8mm depth-0.3mm}}
\newcommand{\hket}{
  \hbox to \columnwidth{\vrule width0.3mm height1.5mm
    \leaders\hrule height0.3mm\hfill
    \vrule width0.3mm height1.5mm}}

\newcommand{\ratio}{.35}

\newenvironment{tdisplay}[1]{\medskip
    \noindent\textbf{\normalsize #1}\\[-.3ex]
    \hbra\\[-.4ex]
  }{\par\hket
  \medskip}

\RequirePackage{scalerel}
\RequirePackage{accsupp}
\newcommand*{\llbrace}{%
  \BeginAccSupp{method=hex,unicode,ActualText=2983}%
    \textnormal{\usefont{OMS}{lmr}{m}{n}\char102}%
    \mathchoice{\mkern-4.05mu}{\mkern-4.05mu}{\mkern-4.3mu}{\mkern-4.8mu}%
    \textnormal{\usefont{OMS}{lmr}{m}{n}\char106}%
  \EndAccSupp{}%
}
\newcommand*{\rrbrace}{%
  \BeginAccSupp{method=hex,unicode,ActualText=2984}%
    \textnormal{\usefont{OMS}{lmr}{m}{n}\char106}%
    \mathchoice{\mkern-4.05mu}{\mkern-4.05mu}{\mkern-4.3mu}{\mkern-4.8mu}%
    \textnormal{\usefont{OMS}{lmr}{m}{n}\char103}%
  \EndAccSupp{}%
}

\newtheoremstyle{DavesFavoriteTheoremStyle}
  {\topsep}
  {\topsep}
  {\rmfamily}
  {0pt}
  {\bfseries}
  {.}
  { }
  {\thmname{#1 \thmnumber{#2} #3}}

\theoremstyle{DavesFavoriteTheoremStyle}

\hypersetup{
 pdfauthor={Mounir Assaf} {David A. Naumann},
 pdftitle={Calculational Design of Information Flow Monitors},
 pdfkeywords={Information Flow} {Monitors} {Security} {Abstract Interpretation},
}
\begin{document}

%

\iftechreport
\title{Calculational Design of Information Flow Monitors (extended version)}
\else
\title{Calculational Design of Information Flow Monitors}
\fi




%
\iftechreport
\author{
 \IEEEauthorblockN{Mounir Assaf}
 \IEEEauthorblockA{
  Stevens Institute of Technology
 }
 \and
 \IEEEauthorblockN{David A. Naumann}
 \IEEEauthorblockA{
     Stevens Institute of Technology
 }
}
\else
\author{
 \IEEEauthorblockN{Mounir Assaf}
 \IEEEauthorblockA{
 Stevens Institute of Technology\\
 Hoboken, NJ 07030, USA\\
 massaf@stevens.edu
 }
 \and
 \IEEEauthorblockN{David A. Naumann}
 \IEEEauthorblockA{
 Stevens Institute of Technology\\
 Hoboken, NJ 07030, USA\\
 naumann@cs.stevens.edu
 }
}
\fi




\maketitle
\pagestyle{plain} 

\begin{abstract}
Fine grained information flow monitoring can in principle address
a wide range of security and privacy goals, for example in web applications.  
But it is very difficult to achieve sound monitoring with acceptable runtime cost and sufficient precision to avoid impractical restrictions on programs and policies.
We present a systematic technique for design of monitors that are correct by construction.
It encompasses policies with downgrading.
The technique is based on abstract interpretation which is a standard basis for static analysis of programs.
This should enable integration of a wide range of analysis techniques, enabling more sophisticated engineering of monitors to address the challenges of precision and scaling to widely used programming languages.
\end{abstract}


%
\IEEEpeerreviewmaketitle

\definecolor{mygray}{gray}{0.6}
\definecolor{dgray}{gray}{0.3}
\newcommand{\highlight}[1]{\textcolor{dgray}{#1}}
\definecolor{Brown}{cmyk}{0,0.81,1,0.60}
\definecolor{OliveGreen}{cmyk}{0.64,0,0.95,0.40}
\definecolor{CadetBlue}{cmyk}{0.62,0.57,0.23,0}
\definecolor{lightlightgray}{gray}{0.9}
\newcommand{\lstmathcomment}[1]{\ensuremath{\smash{\color{mygray}{#1}}}}

\newlength{\mylistingskipamount}
\setlength{\mylistingskipamount}{.5\medskipamount}
\lstdefinestyle{simple}{
escapeinside={@}{@}, float=htp, 
language=C,                             
basicstyle=\footnotesize\sffamily,
commentstyle={\color{gray}}, 
numbers=left,                           
numberstyle=\tiny,                      
stepnumber=1,                           
numbersep=5pt,                          
tabsize=2,                               
captionpos=b,                         
breaklines=true,                      
breakatwhitespace=false,        
showspaces=false,                  
showtabs=false,                       
morekeywords={then,label,output, stop, assume, assert,seed},
literate=%
{<=}{{\tiny$\leq$}}1%
{>=}{{\tiny$\geq$}}1%
{\\agree}{{$\agree\:$}}1%
}

\section{Introduction}\label{sec:intro}

Runtime monitoring can serve to test a program's security or to ensure its
security by detecting violations.  
Monitoring is a good fit for access control policies, which are safety
properties.
A run either does or does not satisfy the policy, and a monitor can be precise
in the sense of raising an alert only when the run is poised to violate the policy. 
Information flow policies are about dependency, e.g., an untrusted
(resp.\ secret) input should not influence a trusted (resp.\ public) output. 
Formal definitions of information flow (\dt{IF}) security are
``hyperproperties''~\cite{CS10} involving multiple runs. 
Suppose an observer classified as ``low'' knows the code, the set of possible
secret inputs, and the low input (from which they can deduce the possible runs
and low outputs).   
A policy specifies what can be learned about the secret upon observing a
particular low output.  Learning means determining a smaller set of possible
values of the secret. 
How is a monitor, acting only on the actual execution, to detect violations of a
property defined with respect to all (pairs of) runs?  
Remarkably, this was shown to be possible~\cite{Lal.07}.
In this paper we show how to design such monitors systematically. 

A popular way to monitor dependency is to tag secret data and propagate tags
whenever tagged data is involved in computing other data.   If an output is not
tagged, we might conclude that in all possible runs, the output would have the
same value, i.e., nothing has been learned about the secret.   
The conclusion is wrong, owing to information channels besides data flow. 
The most pervasive and exploitable such channel is control flow.
If some branch condition depends on a secret, the low observer may learn the
secret from the absence of an observable action that happens in the other
branch. 

Owing to the possibility of such \emph{implicit flow}, sound monitors are not in general precise.
A simple technique is to raise an alert if 
a low assignment is attempted in a high branch,
yielding false positives in cases like this: $\ifcom~ inhi~ \thencom~ outlo:=0~ \elsecom~ outlo:=0$.
Another technique which has been investigated extensively is to rely on static
analysis to determine which locations might have been updated in
executions that do not follow the same branch as the actual execution.
The monitor tags all such locations when the control join point is reached.
These techniques provide monitoring that is provably sound with respect to
idealized semantics that ignores covert channels like timing 
(e.g.,~\cite{Lal.07,ShroffST07,RS10,AF10,Beringer12}).

These and related techniques have been investigated and implemented but had
quite limited practical impact.   
One obvious reason is the difficulty of specifying policies with sufficient
flexibility to capture security goals without excessive restriction.   
Another impediment to practical use is that keeping track of possible
alternate control paths has high performance cost. 
Lowering precision to reduce cost can result in intolerably many false positives.
It is an active area of research to improve monitors for better performance,
better precision, and more subtle policies.  (See \Cref{sec:related} for related work.)

Another impediment to practical IF monitoring is that, if enough is at stake to motivate paying the costs
of policy specification, performance degradation, and possible false positives,
there should be high assurance of correctness.  The complexity of monitoring
grows with the complexity of the programming language and the monitoring 
techniques. While there are machine-checked correctness proofs for theoretical
models, there are few for practical implementations. 
The proofs known to us have been done ``from scratch'', rather than building on
and reusing prior results (though of course one can identify common techniques).
Such proofs are not easily maintained as the monitored language and platform
evolves.

This paper addresses the impediments related to the precision of monitors as well
as the complexity of their design and correctness proofs.

For safety properties, the theory of abstract interpretation \cite{CC79} is well
established and widely used to guide and validate the design of static analyses
\cite{Cal06}.   
Like the best theories in engineering, abstract interpretation allows designs to
be derived from their specification instead of merely helping to justify them after
the fact \cite{Cou99}. 
Abstract interpretation underlies the static analysis part of some IF monitors
\cite{BBJ13,MC11}, but the monitor design and justification remains ad hoc. 
Abstract interpretation has also been used for static analysis of
noninterference \cite{KSF13}.  

Chudnov et al.~\cite{CKN14} suggest that an IF monitor should be viewed as
computing an abstract interpretation to account for alternate runs vis-{\`a}-vis
the monitored run. 
The key observation is that typical IF policies are 2-safety~\cite{CS10}:
a violation has the form of a pair of runs, so the monitored run (\emph{major run})
need only be checked with respect to each alternate (\emph{minor run}) individually.
What needs to be checked about the
minor run is a safety property, defined in terms of the major run (and thus fully
known to the monitor only upon completion of the major run).   
This view offers a path to more sophisticated monitoring and systematic
development of monitors for real world languages and platforms, and modular machine
checked correctness proofs. 
But the paper \cite{CKN14} is devoid of Galois connections or other trappings of
abstract interpretation! It offers only a rational reconstruction of an existing
monitor for the simple while language, augmented with downgrading and
intermediate release policies.   

\begin{list}{}{}
\item[\textbf{Contribution: ideal monitor}]
  We reformulate the idea of ``tracking set'' in Chudnov et al.~\cite{CKN14}  
as a novel variation of the standard notion of collecting semantics \cite{CC77},
which serves as specification for---and basis for deriving---static analyses.  
We generalize collecting semantics to depend on the major run, 
in an \emph{ideal monitor} which we prove embodies checking of
noninterference for the major execution.
\item[\textbf{Contribution: derived monitors}]
  We derive several monitors from
  the monitoring semantics, using techniques of abstract interpretation to show
  the monitors are correct by construction. 
That is, the definitions are obtained by calculation,  disentangling routine
steps from inventive steps and design choices, inspired by Cousot~\cite{Cou99}. 
We identify two main ways in which a monitor can glean information from the major
run and the abstractly interpreted collective minor runs, accounting for
existing monitors as well as showing the way to further advances that can be made in precision and efficiency. 
\end{list}

A key abstraction used in our monitors is one for relational formulas as in~\cite{CKN14}, 
here formulated as a Galois connection.  
In the cited paper, the derived monitor exhibits ad hoc features that reflect implementation details,
e.g., simple agreement relations are represented both by taint tags on variables and by formulas.
Here we refrain from dwelling on implementation and instead explore how some
existing static analyses can be used in monitoring with little or no change. 
For example, one of our monitors uses an interval analysis, known to have good performance in practice.
Other analyses, like constant propagation could as well be incorporated, as we discuss.
As in standard static analysis, the notion of reduced
product~\cite{CC79,Gran89,CCP13} serves to share information between different
analyses, increasing their precision and efficiency.   

These first steps are a proof of principle.
In the future, solid theoretical underpinnings can enable aggressive engineering
of monitors while retaining high assurance. 

\emph{Outline:}
Sec.~\ref{sec:background} introduces the simple language used to present our
ideas, and reviews key notions from abstract interpretation, especially
collecting semantics. 
For commands we choose standard denotational semantics, because it 
facilitates streamlined notations in what follows.
Sec.~\ref{sec:monitoring} presents the \emph{ideal monitor} 
and shows how extant security policies are defined in terms of this semantics.
Sec.~\ref{sec:relform} defines a Galois connection for the lattice of relational formulas.
That connection induces a specification of a monitor that approximates the ideal monitor.
We sketch the derivation, from that specification, of a generic monitor.
Sec.~\ref{sec:derivation} derives several monitors, by refining the generic
monitor to use different abstract interpretations for the minor runs.  These
include a new purely dynamic monitor as well as improvement on prior monitoring
techniques. 
Sec.~\ref{sec:related} discusses related work.
Sec.~\ref{sec:discussion} discusses ideas for monitoring richer languages and for gaining precision by
leveraging existing static analyses.

\iftechreport
An appendix, providing detailed proofs for all results, 
can be found in the end of this technical report.
\else
A technical report, providing detailed proofs for all results, 
can be found on  the authors' web pages.
\fi





\section{Background}\label{sec:background}

\subsection{Language syntax and standard semantics}

To expose the main ideas it suffices to work with the simple imperative language 
with integer variables.
The only non-standard features are the \dt{annotation commands}, assert and assume.
These use relational formulas as in Chudnov~et~al.~\cite{CKN14}, and are
explained in due course.


\begin{tdisplay}{Program syntax}
\vspace*{-5ex}
    \begin{align*}
      e  ::= {}  & n {} \mid {} id {} \mid {} e_1 \oplus e_2 
             {} \mid {} b  \\
      b ::= {} &  e_1 < e_2 {} \mid {} e_1 = e_2 {} \mid {} \neg b {}
             \mid {} b_1 \wedge b_2 \\
      c  ::= {}  &
           {} id := e
           {} \mid {} c_1; c_2 
           {} \mid {} \ifcom~ b~ \thencom~ c_1~ \elsecom~ c_2      
           {} \mid {}  \whilecom~ b~\docom~ c  \\
           & {} \mid {} \skipcom 
             {} \mid {} \assumecom~ \Phi {} \mid {} \assertcom~ \Phi \\
    \Phi ::= {} & \agree e {} \mid {} \bagree b {} \mid {} \bagree b \imp \agree e \quad\text{(basic formulas)} \\
                & {} \mid {} \Phi, \Phi
    \end{align*}
\vspace*{-5ex}
\end{tdisplay}


Expressions are integer-valued. They include constants $n$, variables $id$, 
binary operators (indicated by $\oplus$), and boolean expressions $b$.

A \dt{state} is a mapping from variables $id$ to values $v \in \mathbb{Z}$.
For $\st \in \States$ we define the denotation $\semexp{e}{\st}$ of an expression
$e$ as usual.  
For example, $\semexp{id}{\st} \triangleq \st(id)$.
Boolean expressions evaluate to either integer 0 or 1,
e.g., 
$\semexp{\neg b}{\st}$ is 1 if $\semexp{b}{\st} = 0$.
We omit the details, which are standard, and for simplicity 
we assume that every expression has a value in every state.

We define the set of outcomes $\States_\bot \triangleq \States {} \cup {} \{\bot\}$
where $\bot$ is distinct from proper states (representing divergence).
We denote by $\preccurlyeq$ the
approximation partial order over the flat domain $\States_\bot$, 
and ${} \dotpreccurlyeq {}$ its lifting to functions over outcomes.
Therefore, the denotation of a command $c$ is a function
\[ \semins{c}{}
   \in \States_\bot \to \States_\bot
\]
For background on denotational semantics, see \cite{Win93}.

The clause $\semins{c}{\bot} \triangleq \bot$ indicates that $\semins{c}$ is $\bot$-strict for all $c$ and we ignore $\bot$ in the subsequent cases.
The annotation commands act like skip.
In the clause for if/else, we confuse 1 and 0 with truth and falsity in the metalanguage, to avoid writing $\semexp{b}{\st} = 1$ etc.
Throughout this paper, we denote the least fixpoint of a monotonic function $f
\in A \to A$ that is $\sqsubseteq$-greater than $x \in A$ by
$\operatorname{lfp}_{x}^{\sqsubseteq} f$.


\begin{tdisplay}{Standard semantics of commands \hfill $\semins{-}$}
\vspace*{-3ex}
  \begin{mathpar}
    \semins{c}{\bot} \triangleq \bot 
      
    \semins{id := e}{\st} \triangleq \st[ id \mapsto \semexp{e}{\st }]

    \semins{c_1; c_2}{\st} \triangleq \semins{c_2}{} \comp \semins{c_1}{\st}
    
    \semins{\ifcom~ b~ \thencom~ c_1~ \elsecom~ c_2}{\st} \triangleq
    \begin{cases}
      \semins{c_1}{\st} & \text{if } \semexp{b}{\st} \\
      \semins{c_2}{\st} &  \text{if } \neg \semexp{b}{\st} 
    \end{cases}\\
    
    \semins{\assumecom~\Phi}{\st} \triangleq \st

    \semins{\assertcom~\Phi}{\st} \triangleq \st

    \semins{\skipcom}{\st} \triangleq \st

    \semins{\whilecom~ b~ c}{\st}  
       \begin{array}[t]{l}
         \triangleq (\operatorname{lfp}^{\dotpreccurlyeq}_{(\lambda \st. \bot)} \mathcal{F})(\st) \\
         \text{where } \mathcal{F}(w)(\rho) \triangleq
            \begin{cases}
              \rho & \text{if } \neg \semexp{b}{\rho}  \\
              w \comp \semins{c}{\rho} & \text{otherwise}
            \end{cases}
       \end{array}

  \end{mathpar}
\vspace*{-3ex}
\end{tdisplay}


\subsection{Relational formulas}


Relational formulas relate two states.  The \dt{agreement} formula 
$\agree x$ says the two states have the same value for $x$.
In relational logics, initial and final agreements indicate which variables 
are ``low'', as in the low-indistinguishability relations used to define noninterference~\cite{AB04}.
In this paper we internalize specifications using annotation commands
as in~\cite{CKN14}.  
Given a command $c$, consider the command 
\begin{equation}\label{eq:specSandwich}
 \assumecom~\agree x,\agree y;~c;~\assertcom~\agree z 
\end{equation}
This expresses that the final value of $z$ may depend on the initial values of $x$ and $y$ but not on other variables.  If, from two states that agree on $x$ and $y$, executions of $c$ lead to different values of $z$, the assertion will fail.

Assumptions at intermediate points in the program serve to specify downgrading, similar 
to explicit code annotations in some work \cite{Mye99,jif,BDS13}.  
Although we do not model intermediate output as such, one may model an output
channel as a variable, say $out$; the policy that it is low can be specified by 
asserting agreement over what is assigned to $out$.

Relational formulas also feature a ``holds in both'' operator:
Two states $\st$ and $\stm$ satisfy $\bagree b$ iff.\ they both evaluate
the conditional expression $b$ to 1.
The third basic form is conditional agreement~\cite{AmtoftB07,CKN14},
$\bagree b \imp \agree e$, which can be used to encode multilevel security
policies as well as to encode conditional downgrading (e.g.,~\cite{BS06,BNR08}).



\begin{tdisplay}{Semantics of relational formulas \hfill $\st \shortmid \stm \models \Phi$}
\vspace*{-2ex}
  \begin{mathpar}
  \st \shortmid \stm \models \agree e \text{ iff.\ }
  \semexp{e}{\st} = \semexp{e}{\stm}

  \st \shortmid \stm \models \bagree b \text{ iff.\ }
  \semexp{b}{\st} \text{ and } \semexp{b}{\stm}

  \st \shortmid \stm \models (\bagree b \imp \agree e) 
  \text{ iff.\ } \st \shortmid \stm \models \bagree b \text{ implies }
    \st \shortmid \stm \models \agree e

  \st \shortmid \stm \models \Phi,\Psi 
  \text{ iff.\ } \st \shortmid \stm \models \Phi 
  \text{ and } \st \shortmid \stm \models \Psi 
  \end{mathpar}
\vspace*{-4ex}
\end{tdisplay}

Relational formulas are closed under conjunction, written  $\Phi,\Psi$ 
as a reminder that sometimes we abuse notation and treat a relational formula 
as a set of basic formulas.


  



\subsection{Collecting semantics and abstract interpretation}

Usually,  abstract interpretation-based static analyses introduce a
collecting semantics (also known as a static semantics~\cite[Section 4]{CC77})
aimed at 
formalizing the possible behaviours of a program wrt.\ a property of interest.
This serves as a starting point for the derivation of sound
approximate representations of program behaviours.
The collecting semantics lifts the standard semantics to apply to arbitrary sets of proper states, ignoring the $\bot$ outcome 
that indicates divergence because --- like most work on information flow
monitoring --- we aim for termination-insensitive security~\cite{BR16}. 
\begin{align}
    \seminsc{c}{} & \in \pstates \to \pstates \notag \\
    \seminsc{c}{\St} & \triangleq
    \{ \semins{c}{\st} \mid \st \in \St \mbox{ and } \semins{c}{\st}\neq\bot \} 
  \label{fig:collecting_semantics}
\end{align}
The powerset $\pstates$, with set inclusion as a partial order, is a complete lattice.
The collecting semantics can be given a direct definition that
makes explicit the fixpoint computation over a set of states, rather than relying on the
underlying fixpoint of the functional $\mathcal{F}$ used in the
standard semantics of $\whilecom$.


\begin{tdisplay}{Collecting semantics \hfill $\seminsc{-}{}$ \quad $\guardc{b}$}
\vspace*{-2ex}
  \begin{mathpar}

    \guardc{b}(\St) \triangleq \{ \stm \in \St \mid \semexp{b}{\stm} = true \}    

     \seminsc{id := e}{\St} = \{ \st[ id \mapsto \semexp{e}{\st }]  \mid \st
     \in \St \}

     \seminsc{c_1; c_2}{\St} = \seminsc{c_2}{} \comp \seminsc{c_1}{\St}
     

     \seminsc{\ifcom~ b~ \thencom~ c_1~ \elsecom~ c_2}{\St}  \; =
       \begin{array}[t]{l}
          \; \seminsc{c_1}{} \comp  \guard{b}{\St} \\
          {} \cup {} \seminsc{c_2}{} \comp  \guard{\neg b}{\St}
       \end{array}

     \seminsc{\assumecom~\Phi}{\St} = \St

     \seminsc{\assertcom~\Phi}{\St} = \St

     \seminsc{\skipcom}{\St} = \St

     \seminsc{\whilecom~ b~ c} \St =
    \guard{\neg b}{ \operatorname{lfp}_{\St}^{\subseteq} \seminsc{\ifcom~
        b~ \thencom~ c~ \elsecom~ \skipcom}{} }

  \end{mathpar}
\vspace*{-3ex}
\end{tdisplay}



     





\begin{restatable}{mylem}{lemeqstaticcollecting}
  \label{lem:eq_static_collecting}
The displayed equations define the same semantics as 
\Cref{fig:collecting_semantics}.
\end{restatable}

The proof is by structural induction on
commands.\footnote{While this
  formulation of the collecting semantics  
``is sometimes taken as \textit{standard} in
abstract interpretation works''~\cite{CP10}, Cachera~and~Pichardie~\cite{CP10}
provide the first precise proof relating it to a 
small-step operational semantics.} 


A \dt{Galois connection},
written 
\( (C;\leq) \galois{\alpha}{\gamma} (A;\sqsubseteq) \),
comprises partially ordered sets with monotonic functions $\alpha,\gamma$ 
such that 
\( \alpha(x)\sqsubseteq y \mbox{ iff.\ } x \leq \gamma(y)
\mbox{ for all }x\in C, y\in A  
\).

In case $C$ is $\pstates$ and $A$ is some lattice of abstract states,
a \dt{sound approximation} for command $c$ is 
$t\in A\to A$ such that the following holds (writing $\dot{\sqsubseteq}$ for the
pointwise lift of $\sqsubseteq$): 
\begin{equation}\label{eq:bestTrans}
 \alpha \comp \seminsc{c} \comp \gamma \mathrel{\dot{\sqsubseteq}} t 
\end{equation}
The \dt{best abstract transformer} for $c$ is
$\alpha \comp \seminsc{c} \comp \gamma$. 
It is not computable, in general, 
but it serves to specify the abstract interpretation $t$.
The idea is to derive an abstract semantics $\seminscabs{-}$ so that,   
for all $c$, $\seminscabs{c}$ is a sound approximation of $c$ and can be
implemented efficiently.


\section{Ideal Monitor}\label{sec:monitoring}


We introduce a concrete monitoring semantics which serves as basis to define the
security property by interpreting annotation commands with respect to both the
actual execution (\dt{major run}) and all possible alternatives (\dt{minor
  runs}). 
Readers familiar with Chudnov~et~al.~\cite{CKN14} may see this as a principled 
account of their notion of ``tracking set'', adapted to denotational semantics.
\Cref{sec:relform,sec:derivation} derive monitors as 
abstract interpretations of this ideal monitor.



The main difference between the collecting semantics and the ideal monitor is
that the ideal monitor is parametrised by the current state 
$\st$ of the major run -- we call this a \dt{major state}.
The ideal monitor is responsible for interpreting annotation commands in
order to track and verify the relational formulas satisfied by
all minor states $\stm$ in the tracking 
set $\St$, wrt.\ the major state $\st$.

The ideal monitor also has to signal security violations due
to assertion failures. 
We use the term \dt{fault}, denoted by $\err$.
We define $\pstateserr \triangleq 
\pstates \cup \{\err\}$.
Therefore, the ideal monitor 
\[ \seminsm{c}{}{} \in \States_\bot \to \pstateserr \to \pstateserr \]
is applied to the initial major state and 
maps an input
tracking set $\St$ to an output set $\St'$ or fault $\err$.
We also introduce only one rule
($\seminsm{c}{\st}{\err} \triangleq \err$) to mean that the ideal monitor maps
fault to fault.  

In order to use the framework of abstract interpretation, we provide the set
$\pstateserr$ with a lattice structure. 
To this end, we lift set inclusion, the natural partial order over the
powerset $\pstates$, to the set $\pstateserr$.
Therefore, we let $\err$ be the top element of the set $\pstateserr$ and we
denote by $\concleq$ the lifting of set inclusion $\subseteq$ to the set 
$\pstateserr$:
\[ \forall \St,\St' \in \pstateserr, \quad \St \concleq \St' 
\text{ iff. }
 (\St' = \err
\vee \St \subseteq \St')
\]
Let $\concjoin$ denote the lifting of set union to the set
$\pstateserr$. 


The ideal monitor relies on the collecting semantics for branching commands.
Therefore, we also lift the collecting semantics to the set
$\pstateserr$, by letting $\seminsc{c}{\err} \triangleq \err$.
This guarantees that both the ideal monitor and the
collecting semantics are monotonic.

The ideal monitor can be seen as a hybrid
monitor comprised of a dynamic part and a static part.
It directly handles the dynamic part, but 
delegates the static part to the collecting semantics.
The dynamic part consists of tracking the minor states that follow
the same execution path as the major state, whereas the static part consists in
tracking  the minor states that follow a different execution path.
The monitor semantics of conditional commands best illustrates
the intertwining between the dynamic and static part of this ideal monitor.
When the major state evaluates the conditional guard to true, the monitoring
semantics continues tracking 
all minor states that also evaluate the guard to true.
As for the minor states that evaluate the conditional guard to false, they are
propagated through the else-branch by the collecting semantics.
Notice that the monitor semantics of conditionals
ignores annotation commands in non-executed branches.
We revisit this later.


\begin{tdisplay}{Ideal monitor \hfill $\seminsm{-}{-}{}$
\quad $\guardm{b}{\st}$}
  \vspace*{-2ex}
  \begin{mathpar}
    \operatorname{snd} (\st,\St) \triangleq \St

    \guardm{\st}{b}(\St) \triangleq 
    \{ \stm \in \St \mid \semexp{b}{\stm} = true  \}     

    \seminsm{c}{\st}{\err} \triangleq \err

    \seminsm{id := e}{\st}{\St} \triangleq
    \left\{ \semins{id := e}{\stm} \mid \stm \in \St \right\}

    \seminsm{c_1; c_2}{\st}{\St} \triangleq
    \seminsm{c_2}{\semins{c_1}{\st}}{} \comp \seminsm{c_1}{\st}{\St}

     \seminsm{\skipcom}{\st}{\St} \triangleq  \St
    
    \seminsm{\assumecom~\Phi}{\st}{\St} \triangleq
    \{ \stm \in \St \mid  \quad \st \shortmid \stm \models \Phi \}

    \seminsm{\assertcom~\Phi}{\st}{\St} \triangleq
    \begin{cases}
      \St & \text{if } \forall \stm \in \St, \st \shortmid \stm \models \Phi \\
      \err & \text{otherwise}
    \end{cases}

      
    \begin{array}[t]{l}
    \seminsm{\ifcom~ b~ \thencom~ c_1~ \elsecom~ c_0}{\st}{\St}
    \triangleq \\ \quad 
    \begin{cases}
      \seminsm{c_1}{\st}{} \comp \guardm{\st}{b}{\St} 
      {} \concjoin {}
      \seminsc{c_0}{} \comp \guardc{\neg b}{\St}
      &
      \text{if } \semexp{b}{\st} \\
      \seminsc{c_1}{} \comp \guardc{b}{\St}
      {} \concjoin {}
      \seminsm{c_0}{\st}{} \comp \guardm{\st}{\neg b}{\St}
      & \text{otherwise} 
    \end{cases}
    \end{array}

    
    \seminsm{ \whilecom~ b~\docom~ c}{\st}{\St} \triangleq
    \operatorname{snd} \left(
    \left(\operatorname{lfp}_{\lambda
        (\st,\St).(\bot,\emptyset)}^{\dotpreccurlyeq \times 
      \dotconcleq} \mathcal{G} \right) (\st,\St) \right)  

    \begin{array}[t]{l}
      \mathcal{G}(w)(\st,\St)  \triangleq 
    \begin{cases}
      ( \st,
      \seminsc{\whilecom~b~\docom~c}{\St} )
      & \!\! \text{if }  \neg \semexp{b}{\st} \\
      w \left( \semins{c}{\st},
      \seminsm{ \ifcom~ b~\thencom~ c~\elsecom~ \skipcom}{\st}{\St} \right)
      & \!\! \text{otherw.}
    \end{cases}
    \end{array}

  \end{mathpar}
  \vspace*{-6ex}
\end{tdisplay}

For $\assumecom~\Phi$, the ideal monitor reduces the initial 
tracking set $\St$ to the set of minor states $\stm$ whose pairing with the
major state $\st$ satisfy the relational formula $\Phi$.
The monitor rules out all alternative executions on the same
control path that do not comply with the assumption $\Phi$.
These are termed ``assumption failures'' in~\cite{CKN14}.

For $\assertcom~\Phi$, the ideal monitor checks whether all minor states in the
initial tracking 
set $\St$ satisfy the relational formula $\Phi$ when paired with the current
major state $\st$. 
If so, the monitor returns $\St$.
Otherwise, 
the monitor concludes that one of the
alternative executions --- that satisfies all assumptions encountered so far ---
falsifies the assertion when paired with the major state.
So the semantics signals a security violation 
by returning fault $\err$.

For while loops, the ideal monitor is defined as the least fixpoint of a
functional $\mathcal{G}$ that formalises the simultaneous evaluation on both the
major state and the tracking set.
Erasing operations related to tracking sets in the functional $\mathcal{G}$
yields the functional $\mathcal{F}$ used in the standard
denotational semantics.  
Along the iteration of the major state on the loop body, the ideal monitor
iterates the tracking set on a conditional --- ensuring only minor states that
have not yet reached a fixpoint go through an additional iteration of the loop body.
When the major state reaches a fixpoint (i.e.\ $\neg \semexp{b}{\st}$), the
tracking set is fed to the collecting semantics before exiting the loop ---
ensuring all minor states reach their fixpoint.


Like the collecting semantics, the ideal monitor is defined
over concrete executions. 
Therefore, there is no loss of information, in the sense that the
ideal monitor returns a fault iff.\ there is a minor state
that falsifies the assertion.

The semantics of the ideal monitor is the concrete specification
for a more abstract monitor whose transfer functions are computable.
We will rely on the framework of abstract interpretation in order
to derive sound monitors by approximating both the concrete collecting
semantics and the ideal monitor.

The ideal monitor and the collecting semantics are
equivalent for annotation-free commands, as long as the major run terminates.

\begin{restatable}[
  ]{mylem}{lemmonstaticeq} 
  \label{lem:mon_static_eq}
For all annotation-free commands $c$, all $\st \in
\States$ such that $\semins{c}{\st} \neq \bot$, and all sets $\St \subseteq
\States$, it holds that 
\[ \seminsm{c}{\st}{\St} = \seminsc{c}{\St}\]
\end{restatable}
In fact $\seminsm{c}{\st}{\St} \subseteq \seminsc{c}{\St}$ as long as $c$ 
is assertion-free.
The proof is by structural induction.

If the major run diverges, the
least fixpoint of the functional 
$\mathcal{G}$ yields an undefined outcome $\bot$, paired with an empty set
$\emptyset$ of minor states.


Assuming a set $in$ of variables considered  low inputs, 
and a set $out$ considered  low outputs, let us use the notation
$\agree in$ and $\agree out$ to abbreviate the conjunction of 
basic agreements for these variables.
\Cref{theo:mon_collecting_sound} states that,
for $c$ annotated following the pattern of \Cref{eq:specSandwich},
the ideal monitor,
parametrised with a major state $\st_1$,
does not result in a fault $\err$ iff.\ 
the standard notion of termination-insensitive noninterference (\dt{TINI}) holds
for $c$ and $\st_1$. 

\begin{restatable}[%
]{mytheo}{theomoncollectingsound}
  \label{theo:mon_collecting_sound}
   Let $in$ and $out$ be two sets of variables and $c$ be an
   annotation-free command.   
   Let command $\hat{c}$ be defined as 
   $\assumecom~\agree in;~c~; \assertcom~\agree out$.
   For all
  $\st_1,\st_1' \in \States$ such that $\semins{c}{\st_1} = \st_1'$,
  we have $\seminsm{\hat{c}}{\st_1}{\States} \neq \err$ iff.\
  \[ \forall \st_2,\st_2' \in \States, 
     \semins{c}{\st_2}=\st_2' \mathrel{\land}
          \st_1 =_{in} \st_2 \mathord{\implies} \st_1' =_{out} \st_2' \]
\end{restatable}
\noindent Here we write $=_{in}$ to indicate agreement on the variables $in$.

The security property in \cite{CKN14} allows intermediate annotations, but disallows them in high branches,
to ensure robustness of declassification etc.
In an Appendix we provide an alternative ideal monitor and an alternative
collecting semantics, both of which fault $\err$ if an assertion or assumption
occurs in a \dt{high conditional} --- one for which some minor states do not agree with the major state on its conditional guard.

\textbf{Strong conjecture:} for terminating executions, the security property in
\cite{CKN14} holds iff.\ the alternative ideal monitor does not fault.

The alternative  ideal monitor is a simple variation, but with the notational
complication of threading an additional parameter through the definitions. So we
do not use it in the body of the paper.   
However, for each of the derived monitors, there is a very similar one derived
from the alternate ideal monitor, and therefore sound with respect to the
security property in \cite{CKN14}. 
 
\bigskip

In summary, the idea is that from an initial state $\st$,
a monitored execution evaluates $\semins{c}{\st}$ and in parallel 
should evaluate an abstraction of $\seminsm{c}{\st}{}$, written
$\seminsmabs{c}{\st}{}$.  
The monitored execution yields $\st'$, with $\semins{c}{\st} = \st'$, if it can
guarantee that $\seminsmabs{c}{\st}{\States} \neq \err$. 
Otherwise, there is a potential security violation.
This parallel evaluation can be formalized, as it is 
in the functional $\mathcal{G}$ for the monitor semantics of loops, 
but to streamline notation in the rest of the paper 
we focus on what is returned by the ideal monitor.


\section{Lattice of Relational Formulas}\label{sec:relform}
The derived monitors use abstract interpretations based on relational formulas.
This section defines the abstraction and uses it to derive a generic
monitor that is refined in \Cref{sec:derivation}.

First we define the lattice of relational formulas.
To make it finite, expressions in relational formulas are restricted 
to those that occur in the program to be monitored, 
as well as their negations to facilitate precision in the monitors we derive.

A set of formulas is interpreted conjunctively.  To streamline notation we
confuse a conjunctive formula, say ``$\agree x, \agree y$'', with the set $\{
\agree x, \agree y \}$. 
That is why the lattice is defined in terms of basic formulas.

\begin{tdisplay}{Lattice of relational formulas \hfill
$\relform$ \quad $\prelformerr$}
\textbf{Assumption:} for a given command $c$, 
let $\relform$ be a finite set of basic relational formulas 
that is closed under negation of boolean expressions 
and which contains at least $\bagree b$, $\agree e$, and
$\bagree b\imp \agree e$ for every $b$ and $e$ that
occur in $c$.

We use the powerset $\prelform$ as a lattice ordered by $\supseteq$ with
$\emptyset$ on top and intersection as join:
  \( (\prelform;\supseteq,\mathcal{L},\emptyset,\cap,\cup)\).
Let $\prelformerr$ be $\prelform \cup \{\err\}$.
Let $\absleq$ be the lifting of the 
  partial order $\supseteq$ such that $\err$ 
  is the top element of the lattice $\prelformerr$:
  \[ (\prelformerr;\absleq,\mathcal{L},\err,\absjoin,\absmeet)\]
We also let $\absjoin$ (resp. $\absmeet$) denote the lifting of set intersection
$\cap$ (resp. the lifting of set union $\cup$) to the lattice $\prelformerr$.
\end{tdisplay}

The notation elides dependence of $\relform$ on $c$ because $c$ should 
be the main program to be monitored; a fixed $\relform$ will be 
used in the context of the monitor semantics which is recursively
applied to sub-programs of $c$.



The monitor will maintain an over-approximation of the relational
formulas satisfied by the major state $\st$ and every minor state
$\stm$ of interest.  A set of formulas is interpreted to mean 
all the formulas hold for every such pair $(\st,\stm)$.
The empty set indicates no relations are known, whereas $\err$ 
serves to indicate that some required relation fails to hold.

Given a major state $\st$, we aim to define an approximation of a tracking set
$\St$, in order to account for relational formulas satisfied by the major state
$\st$ and each minor state $\stm \in \St$.
Subsequently, we lift this abstraction in order to approximate the
monitoring collecting semantics and obtain sound computable abstract
transfer functions of a monitor tracking relational formulas.
We formalise this abstraction of the tracking set $\St$ as a
function $\alphast$ that is parametrised by a state $\st$:\looseness=-1
\begin{align*}
  \alphast & \in  \pstateserr \to \prelformerr \\
  \alphast(\St) & \triangleq
  \begin{cases}
    \err & \text{if } \St = \err \\
    \{ \Phi \mid \forall \stm \in \St,
  \st \shortmid \stm \models \Phi \} & \text{otherwise}
  \end{cases}
\end{align*}
The associated concretisation function $\gammast$ is also parametrised by a
major state $\st$.
The concretisation of a set $\Delta \in \pset{\mathcal{L}}$ of
relational formulas yields a set $\St$ of minor states, such that every minor
state $\stm \in \St$ and the major state $\st$ satisfy all relational formulas
$\Phi \in \Delta$. 
\begin{align*}
  \gammast & \in \prelformerr \to \pstateserr \\
  \gammast(\Delta) & \triangleq
  \begin{cases}
    \err & \text{if } \Delta = \err \\
  \{ \stm \in \States \mid \forall \Phi \in \Delta, \st \shortmid \stm \models
  \Phi \} & \text{otherwise}
  \end{cases}
\end{align*}


\begin{restatable}[
  ]{mylem}{lemalphastgammast}
  \label{lem:alphastgammast}
  For all $\st \in \States$, the pair $(\alphast,\gammast)$ is a
  Galois connection:  
\(
(\pstateserr;\concleq)
\galois{\alphast}{\gammast}
(\prelformerr;\absleq) 
\).
That is, $\forall \St \in \pstateserr, \forall \Delta \in \prelformerr$:
\[
  \alphast(\St) \concleq \Delta 
  \iff 
  \St \absleq \gammast(\Delta)
\]
\end{restatable}

\Cref{fig:best_transformer} illustrates the best abstraction of the monitoring
collecting semantics $\seminsm{c}{\st}{}$.
If a set $\St$ of minor states is abstracted wrt.\ a major state
$\st$ by a set $\Delta$ of relational formulas
($\St \concleq \gammares{\st}(\Delta)$), then the resulting set
$\St' = \seminsm{c}{\st}{\St}$ of minor states is abstracted wrt.\ the
resulting major state $\st' = \semins{c}{\st}$ by the set
$\alphares{\st'} \comp \seminsm{c}{\st}{} \comp \gammares{\st}(\Delta)$ of
relational formulas $(\St' \concleq \gammares{\st'} ( \alphares{\st'} \comp
\seminsm{c}{\st}{} \comp \gammares{\st}(\Delta) ))$.  
\begin{figure}[hbp]
  \centering
\iftechreport
   \includegraphics{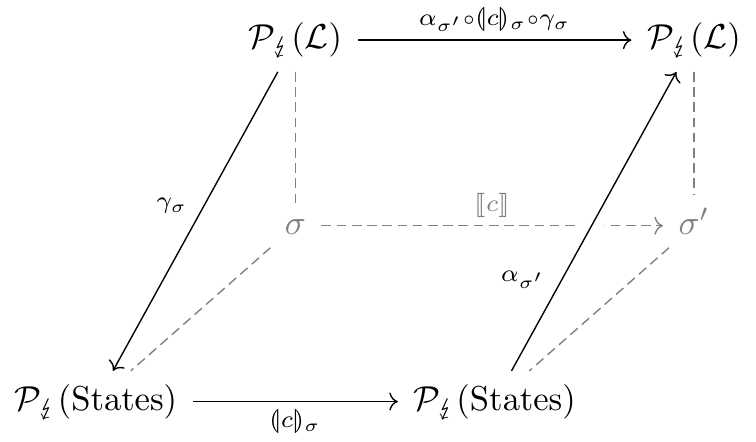}
\else
  \scalebox{0.89}{
  \begin{tikzcd}[ampersand replacement=\&,row sep=huge,column sep=small]
    {} \&
    \prelformerr \arrow[ddl,swap,"\gammares{\st}"]
    \arrow{rr}{\alphares{\st'} \comp \seminsm{c}{\st}{} \comp \gammares{\st}}
    \arrow[d,dash,dashed,color=gray]
    \&
    \&
    \prelformerr
    \arrow[d,dash,dashed,color=gray]
    \\
    \&
    \color{gray}{\st} \arrow[rr,dashed,"{\semins{c}{}}",color=gray]
    \arrow[dl,dash,dashed,color=gray]
    \&
    \&
    \color{gray}{\st'}
    \arrow[dl,dash,dashed,color=gray]
    \\
    \pstateserr \arrow{rr}[swap]{\seminsm{c}{\st}{}}
    \&
    \&
    \pstateserr \arrow[uur,crossing over,"\alphares{\st'}"' near start,swap] 
    \& {}
  \end{tikzcd}}
\fi
  \caption{Best abstract transformer}
  \label{fig:best_transformer}
\end{figure}
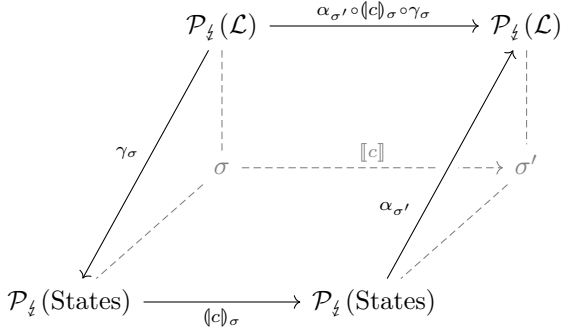
This best abstract transformer is not computable in general.
An abstract interpretation $\seminsmabs{c}{\st}{}$ 
is \dt{sound} if it satisfies the following condition:
\begin{equation}
  \label{eq:sound_abs_mon}
\alphares{\st'} \comp \seminsm{c}{\st}{} \comp \gammast 
\dotabsleq
\seminsmabs{c}{\st}{}
\quad\mbox{where } \st'=\semins{c}{\st}
\end{equation}
Note that we denote by $\dotconcleq$  (resp. $\dotabsleq$)  the pointwise
lifting to functions of the partial order $\concleq$ over $\pstateserr$
(resp. of the partial order $\absleq$ over $\prelformerr$).

\begin{restatable}[Soundness conditions]{mylem}{lemsoundabsmon}
  \label{lem:soundabsmon}
  Consider any $c,\st,\st'$ such that $\st' =\semins{c}{\st}$.
  \Cref{eq:sound_abs_mon} is equivalent to each of the following:
  \[ 
     \seminsm{c}{\st}{} \comp \gammares{\st} {} \dotconcleq  {} \gammares{\st'}
     \comp \seminsmabs{c}{\st}{}  
     \quad\mbox{and}\quad
     \alphares{\st'} \comp \seminsm{c}{\st}{} {} \dotabsleq {}
     \seminsmabs{c}{\st}{} \comp \alphares{\st}   
  \]
\end{restatable}

In the process of deriving a sound abstract monitoring semantics
$\seminsmabs{c}{\st}{}$ approximating the monitoring collecting semantics 
$\seminsm{c}{\st}{}$, we also have to derive a sound abstract static semantics 
$\seminscabs{c}{} \in \prelformerr \to \prelformerr$ approximating the static
collecting semantics 
$\seminsc{c}{} \in \pstateserr \to \pstateserr$.


\Cref{eq:bestTrans} provides a notion of soundness for static semantics, whereas
\Cref{eq:sound_abs_mon} provides a notion of soundness for monitoring
semantics. 
While approximating the monitoring collecting semantics, we will find good ways
for the abstract static and monitoring semantics to interact. 
In particular, the abstract static analyses we propose will account for that
interaction by additional parameters.
Additionally, we will also prove soundness results in
\Cref{lem:soundabsstaticpured,lem:soundabsstatichybrid,lem:soundabsstaticinterval}
that embody not only the soundness condition of \Cref{eq:bestTrans}, but
also variations on that property that take into account this interaction
between the dynamic and static analyses.

The derivation of a sound abstract monitoring semantics is by structural
induction over commands. 
As an example, consider the case of a conditional command
$c \triangleq \ifcom~b~\thencom~c_1~\elsecom~c_0$, an initial state $\st$ and a
final state $\st'$ such that $\st'=\semins{c}{\st}$. 
Let us consider the case where the guard evaluates to true ($\semexp{b}{\st} =
1$), so that $\st' = \semins{c_1}{\st}$.
Then, assuming a sound approximation $\seminscabs{c}{}$ of the
static collecting semantics, we can derive a generic abstract monitoring
semantics by successive approximation, beginning as follows:
\begin{align}
    & \alphares{\st'} \comp
  \seminsm{\ifcom~ b~ \thencom~ c_1~ \elsecom~ c_0}{\st}{}
  \comp \gammast(\Delta)  \notag \\ 
  & = \alphares{\st'}\big(
  \seminsm{c_1}{\st}{} \comp \guardm{\st}{b}{} \comp \gammast(\Delta) {}
  \concjoin {} 
  \seminsc{c_0}{} \comp \guardc{\neg b}{} \comp \gammast(\Delta) \big)
  \notag\\
  & =  \text{$\Lbag$Since $\alphares{\st'}$ is additive:
    $\alphares{\st'}(\St \cup \St') = \alphares{\st'}(\St) \absjoin
    \alphares{\st'}(\St')\Rbag$} \notag\\
  & \qquad
  \alphares{\st'} \comp
  \seminsm{c_1}{\st}{} \comp \guardm{\st}{b}{} \comp \gammast(\Delta) {}
  \absjoin {} \notag\\ 
  & \qquad\qquad
  \alphares{\st'} \comp
  \seminsc{c_0}{} \comp \guardc{\neg b}{} \comp \gammast(\Delta) \notag\\
  & \absleq \text{$\Lbag\alphares{\st'}$, $\seminsm{c_1}{\st}{}$, and
    $\seminsc{c_1}{}$ are monotone, $\gammast \comp \alphast$ is} \notag\\
  &  \quad \text{extensive:
    $\lambda \St. \St \dotconcleq \gammast \comp \alphast\Rbag$} \notag\\
  & \qquad
  \alphares{\st'} \comp
  \seminsm{c_1}{\st}{} \comp \gammast \comp \alphast \comp
  \guardm{\st}{b}{} \comp \gammast(\Delta) {} \absjoin {} \notag\\
  & \qquad\qquad
  \alphares{\st'} \comp
  \seminsc{c_0}{} \comp \gammast \comp \alphast \comp
  \guardc{\neg b}{} \comp \gammast(\Delta) \notag\\
  &  \absleq \text{$\Lbag\seminsmabs{c_1}{\st}$ is sound
    by induction hypothesis of  \Cref{eq:sound_abs_mon}$\Rbag$} \notag\\
  & \qquad 
  \seminsmabs{c_1}{\st}{} \comp \alphast \comp \guardm{\st}{b}{} \comp
  \gammast(\Delta) {} \absjoin {} 
  \notag\\ 
  & \qquad\qquad
  \alphares{\st'} \comp
  \seminsc{c_0}{} \comp
   \gammast \comp \alphast \comp
   \guardc{\neg b}{} \comp \gammast(\Delta) \notag\\
   & \absleq \text{$\Lbag$The static analysis is sound: $\alphares{\st'} \comp
     \seminsc{c_0}{} \comp 
     \gammast \dotabsleq \seminscabs{c_0}{}\Rbag$} \notag\\
  & \qquad 
  \seminsmabs{c_1}{\st}{} \comp \alphast \comp \guardm{\st}{b}{} \comp
  \gammast(\Delta) {} \absjoin {} 
  \notag\\ 
  & \qquad\qquad
   \seminscabs{c_0}{} \comp \alphast \comp
   \guardc{\neg b}{} \comp \gammast(\Delta) \label{eq:gen_approx_cond}
\end{align}
The above derivation is a routine use of abstract interpretation techniques.
The last step uses a soundness property for the static part of the monitor that is
similar to \Cref{eq:bestTrans}.
So far, we relied on an approximation of  the monitoring collecting semantics of
command $c_1$ and an approximation of the static
collecting semantics of command $c_0$. 
To continue this generic derivation, we need to derive a sound approximation for
the guard operators $\guardm{\st}{b}$ and $\guardc{\neg b}$.

The approximation of $\guardm{\st}{b}$ proceeds as follows:
\begin{align}
  & \alphast \comp \guardm{\st}{b}{} \comp \gammast(\Delta) \notag\\
  &  = \alphast( \{ \stm \in \gammast(\Delta) \mid \semexp{b}{\stm} = 1 \} )
  \notag\\ 
  &  = \text{$\Lbag$since the major state evaluates $b$ to true$\Rbag$} 
  \notag\\ 
  &   \qquad
  \alphast( \{ \gammast(\Delta) \cap \{ \stm \in \States \mid \st
  \shortmid \stm \models \bagree b \} ) \notag \\
  & = 
    \alphast \left( \{ \gammast(\Delta) \cap \gammast(\{\bagree b \}) \right)
    \notag\\ 
  & =
  \text{$\Lbag\gammast$ is multiplicative: $\gammast(\Delta \absmeet
    \Delta') = \gammast(\Delta) \concmeet \gammast(\Delta') \Rbag$} \notag\\ 
  & \qquad
  \alphast \comp \gammast(\Delta \absmeet \{ \bagree b\})  \notag\\
  & \absleq \text{$\Lbag\alphast \comp \gammast$ is reductive:
    $ \alphast\comp\gammast \dotabsleq \lambda \Delta. \Delta \Rbag$} \notag\\
  & \qquad \Delta \absmeet \{ \bagree b\} \label{eq:approx_guardm}
\end{align}

As for the approximation of $\guardc{b}{}$, we have:
\begin{align}
    \alphast \comp \guardc{b}{} \comp \gammast(\Delta) 
  & \absleq \text{$\Lbag\alphast$ is monotone and $\guardc{b}{} \dotconcleq
    \lambda \St.\St \Rbag$} \notag
  \\ 
  & \qquad \alphast \comp \gammast(\Delta) \notag \\
  & \absleq \text{$\Lbag$since $\alphast \comp \gammast$ is reductive$\Rbag$}
  \notag\\ 
  & \qquad \Delta \label{eq:approx_guardc}
\end{align}
To sum up, we rewrite the approximations of operators $\guardm{\st}{b}{}$ and
$\guardc{b}{}$ of 
\Cref{eq:approx_guardm,eq:approx_guardc} in the intermediate 
abstraction of conditionals obtained in \Cref{eq:gen_approx_cond}, 
so we can conclude:
\begin{align}
    & \alphares{\st'} \comp
  \seminsm{\ifcom~ b~ \thencom~ c_1~ \elsecom~ c_0}{\st}{}
  \comp \gammast(\Delta)  \notag \\
  & \quad \absleq \left(
  \seminsmabs{c_1}{\st}{} (\Delta \absmeet \{ \bagree b \}) \right)
    {} \absjoin {} 
   \left( \seminscabs{c_0}{} \Delta \right)  
   \label{eq:gen_monitor_cond}
\end{align}
This equation yields one of the generic approximations of conditional
commands.
It says that when the conditional guard is true, monitor the then branch and
statically analyse the else branch.\looseness=-1 

In the following section, we specialise this generic approximation by relying on
different approximations of the static collecting semantics.


Given the importance of relational formulas as an abstraction, it should be 
no surprise that some monitors rely on entailments among formulas.
Precision can be improved by providing the monitor with strong means of
logical deduction, but there is a cost in performance.  
This engineering trade-off can be left open, by just specifying what we need.
We use the notation $\Delta\entail\Phi$, which is either true or false, as follows (cf.~\cite{CKN14}).


\begin{tdisplay}{Approximate entailment \hfill $\Delta\entail\Phi$}
\textbf{Assumption:} For any $\Delta,\Phi$, if $\Delta\entail\Phi$
then the implication is valid.  That is,
for all $\st,\stm \in \States$, if 
$\st\shortmid\stm\models\Delta$ then $\st\shortmid\stm\models\Phi$. 
\end{tdisplay}


\section{Monitor Derivation}\label{sec:derivation}

We derive three different monitors from the ideal monitoring semantics. 
By relying on the framework of abstract interpretation, these monitors are
correct by construction.
Unless it is clear from context, we will differentiate the three abstract semantics
of these monitors by prefixing them with the letters $\pured$, $\hybmod$ and
$\hybint$.

These three monitors illustrate different ways of reasoning about relational
formulas in high branching commands.
The first one treats high branching commands pessimistically by
forgetting all known formulas.
The second one relies on an approximation of the modified variables in order to
determine which relational formulas cannot be falsified, a usual technique in
hybrid monitors. 
The third one deduces new relational formulas by comparing the results of an
interval value analysis to the current values in the monitored execution.

\subsection{Purely-Dynamic Monitor}\label{sec:derivation_pured}

Let us start by deriving a purely-dynamic monitor tracking relational
formulas.
This monitor is an instance of EM mechanisms~\cite{Sch00}.
Thus, it must observe only the execution steps of the major state.
In particular, this monitor should not look aside~\cite{RS10}, meaning that it
does not rely on information about minor states
that do not follow the same execution path as the major state.
This scenario corresponds to approximating the collecting
semantics $\seminsc{c}{}$ by an abstract static semantics
$\pured\seminscabs{c}{}$ that returns the top element of the lattice
$\prelform$, providing no information about minor states that do not follow the same
execution path as the major state. 

\Cref{def:seminscabs_purely_dynamic} introduces such an abstract static
semantics.
We ensure it is strict by mapping the bottom element $\relform$ of the
lattice $\prelform$ to itself. 




\begin{restatable}[
]{mydef}{defabsstaticpured}
  \label{def:seminscabs_purely_dynamic}
  The abstract static semantics $\pured\seminscabs{c}{}$ for a
  purely-dynamic monitor is given by: 
\begin{align*}
  \pured\seminscabs{c}{} & \in \prelformerr \to \prelformerr \\
  \pured\seminscabs{c}{\Delta} & \triangleq
  \begin{cases}
    \err & \text{if } \Delta = \err \\
    \relform & \text{if } \Delta = \relform \\
    \emptyset & \text{otherwise } 
  \end{cases}
\end{align*}
\end{restatable}



\begin{restatable}[
soundness of $\pured\protect\seminscabs{-}{}$
  ]{mylem}{lemsoundabsstaticpured}  
  \label{lem:soundabsstaticpured}
 For all $c$, $\st,\st' \in \States$, it holds that:
 \(\alphares{\st'} \comp \seminsc{c}{} \comp \gammast
   \dotabsleq \pured\seminscabs{c}{}.
\)
\end{restatable}
Mapping $\relform$ to $\relform$ is sound 
because the bottom element $\relform$ is concretised to the empty set of states
(since $\relform$ is closed under negation, thus it contains $\bagree b$ and
$\bagree \neg b$ for some $b$).  

\iftechreport
A detailed derivation of the purely dynamic monitor, as well as subsequent
derivations, are found in the Appendix.
\else
A detailed derivation of the purely dynamic monitor, as well as subsequent
derivations, are found in the accompanying technical report.
\fi

\Cref{fig:seminsmabs_purely_dynamic} presents the derived monitor.
In it we use the notation $\Delta\entail\Phi$ for the approximate
implication specified  at the end of \Cref{sec:relform}.
The semantics for conditional commands is obtained
from the generic derivation for conditionals, presented in \Cref{eq:gen_monitor_cond},
by unfolding the definition of the static analysis $\pured\seminscabs{c}{}$.


\begin{figure*}[thp]
  \fbox{
    \begin{mathpar}
  \seminsmabs{c}{\st}{\err} \triangleq \err
  
  \seminsmabs{\skipcom}{\st}{\Delta} \triangleq \Delta
  
  \seminsmabs{c_1; c_2}{\st}{\Delta} \triangleq
  \seminsmabs{c_2}{\semins{c_1}{\st}}{} \comp \seminsmabs{c_1}{\st}{\Delta}  
 
  \seminsmabs{ id := e}{\st}{\Delta} \triangleq 
  \{ \Phi \in \Delta \mid id \not \in \fv{\Phi} \}  \absmeet 
  \begin{cases}
    {\agree id} & \text{if } \Delta \entail \agree e \\
    \emptyset & \text{otherwise}
  \end{cases}
  
  \seminsmabs{\assumecom~\Phi}{\st}{\Delta} \triangleq \Delta \absmeet \{ \Phi \}

  \seminsmabs{\assertcom~\Phi}{\st}{\Delta} \triangleq
  \begin{cases}
    \Delta \absmeet \{ \Phi \} & \text{if } \Delta \entail \Phi \\
    \err & \text{otherwise}
  \end{cases}
  
  \seminsmabs{ \ifcom~ b~ \thencom~ c_1~ \elsecom~ c_0}{\st}{\Delta}
  \triangleq
  \begin{cases}
   \begin{cases}
    \seminsmabs{c_1}{\st}{(\Delta \absmeet \{\bagree b \})} & \text{if } \Delta
    \entail \agree b \\
    \emptyset & \text{otherwise}
   \end{cases} & \text{if $\semexp{b}{\st}$
   } \\
   \begin{cases}
     \seminsmabs{c_0}{\st}{(\Delta \absmeet \{\bagree \neg b \})} &
     \text{if } \Delta \entail \agree b \\
    \emptyset & \text{otherwise}
   \end{cases}
   & \text{if $\neg \semexp{b}{\st}$ 
   }
  \end{cases}

  \seminsmabs{\whilecom~ b~\docom~ c}{\st}{\Delta} 
    \begin{array}[t]{l}
      \triangleq
      \operatorname{snd} \left(
          (\operatorname{lfp}_{\lambda
            (\st,\Delta).(\bot,\relform)}^{\dotpreccurlyeq \times \dotabsleq}  
          \mathcal{G}^\sharp)  (\st,\Delta)
          \right)
      \\ 
      \text{where }
     \mathcal{G}^\sharp(w^\sharp)(\st,\Delta) \triangleq 
          \begin{cases}
         \st, \err
       \qquad \qquad \qquad \text{ if }  \neg \semexp{b}{\st} \text{ and } \Delta =
       \err \\ 
       \st, \Delta \absmeet \{ \bagree \neg b \}
       \quad \quad \text{if }  \neg \semexp{b}{\st} \text{ and } \Delta \entail \agree b
       \\
       \st, \{ \bagree \neg b \}
       \qquad \qquad \text{ otherwise if }  \neg \semexp{b}{\st} \\
      w^\sharp  
        \big( \semins{c}{\st}, 
        \seminsmabs{ \ifcom~ b~\thencom~c~\elsecom~
          \skipcom}{\st}{\Delta}
        \big) 
      \quad \text{ otherwise}
       \end{cases}
    \end{array}
\end{mathpar}}
\caption{Purely-dynamic monitor
$\pured\seminsmabs{-}{-}{}$,
as abstract semantics derived from the ideal monitor}
\label{fig:seminsmabs_purely_dynamic}
\end{figure*}

The abstract monitoring semantics of any command maps fault to fault
($\seminsmabs{c}{\st}{\err} \triangleq \err$). 
For assignments $id:=e$, 
the monitor invalidates all relational formulas that involve variable $id$. 
It also deduces that all resulting minor states agree with the
major state on variable $id$, if $\Delta \entail \agree e$.
For a sequence $c_1;c_2$,
the initial state $\st$ is used to monitor $c_1$, 
and then the resulting major state
$\semins{c_1}{\st}$ is used to monitor $c_2$. 

For assumptions, the  monitor adds the assumed
formula to the input set.
For asserts, the monitor returns fault $\err$ if it
cannot determine that the known $\Delta$ implies the asserted $\Phi$.

For conditionals, if the monitor is unable to determine that all minor states
agree with the major state on the value of the conditional guard, 
it must treat it as a ``high conditional''.
So the purely-dynamic monitor conservatively forgets all known relational formulas.

Similarly to the standard denotational semantics and the
ideal monitor semantics, the
abstract monitoring semantics of loops is defined as a fixpoint of an abstract
functional $\mathcal{G}^\sharp$.
This abstract fixpoint behaves as a finite sequence of conditionals.
To each iteration of the major state through the loop body corresponds a
simultaneous iteration of the monitor on a conditional.
It is important that the monitor treats each iteration
of the loop body as a conditional, in order to soundly track formulas satisfied
by minor states that exit the loop before the major state.
When the major state exits the loop, the monitor relies
on the static analysis in order to account for minor states that may continue
iterating through the loop. 


\begin{restatable}[
]{mytheo}{lemsoundabsmonitorpured}
  \label{lem:soundabsmonitorpured}
  The abstract monitoring semantics $\pured\seminsmabs{c}{\st}{}$ of the
  purely-dynamic monitor is sound.
  For all commands $c$, for all $\st, \st' \in \States$ such that
  $\st' = \semins{c}{\st}$, it holds that:
  \[ \alphares{\st'} \comp \seminsm{c}{\st}{} \comp \gammast \dotabsleq  
\pured\seminsmabs{c}{\st}{} \] 
\end{restatable}

\begin{lstlisting}[label={lst:nsu}, style=simple, 
caption={Example program},float=tp,firstnumber=0]
// @$\lstmathcomment{\st_0 \triangleq [\text{secret} \to 1; \text{public} \to 0], \Delta_0 = \emptyset}$@
assume @$\agree \text{public}$@; // @$\lstmathcomment{[\text{secret} \to 1; \text{public} \to 0], \{\agree \text{public}\}}$@
if (secret > 0) then {
  public := public + 1;
} // @$\lstmathcomment{[\text{secret} \to 1; \text{public} \to 1],}$@ @$\lstmathcomment{\{\agree \text{public}, \bagree (\text{secret} > 0)\}}$@
else {
  skip;
} // @$\lstmathcomment{[\text{secret} \to 1; \text{public} \to 1],\emptyset}$@
y := 0; // @$\lstmathcomment{[\text{secret} \to 1; \text{public} \to 1, \text{y}\to 0], \{ \agree y \}}$@
assert @$\agree \text{y}$@; @$\lstmathcomment{[\text{secret} \to 1; \text{public} \to 1, \text{y}\to 0], \{ \agree y \}}$@
\end{lstlisting}
Similarly to the No-Sensitive Upgrade (\dt{NSU}) approach~\cite[Section
3.2]{Zda02}\cite{AF09}, the purely-dynamic monitor of
\Cref{fig:seminsmabs_purely_dynamic}  
is relatively efficient, at the cost of precision (i.e., it rejects many secure executions). 
However, notice that both purely-dynamic approaches are incomparable. 
Indeed, consider for instance the program in \Cref{lst:nsu} where only variable
{\ttfamily secret} is high.
An NSU approach would stop the program when the assignment to public is
executed, signaling a possible security violation.
In contrast, our purely dynamic monitor simply forgets all known relational
formulas at the merge point of the conditional;
in the case of simple agreements, this is tantamount to labelling all variables
as high. 
Thus, at the assignment to variable {\ttfamily y}, our monitor deduces that all
minor states agree with the major state on the value of {\ttfamily y}, which
means that no security violation is raised since the relational assertion is
satisfied.
Notice that if this example program asserts $\agree \text{{\ttfamily public}}$
instead of $\agree \text{{\ttfamily y}}$, our monitor would always signal a
violation $\err$, whereas a NSU approach would not signal a fault when
the conditional guard evaluates to false ($\text{\ttfamily secret} \leq 0$).

\subsection{Hybrid Monitor with the Modified Variables}
\label{sec:derivation_hybmod}

To achieve more precision, we need non-trivial static analysis to provide
information about minor runs 
that do not follow the same execution path as the major run (``high branches'').
The next monitor relies on an over-approximation of
variables that are modified by a command.

\begin{tdisplay}{Modifiable variables \hfill $\Mod$}
\textbf{Assumption:}
$\Mod \in Com \to \pset{Var}$ 
satisfies the following: for all $c,id$,
if there is $\st$ with 
$\semins{c}{\st}\neq \bot$ and $\semins{c}{\st}(id) \neq \st(id)$ then $id \in
\Mod(c)$. 
\end{tdisplay}

For our simple language, the obvious implementation is to return the set of assignment targets.

Instead of conservatively forgetting about all known relational formulas
at the merge point of high conditionals, the monitor will be able to retain
information about variables that are modified in neither conditional branches,
similarly to existing hybrid monitors~\cite{Lal.07,RS10}.\looseness=-1

Guided by the soundness condition, we derive
an abstract static semantics that leverages 
modified variables.

\begin{restatable}[]{mydef}{defabsstatichybrid}
  \label{def:seminscabs_hybrid}
  For all commands $c$ and $c'$,
  the abstract static semantics $\hybmod\seminscabshybmod{c}{}{c'}$ 
for modified variables is
  given by: 
\begin{align*}
  \hybmod\seminscabshybmod{c}{}{c'} & \in \prelformerr \to \prelformerr \\
  \hybmod\seminscabshybmod{c}{\Delta}{c'} & \triangleq
  \begin{cases}
    \err & \text{if } \Delta = \err \\
    \mathcal{L} & \text{if } \Delta = \mathcal{L} \\
    \big\{ \Phi \in \Delta \mid \forall id \in \fv{\Phi},  & \\
     \quad  id \not \in \Mod(c) \cup
    \Mod(c') \big\} & \text{otherwise }  
  \end{cases}
\end{align*}
\end{restatable}

\begin{figure*}[htp]
  \fbox{
\begin{mathpar}
  
  
  


  \begin{array}{l}
  \seminsmabs{ \ifcom~ b~ \thencom~ c_1~ \elsecom~ c_0}{\st}{\Delta}
  \triangleq \\
  \qquad
  \begin{cases}
     \begin{cases}
      \seminsmabs{c_1}{\st}{(\Delta \absmeet \{ \bagree b \})} & \text{if } \Delta 
      \entail \agree b \\
      \seminsmabs{c_1}{\st}{(\Delta \absmeet \{ \bagree b \})}
                 \absjoin
      \{ \Phi \in \Delta \mid \fv{\Phi} \cap
       ( \Mod(c_1) \cup
      \Mod(c_0)) = \emptyset \}
      & \text{otherwise} 
   \end{cases} & \text{if $\semexp{b}{\st}$
   } \\
   \begin{cases}
    \seminsmabs{c_0}{\st}{(\Delta \absmeet \{ \bagree \neg b \})} & \text{if }
    \Delta  \entail
      \agree b \\
      \seminsmabs{c_0}{\st}{(\Delta \absmeet \{ \bagree \neg b \})}
                \absjoin
      \{ \Phi \in \Delta \mid \fv{\Phi} \cap
       ( \Mod(c_1) \cup
      \Mod(c_0)) = \emptyset \}
      & \text{otherwise}
   \end{cases}
   & \text{if $\neg \semexp{b}{\st}$
   }
  \end{cases}
  \end{array}

  \seminsmabs{\whilecom~ b~\docom~ c}{\st}{\Delta} \triangleq
       \operatorname{snd} \left(
     (\operatorname{lfp}_{\lambda (\st,\Delta).(\bot,\relform)}^{\dotpreccurlyeq
        \times \dotabsleq} 
       \mathcal{G}^\sharp)  (\st,\Delta) \right) 

       
     \mathcal{G}^\sharp(w^\sharp)(\st,\Delta) \triangleq 
       \begin{cases}
         \st, \err
       & \text{ if }  \neg \semexp{b}{\st} \text{ and } \Delta = \err \\
       \st, \Delta \absmeet \{ \bagree \neg b \}
        & \text{ if }  \neg \semexp{b}{\st} \text{ and }
        \Delta \entail \agree b 
       \\ 
       \st, 
       \big( \{ \Phi \in \Delta \mid \fv{\Phi} \cap \Mod(c)=  \emptyset \}
           {} \absjoin {}
           \Delta  \big) {} \absmeet {}  \{ \bagree \neg b \} 
      & \text{ otherwise if }  \neg \semexp{b}{\st} \\
      w^\sharp 
      \big( \semins{c}{\st},
        \seminsmabs{ \ifcom~ b~\thencom~c~\elsecom~
          \skipcom}{\st}{\Delta}
        \big) 
      & \text{ otherwise}
       \end{cases}
\end{mathpar}}
\caption{Hybrid monitor $\hybmod\seminsmabs{-}{-}{}$ using modified variables}
\label{fig:seminsmabs_hybrid}
\end{figure*}

The soundness condition is similar to
\Cref{lem:soundabsstaticpured}, adapted to the extra parameter.

\begin{restatable}[soundness
of $\hybmod\protect\seminscabshybmod{-}{}{-}$
]{mylem}{lemsoundabsstatichybrid}
  \label{lem:soundabsstatichybrid}
  For all $c,c',\st,\st'$ such that 
 $\st' = \semins{c'}{\st}$, it holds that:
  \(
  \alphares{\st'} \comp \seminsc{c}{} \comp \gammast
   \dotabsleq \hybmod\seminscabshybmod{c}{}{c'} 
  \).
\end{restatable}

For a given set $\Delta$ of relational formulas that hold between an initial
major state $\st$ and every initial minor 
state $\stm \in \St$, this abstract static semantics deduces a set $\Delta'$ of
relational formulas that hold between an output major state $\st' =
\semins{c'}{\st}$ and every output minor 
state $\stm' = \semins{c}{\stm}$.
Intuitively, the set $\Delta'$ is deduced from $\Delta$ by keeping only the
relational formulas that cannot be falsified since their free variables are not
modified.

\Cref{fig:seminsmabs_hybrid} introduces the abstract semantics
$\hybmod\seminsm{c}{\st}{}$ of the hybrid monitor relying on a static analysis
of modified variables. 
Most transfer functions are essentially the same as the ones introduced for the
purely-dynamic monitor.
Thus, we refer to \Cref{fig:seminsmabs_purely_dynamic} and redefine only the ones
that are different, namely conditionals and loops. 
The main difference compared to the purely-dynamic monitor resides in the
treatment of high branchings. 
For a high conditional, the abstract static analysis of modified variables
enables the hybrid monitor to deduce that if a relational formula $\Phi$ holds
before the conditional command and if its free variables are modified in
neither conditional branches, 
then $\Phi$ also holds after the execution of the conditional
command. 
This behaviour is similar to the treatment of high conditionals by existing 
hybrid information flow monitors~\cite{Lal.07,RS10}.
However, our hybrid monitor does not rely
on labelling the program counter with a security context in order to track
implicit flows.
This is similar to Besson~et~al.'s approach~\cite{BBJ13}; it facilitates better precision,  as they show  and we see in the next subsection. 

\begin{restatable}[soundness of $\hybmod\seminsmabs{-}{-}{}$
]{mytheo}{lemsoundabsmonitorhybrid}
  \label{lem:soundabsmonitorhybrid}
  The hybrid monitor $\hybmod\seminsmabs{c}{\st}{}$ 
is sound in the sense of \Cref{eq:sound_abs_mon}:
for all $c,\st$:
\[ \alphares{\st'} \comp \seminsm{c}{\st}{} \comp \gammast \dotabsleq
  \hybmod\seminsmabs{c}{\st}{} 
\quad\mbox{ where } \st' = \semins{c}{\st}
\] 
\end{restatable}

The derivation proof of this monitor 
is similar to the derivation of the previous purely-dynamic
monitor. 
It mostly leverages the abstract static analysis of
\Cref{def:seminscabs_hybrid} and \Cref{lem:soundabsstatichybrid}, 
in order to treat more precisely high branching commands.
 
Although we avoid relying on explicit tracking of a security context, for
reasons of precision, that is useful for another purpose: 
enforcing security policies such as robust declassification~\cite{ZdancewicMyers01}.
Recall the strong conjecture following \Cref{theo:mon_collecting_sound}.
One can replay the derivation of all three of our monitors, starting from 
the alternative ideal monitor in \Cref{sec:alt_monitoring_collecting}, with minimal
changes.  We thus obtain three alternative monitors that track a security context
and are conjectured to be sound for the semantics in \cite{CKN14}.

\subsection{Hybrid Monitor with Intervals}
\label{subsec:mon_hybint}
We now derive a hybrid monitor that relies on a static analysis
approximating the range of values each variable may take.
This allows to infer agreements even for locations that are modified in high branches.




\begin{tdisplay}{Interval analysis \hfill $\seminscabsint{-}{}$}
\textbf{Assumptions:}
$\statesint$ is a  set of abstract environments mapping variables 
to intervals
(and $\statesinterr \triangleq \statesint {} \cup {} \{ \err \}$). \\
$\seminscabsint{c}{} \in  \statesinterr \to  \statesinterr$ is an interval
static analysis satisfying:
$\alphaint \comp \seminsc{c}{} \comp \gammaint \dotleqintervals
\seminscabsint{c}{}$, with  $(\alphaint,\gammaint)$ being the Galois connection
enabling its derivation:
$(\pstateserr; \concleq) 
\galois{\alphaint}{\gammaint}
(\statesinterr;\leqintervals)$
\end{tdisplay}

Interval static analysis is standard~\cite{CC77};
we present one in full detail in the long version of the paper. 

Unlike the previous hybrid monitor relying on the modified variables, this
monitor reasons on the values variables may take, in order to
establish relational formulas  in the case of high branching commands.
Similarly to the condition stated in
\Cref{lem:soundabsstaticpured,lem:soundabsstatichybrid}, the abstract static
analysis $\seminscabs{c}{}$ must satisfy:
\[ \alphares{\st'} \comp \seminsc{c}{} \comp \gammast \dotabsleq
\seminscabs{c}{} \]
Let us derive such an abstract static analysis by relying on the
interval analysis $\seminscabsint{c}{}$.
Observe that
\begin{align}
  & \alphares{\st'} \comp \seminsc{c}{} \comp \gammast
  \notag
  \\
  & \quad \dotabsleq \text{$\Lbag$since $\alphares{\st'}$, $\seminsc{c}{}$ are
    monotone, $\gammaint \comp \alphaint$ is extensive$\Rbag$}
  \notag
  \\
  & \quad \qquad
  \alphares{\st'} \comp \gammaint \comp \alphaint \comp \seminsc{c}{} \comp
  \gammaint \comp \alphaint \comp \gammast
  \notag
  \\
  & \quad \dotabsleq \text{$\Lbag$since $\alphares{\st'}, \gammaint$ are
    monotone, $\seminscabsint{c}{}$ is sound$\Rbag$} 
  \notag
  \\
  & \quad \qquad 
  \alphares{\st'} \comp \gammaint \comp \seminscabsint{c}{} \comp
  \alphaint \comp \gammast
  \label{eq:soundness_static_interval}
\end{align}

Consequently, we can leverage an interval static analysis to derive a
monitor tracking relational formulas, provided that we derive
an interface between the two abstractions.
First, we have to approximate the operator $\alphaint \comp \gammast$ which
translates relational formulas --- that holds wrt.\ a major state $\st$ --- to
interval constraints over variables.
Second, we need to approximate the operator
$\alphares{\st'} \comp \gammaint$ which translates interval constraints over
variables to relational formulas --- that holds wrt.\ a major state $\st'$.

These two operators $\alphaint \comp \gammast$ and
$\alphast \comp \gammaint$ are similar to Granger's reduced product
operators~\cite{Gran92,CCP13} over the Cartesian product of both intervals
and relational formulas, as we shall explain.

Since combining different static analyses independently by relying on a
Cartesian product does not yield optimal results in general, abstract
interpretation relies on a notion of a reduced products~\cite{CC79},
in order to enable the sharing of information between the different abstractions
and gain more precision.   

Let $(\alphastint,\gammastint)$ be the Galois connection associated with the
Cartesian product of both abstractions:\\
\mbox{\(
\gammastint(\imath,\Delta)  \triangleq \gammaint(\imath) \concmeet
\gammast(\Delta)
\) 
and 
\( \alphastint(\St)  \triangleq (\alphaint(\St),
\alphast(\St))
\).}


\begin{restatable}[
  ]{mydef}{defgrangerprod}
  \label{def:grangerprod}
A Granger's reduced product for the Cartesian abstraction
$\statesinterr {} \times {} \prelformerr$ 
is a pair of operators
$\redtointst \in \statesinterr \times \, \prelformerr \to \statesinterr$ and
$\redtoprelformst \in \statesinterr \times \, \prelformerr \to \prelformerr$,
parametrised by a state $\st \in \States$ and
satisfying two conditions:
\begin{itemize}
\item Soundness:
  \(\begin{array}[t]{l}
\gammastint\left(\redtointst(\imath,\Delta),\Delta\right) =  \gammastint\left(\imath,\Delta\right) \text{ and }  \\ \gammastint\left(\imath,\redtoprelformst(\imath,\Delta)\right)  \gammastint\left(\imath,\Delta\right)
  \end{array}\)
\item Reduction:
  \( \redtointst(\imath,\Delta) \leqintervals \imath \text{ and }
  \redtoprelformst(\imath,\Delta) \absleq \Delta \)
\end{itemize}
\end{restatable}



\begin{restatable}[
    ]{mylem}{lemgrangerprod}
  For all $\st \in \States$, any pair of operators that is a  Granger's reduced
  product  $(\redtointst,\redtoprelformst)$ for the Cartesian abstraction
  $\statesinterr {} \times {} \prelformerr$ provides a sound 
  approximation of the interface between intervals and relational formulas:  
  \label{lem:grangerprod}
  \begin{align*}
  \alphaint \comp \gammast & {} \dotleqintervals {}
  \lambda \Delta. 
  \redtointst(\err, \Delta)
  \text{ and } \\
  \alphast \comp \gammaint & {} \dotabsleq {} \lambda
  \imath.  
  \redtoprelformst(\imath,\err) 
  \end{align*}
\end{restatable}
The proof of this result applies to any abstraction and is not limited to an interval
analysis.
In a nutshell, for any off-the-shelf static analysis, we can define a Granger's
reduced product for its Cartesian product with the relational formulas
in order to interface this analysis with our monitor, and guarantee soundness by
\Cref{eq:soundness_static_interval}. 
Therefore, we introduce in the following a pair of operators
$(\redtointst,\redtoprelformst)$, that  we prove defines a Granger's reduced product
for the Cartesian product of intervals and relational 
formulas, in \Cref{lem:redprodintrelform}.


\begin{tdisplay}{A Granger's reduced product \hfill
    $\redtointst$ \quad  $\redtoprelformst$}
  \begin{mathpar}
    \redtointst(\imath,\err) \triangleq \imath

    \redtoprelformst(\err,\Delta) \triangleq \Delta

    \redtointst(\imath,\Delta) \triangleq
    \mathop{\capintervals}\limits_{\Phi \in \Delta}
    \redtointst\left(\imath,\{ \Phi \}\right)

    \redtointst(\imath,\{ \agree e \}) \triangleq
    \imath {} \capintervals {}
    \guardop_{e = v}^{\sharp,\intervals}(\imath) \quad \text{ with } 
    v \triangleq \semexp{e}{\st} 

    \redtointst(\imath, \{ \bagree b\}) 
    \triangleq
    \imath {} \capintervals {}
    \guardop_{b}^{\sharp,\intervals}(\imath)

    \max([a,b]) \triangleq b

    \redtointst(\imath, \{ \bagree b \entail \agree e \}) 
    \triangleq
    \imath

    \min([a,b]) \triangleq a 

    \redtoprelformst(\imath,\Delta) \triangleq
    \Delta \absmeet
    \{ \agree x \mid
    \min(\imath(x)) = \max(\imath(x)) = \st(x) \}
  \end{mathpar}
\end{tdisplay}

\begin{restatable}[
  ]{mylem}{lemsoundredprod}
  \label{lem:redprodintrelform}
  For all $\st \in \States$, the pair of operators
  $\redtointst,\redtoprelformst$ is a Granger's reduced product.  
\end{restatable}

The proof derives the definition from the required properties.

The operator $\redtointst$ reduces an interval environment $\imath$
by accounting for the additional constraint that a set of relational formulas
must hold wrt.\ to a state $\st$.
For instance, reducing $\imath$ to account for a set of constraints $\Delta$
amounts to computing the intersection over the reduced interval environments
wrt.\ each formula $\Phi \in \Delta$. 
Also, reducing $\imath$ to account for the constraint that $\agree e$ holds
wrt.\ to a major state $\st$ amounts to reducing $\imath$ to account for the
additional constraint that expression $e$ evaluates to a particular
value $\semexp{e}{\st}$ that is determined by the major state $\st$.
This reduction can be achieved by using an abstract operator
$\guardint{b}$ that over-approximates the concrete operator
$\guardc{b}$.
The soundness of $\guardint{b}$ guarantees the soundness
of operator $\redtointst$, and the abstract meet with the
initial interval environment also guarantees the reduction condition stated in
\Cref{def:grangerprod}.

The operator $\redtoprelformst$ reduces interval constraints over
variables to a set of relational formulas that hold wrt.\ a major state $\st$.
Whenever an interval environment $\imath$ maps a variable
$x$ to a singleton value that matches $\st(x)$, we can deduce that all
the minor states  abstracted by $\imath$ agree with $\st$ on the value of
$x$.

Notice that $\redtoprelformst$ does not 
iterate over the finite lattice $\prelform$,
though in theory that can be done in order to 
determine all relational formulas that hold.
A smarter way to improve the precision of $\redtoprelformst$ would take hints from the
monitored branch, by trying to prove
that some particular relational formulas are satisfied wrt.\ the major state.
This would improve precision for the statically analysed branch.

Having derived a sound approximation of the interface between intervals and
relational formulas,  
we can resume the derivation started in 
\Cref{eq:soundness_static_interval},
using \Cref{lem:redprodintrelform,lem:grangerprod},
to obtain a sound abstract static semantics.


\begin{restatable}[
    ]{mydef}{defabsstaticinterval}
    \label{def:seminscabs_interval}
  For all $\st,\st' \in \States$, 
the abstract static semantics 
  $\hybint\seminscabshybint{c}{}{\st}{\st'}$ 
based on $\seminscabsint{c}{}$ is given  by: 
  \begin{align*}
    \hybint\seminscabshybint{c}{}{\st}{\st'} & \in \prelformerr \to \prelformerr
    \\ 
    \hybint\seminscabshybint{c}{\Delta}{\st}{\st'} & \triangleq
    \begin{cases}
      \err & \text{if } \Delta = \err \\
      \mathcal{L} & \text{if } \Delta = \relform \\
       \lambda \imath.\redtoprelformres{\st'}(\imath,\err)  
       \comp & \\
       \qquad \seminscabsint{c}{}
      \comp \redtointst(\err, \Delta)
      & \text{otherwise}
    \end{cases}
  \end{align*}
\end{restatable}

\begin{restatable}[
soundness of $\hybint\protect\seminscabshybint{-}{}{-}{-}$ 
    ]{mylem}{lemabsstaticinterval}
  \label{lem:soundabsstaticinterval}
For all $c,c'$ and $\st,\st'$ in $\States$
it holds that:
  \[
  \alphares{\st'} \comp \seminsc{c}{} \comp \gammast
   \dotabsleq \hybint\seminscabshybint{c}{}{\st}{\st'}
\quad \mbox{where } \st' = \semins{c'}{\st}
  \]
\end{restatable}

\begin{figure*}[htp]
  \fbox{
    \begin{mathpar}
      
  

  


  \begin{array}{l}
  \seminsmabs{\ifcom~ b~ \thencom~ c_1~ \elsecom~ c_0}{\st}{\Delta}
  \triangleq \\ 
  \qquad
  \begin{cases}
     \begin{cases}
      \seminsmabs{c_1}{\st}{(\Delta \absmeet \{ \bagree b \})} & \text{if } 
      \Delta \entail \agree b \\
      \left( \seminsmabs{c_1}{\st}{(\Delta \absmeet \{ \bagree b \})} \right)
      {} \absjoin {}
      \lambda \imath.\redtoprelformres{\semins{c_1}{\st}}(\imath,\err)
      \comp \seminscabsint{c_0}{}
      \comp  \guardint{ \neg b} \comp
       \redtointst(\err,\Delta)
      & \text{otherwise} 
   \end{cases} & \text{if $\semexp{b}{\st}$
   } \\
   \begin{cases}
    \seminsmabs{c_0}{\st}{(\Delta \absmeet \{ \bagree \neg b \})} & \text{if }
    \Delta \entail \agree b \\
      \left( \seminsmabs{c_0}{\st}{(\Delta \absmeet \{ \bagree \neg b \})} \right)
        {} \absjoin {}
        \lambda \imath.\redtoprelformres{\semins{c_0}{\st}}(\imath,\err)
        \comp \seminscabsint{c_1}{} \comp
        \redtointst(\err,\Delta)
      & \text{otherwise}
   \end{cases}
   & \text{if $\neg \semexp{b}{\st}$
   }
  \end{cases}
  \end{array}
  
  \seminsmabs{\whilecom~ b~\docom~ c}{\st}{\Delta} \triangleq
       \operatorname{snd} \left(
     (\operatorname{lfp}_{\lambda (\st,\Delta).(\bot,\relform)}^{\dotpreccurlyeq
        \times \dotabsleq} 
    \mathcal{G}^\sharp)  (\st,\Delta) \right) 

     \mathcal{G}^\sharp(w^\sharp)(\st,\Delta) \triangleq 
    \begin{cases}
         \st, \err & \text{ if }  \neg \semexp{b}{\st} \text{ and } \Delta = \err \\
         \st, \Delta \absmeet \{ \bagree \neg b \}
         & \text{ if }  \neg \semexp{b}{\st} \text{ and } \Delta \entail \agree b
         \\
         \st,
         \left(\Delta \absjoin
         \lambda \imath.\redtoprelformres{\st}(\imath,\err)
         \comp \seminscabsint{\whilecom~b~\docom~c}{} 
         \comp
         \redtointst(\err, \Delta)
         \right) \absmeet \{ \bagree \neg b \}
         & \text{ otherwise if }  \neg \semexp{b}{\st} \\
             w^\sharp 
             \big( \semins{c}{\st},
             \seminsmabs{ \ifcom~ b~\thencom~c~\elsecom~
          \skipcom}{\st}{\Delta}
             \big) 
           &  \text{ otherwise}
    \end{cases}
\end{mathpar}}
\caption{Hybrid monitor $\hybint\seminsmabs{-}{-}{}$ with an interval analysis} 
\label{fig:seminsmabs_interval}
\end{figure*}

\Cref{fig:seminsmabs_interval} introduces the abstract monitoring semantics we
derive for the hybrid monitor relying on intervals.
Most abstract transfer functions are similar to the ones introduced for the
purely-dynamic monitor.
Thus, we refer to \Cref{fig:seminsmabs_purely_dynamic} and redefine only the ones
that are different, namely conditionals and loops.
The main difference concerns branching commands, since we rely on the novel
abstract static analysis of \Cref{def:seminscabs_interval}.


\begin{restatable}[]{mytheo}{lemsoundabsmonitorinterval}
  \label{lem:soundabsmonitorinterval}
  The abstract monitoring semantics $\hybint\seminsmabs{c}{\st}{}$ of the hybrid
  monitor, introduced in 
  \Cref{fig:seminsmabs_interval}, is sound:
  For all $\st, \st' \in \States$ such that $\st' = \semins{c}{\st}$, it holds
  that:
  \[ \alphares{\st'} \comp \seminsm{c}{\st}{} \comp \gammast \dotabsleq
  \hybint\seminsmabs{c}{\st}{} \] 
\end{restatable}

This monitor is sensitive to runtime values. 
It deduces new relational formulas for high branching commands, by comparing the
major state with the results of an interval static analysis.   
Consider for instance the program in \Cref{lst:seed}, that is inspired
by M{\"u}ller~and~al.~\cite{MKS15}. 
Despite being modified in conditional branches that depend on a high guard,
variable {\ttfamily seed} does not leak sensitive data.
Unlike the hybrid flow-sensitive monitors of Le~Guernic et al.~\cite{Lal.07} and
Russo and Sabelfeld~\cite{RS10}, as well as the ones we introduce previously in
\Cref{sec:derivation_pured,sec:derivation_hybmod}, our monitor relying on
an interval analysis determines that all minor states agree with the  major
state on the value of variable {\ttfamily seed} --- at the merge point of the
conditional, which corresponds to labelling this variable as low.
Our hybrid monitor is similar in spirit to the hybrid monitor of
Besson~et~al.~\cite{BBJ13} that is also able to deduce that 
variable {\ttfamily seed} does not convey any knowledge about sensitive data,
by relying on a constant propagation static analysis.
\begin{lstlisting}[label={lst:seed}, style=simple, 
caption={Example program inspired by~\cite{MKS15}},float=htp,firstnumber=0]
// @$\lstmathcomment{\st_0 \triangleq [\text{seed}\to3;\text{secret\_conf} \to 1\ldots], \Delta_0 = \emptyset}$@ 
assume @$\agree \text{seed}$@;// @$\lstmathcomment{[\text{seed}\to3;\text{secret\_conf} \to 1\ldots], \{\agree \text{seed}\}}$@
a := secret_base; 
if (secret_conf) then {
  b := secret_number; //@$\lstmathcomment{[\text{seed}\to3\ldots], \{\agree\text{seed}, \bagree \text{secret\_conf}\}}$@
  //Complicated hash computation on the seed, a & b.
  r := seed * a * b;
  seed := 1 + seed; //@$\lstmathcomment{[\text{seed}\to4\ldots], \{\agree\text{seed}, \bagree \text{secret\_conf}\}}$@
}
else { //@$\lstmathcomment{\text{value-sensitivity: seed is initialised to a singleton interval}}$@ //@$\lstmathcomment{\imath \triangleq \big[\text{seed}\to[3,3];\text{r}\to[-\infty,+\infty]\ldots\big]}$@
  //Complicated hash computation on the seed & a.
  r := seed * a * 42; //@$\lstmathcomment{\big[\text{seed}\to[3,3];\text{r}\to[-\infty,+\infty]\ldots\big]}$@
  seed := 1 + seed; //@$\lstmathcomment{\big[\text{seed}\to[4,4];\text{r}\to[-\infty,+\infty]\ldots\big]}$@
} //@$\lstmathcomment{[\text{seed}\to4\ldots], \{\agree\text{seed}\}}$@
assert @$\agree \text{seed}$@; //@$\lstmathcomment{[\text{seed}\to4\ldots], \{\agree\text{seed}\}}$@
\end{lstlisting}


\section{Related Work}\label{sec:related}

Abstract interpretation has been used for static analysis of noninterference 
in a number of works including 
Kov{\`a}cs~et~al.~\cite{KSF13}, where security is explicitly formulated as 2-safety.
Giacobazzi and Mastroeni show how abstract interpretation can be used to reason about 
downgrading policies and observational power of the attacker~\cite{GiacobazziM10}. 
Here we focus on related work that addresses the challenge areas  
for IF monitoring identified in Sec.~\ref{sec:intro}:
expressive policy, precision versus performance, and assurance of correctness.

To facilitate expression of the range of practical IF policies, researchers have proposed language based approaches that use types and other program annotations to label channels and for downgrading 
directives~\cite{VIS96,Mye99,ZdancewicMyers01,BS06,AskarovS07,BNR08,SS09,BDS13}.
In various ways, policies can refer to meaningful events and conditions in terms of program control and data state
(including instrumentation to express policy~\cite{BS06}).
Relational Hoare logic features assertions that express agreement or ``low indistinguishability'', 
enabling direct specification of conditional and partial dependency 
properties~\cite{Benton04,AB04,AmtoftHRRHG08,NanevskiBG13}.
A strength of the logic approach is that it can offer precise reasoning about complex data and 
control structures~\cite{AmtoftHR10,NanevskiBG13,BeckertBrunsKlebanovEtAl14}.
For end-to-end semantics of IF policies, epistemic formulations are effective~\cite{AskarovS07,AskarovSabelfeld09,AskarovMyers11,BalliuDG11} 
and have been connected with relational logic~\cite{BNR08}.  
Such a connection is evident in the IF property of~\cite{CKN14}, although it is not formalized there.  

The generality and expressiveness of relational logic comes at the cost that it does
not inherently enforce desirable constraints on policy.  For example, consider the policy that 
the initial value of card number is secret, but the low four digits may be released upon successful authentication between merchant and customer.   The relevant condition could be asserted at a point in the code where the release takes place, together with assumed agreement on the expression $ccnum\%10000$, but the policy analyst could also assert that $ccnum$ refers to its initial value 
at this point --- $\assertcom~ \bagree (ccnum = oldcc)$ where $oldcc$ is a variable set to the initial value of $ccnum$ and not changed.


Concerning precision of purely dynamic monitors, the basic technique of No Sensitive Upgrade (NSU)~\cite{Zda02,AF09} has been refined~\cite{AF10,BichhawatRGH14} and several 
implementations exist~\cite{BichhawatRGH14POST,IFM-inlining-CoreJS,HedinBS15,ChudnovN15}.
Experience suggesting NSU is too imprecise --- rejecting many secure executions ---  led Hedin et al.~\cite{HedinBS15} to augment
their monitor with static analysis of modified locations that heuristically suggests label upgrades,
while relying on  NSU for soundness.
Hybrid monitors typically feature static analysis of modified variables.
For the simple while language, a simple approximation is to find all
assignment targets~\cite{Lal.07,RS10}.
For more complex data structure this requires memory abstraction~\cite{MC11}.
For more precision it is better to take into account the actual low values~\cite{Leg08}.  
The term ``value-sensitive'' is used by Hedin et al.~\cite{HedinBS15}
and Bello et al.~\cite{BelloHS15}
for the use of low values in the major run to determine the observable modifications. 

As our third monitor shows, there is a second important way in which 
precision can be gained by sensitivity to low values.
This is also done by Besson et al.~\cite{BBJ13} who propose a generic hybrid monitor for quantitative information
flow by tracking the knowledge about sensitive data that is stored in each program
variable.
Their monitor is also parametrised by a static analysis.
The monitors we derive by relying on abstract interpretation share some
similarities with their monitors.
Indeed, they also model a purely-dynamic monitor by a static analysis that
replies with top, forcing the monitor to treat high branching
commands pessimistically.
Additionally, their monitors do not rely on labelling the program counter with a
security context.

For formal assurance of IF monitor correctness, several works provide detailed formal 
proofs~\cite{IFM-inlining-CoreJS,HedinBS15}; 
some are machine checked~\cite{Beringer12,AmorimCDDHPPPT14}, including 
Besson et al.~\cite{BBJ13}.
Our work relies on the framework of abstract interpretation for the systematic
design and derivation of security monitors.
We hope our work will pave the way for the formal verification of practical
security monitors, by reusing some of the recent developments in formal
verification of abstract interpretation analysers~\cite{Pic06,Jal.15,DaraisMH15}.

\section{Discussion and Conclusion}\label{sec:discussion}


In this paper we propose an ideal monitor as a variation of a collecting
semantics.
We prove that this monitor enforces noninterference, for policies expressed using relational formulas,
to which many other policy formalisms can be translated.
We also conjecture (following \Cref{theo:mon_collecting_sound}) a relation with the security property proposed in \cite{CKN14}
and taking into account intermediate assumptions (for downgrading) 
and assertions (for modularity and intermediate output).
We would like to prove that conjecture, and strengthen it to 
account for intermediate assertions in divergent major runs.  
We believe this can be done using either transition semantics
or trace based denotational semantics.

The ideal monitor is a specification that serves for deriving monitors that are sound by construction. 
We derive three monitors that illustrate various ways of tracking low indistinguishability and other relational formulas between states for high branchings.
Although we provide a systematic approach by which precision can be fine tuned, we do not systematically evaluate the precision of the derived monitors.  Several notions of evaluation have been proposed in the literature, using terms such as permissiveness and transparency.  
Bielova and Rezk~\cite{BR16} disentangle these notions.
In their terms, precision in our sense --- allowing more secure executions --- is termed ``true transparency''.  

A benefit of monitoring, relative to type systems or other static analyses for security, is the potential to leverage runtime values for precision.  Almost all prior work uses value-sensitivity, if at all, as a means of improving precision to determine modifiable locations.  
An exception is Besson et al.~\cite{BBJ13} who rely on constant propagation to delimit implicit flows based on actual values.
A similar result is provided by our value-sensitive hybrid monitor using interval analysis.
We rely on a classical notion in abstract interpretation, reduced
product~\cite{CC79}, to formalize the interactions between both the dynamic and
the static part of the monitor. 
\Cref{lem:grangerprod} and \Cref{eq:soundness_static_interval} in particular are
key results.
By defining a Granger's reduced product, we can immediately  
leverage other off-the-shelf static analyses, such as
polyhedra~\cite{CH78}, trace partitioning~\cite{MR05}, and constant propagation.
Existing monitors already incorporate such complex static analyses that are
spawned during or before the monitored execution~\cite{Mal.11,Mal.13,BBJ13}.
Two challenges remain for the adoption of IF monitors: scaling them to
complex and richer languages, and lowering the incurred overhead.
We believe this paper makes a dent wrt.\ the first dimension, by linking the
design of information flow monitors to the design of static analyses by the
well-established theory of abstract interpretation~\cite{CC77}.
As to the second challenge, we would like to investigate what static information 
can be pre-computed and how a monitor can take advantage of such information
(beyond the easy case of modified variables in a toy language). 

\bigskip

\paragraph*{Acknowledgments}

Thanks to Anindya Banerjee, Andrey Chudnov, and the anonymous reviewers for helpful feedback.
The authors were partially supported by NSF award CNS-1228930.



\bibliographystyle{IEEEtranS}
\bibliography{../biblio}

\begin{thebibliography}{10}
\providecommand{\url}[1]{#1}
\csname url@samestyle\endcsname
\providecommand{\newblock}{\relax}
\providecommand{\bibinfo}[2]{#2}
\providecommand{\BIBentrySTDinterwordspacing}{\spaceskip=0pt\relax}
\providecommand{\BIBentryALTinterwordstretchfactor}{4}
\providecommand{\BIBentryALTinterwordspacing}{\spaceskip=\fontdimen2\font plus
\BIBentryALTinterwordstretchfactor\fontdimen3\font minus
  \fontdimen4\font\relax}
\providecommand{\BIBforeignlanguage}[2]{{%
\expandafter\ifx\csname l@#1\endcsname\relax
\typeout{** WARNING: IEEEtranS.bst: No hyphenation pattern has been}%
\typeout{** loaded for the language `#1'. Using the pattern for}%
\typeout{** the default language instead.}%
\else
\language=\csname l@#1\endcsname
\fi
#2}}
\providecommand{\BIBdecl}{\relax}
\BIBdecl

\bibitem{AB04}
T.~Amtoft and A.~Banerjee, ``{Information Flow Analysis in Logical Form.}''
  \emph{Static Analysis Symposium}, pp. 100--115, 2004.

\bibitem{AmtoftB07}
------, ``Verification condition generation for conditional information flow,''
  in \emph{ACM Workshop on Formal Methods in Security Engineering}, 2007, pp.
  2--11.

\bibitem{AmtoftHR10}
T.~Amtoft, J.~Hatcliff, and E.~Rodr\'{\i}guez, ``Precise and automated
  contract-based reasoning for verification and certification of information
  flow properties of programs with arrays,'' in \emph{European Symposium on
  Programming}, ser. LNCS, vol. 6012, 2010.

\bibitem{AmtoftHRRHG08}
T.~Amtoft, J.~Hatcliff, E.~Rodr\'{\i}guez, Robby, J.~Hoag, and D.~Greve,
  ``Specification and checking of software contracts for conditional
  information flow,'' in \emph{Formal Methods}, ser. LNCS, vol. 5014, 2008.

\bibitem{AskarovMyers11}
A.~Askarov and A.~C. Myers, ``Attacker control and impact for confidentiality
  and integrity,'' \emph{Logical Methods in Computer Science}, vol.~7, no.~3,
  2011.

\bibitem{AskarovS07}
A.~Askarov and A.~Sabelfeld, ``Gradual release: Unifying declassification,
  encryption and key release policies,'' in \emph{IEEE Symposium on Security
  and Privacy}, 2007.

\bibitem{AskarovSabelfeld09}
------, ``Tight enforcement of information-release policies for dynamic
  languages,'' in \emph{IEEE Computer Security Foundations Symposium}, 2009.

\bibitem{AF09}
T.~H. Austin and C.~Flanagan, ``{Efficient purely-dynamic information flow
  analysis},'' in \emph{ACM Workshop on Programming Languages and Analysis for
  Security}, vol.~44, no.~8, Aug. 2009, pp. 20--31.

\bibitem{AF10}
------, ``{Permissive Dynamic Information Flow Analysis},'' in \emph{ACM
  Workshop on Programming Languages and Analysis for Security}.\hskip 1em plus
  0.5em minus 0.4em\relax ACM, 2010, pp. 1--12.

\bibitem{BalliuDG11}
M.~Balliu, M.~Dam, and G.~Le~Guernic, ``Epistemic temporal logic for
  information flow security,'' in \emph{ACM Workshop on Programming Languages
  and Analysis for Security {(PLAS)}}, 2011.

\bibitem{BNR08}
A.~Banerjee, D.~A. Naumann, and S.~Rosenberg, ``{Expressive Declassification
  Policies and Modular Static Enforcement},'' in \emph{IEEE Symposium on
  Security and Privacy}.\hskip 1em plus 0.5em minus 0.4em\relax IEEE, 2008, pp.
  339--353.

\bibitem{BeckertBrunsKlebanovEtAl14}
B.~Beckert, D.~Bruns, V.~Klebanov, C.~Scheben, P.~H. Schmitt, and M.~Ulbrich,
  ``Information flow in object-oriented software,'' in \emph{Logic-Based
  Program Synthesis and Transformation {LOPSTR}}, ser. LNCS, no. 8901, 2014.

\bibitem{BelloHS15}
L.~Bello, D.~Hedin, and A.~Sabelfeld, ``Value sensitivity and observable
  abstract values for information flow control,'' in \emph{Logic for
  Programming, Artificial Intelligence, and Reasoning {(LPAR)}}, 2015, pp.
  63--78.

\bibitem{Benton04}
N.~Benton, ``{Simple relational correctness proofs for static analyses and
  program transformations},'' in \emph{ACM Symposium on Principles of
  Programming Languages}, 2004.

\bibitem{Beringer12}
L.~Beringer, ``End-to-end multilevel hybrid information flow control,'' in
  \emph{Asian Symposium on Programming Languages and Systems {(APLAS)}}, ser.
  LNCS, 2012, vol. 7705, pp. 50--65.

\bibitem{BBJ13}
F.~Besson, N.~Bielova, and T.~Jensen, ``Hybrid information flow monitoring
  against web tracking,'' in \emph{IEEE Computer Security Foundations
  Symposium}.\hskip 1em plus 0.5em minus 0.4em\relax IEEE, 2013, pp. 240--254.

\bibitem{BichhawatRGH14}
A.~Bichhawat, V.~Rajani, D.~Garg, and C.~Hammer, ``Generalizing
  permissive-upgrade in dynamic information flow analysis,'' in \emph{ACM
  Workshop on Programming Languages and Analysis for Security {(PLAS)}}, 2014.

\bibitem{BichhawatRGH14POST}
------, ``Information flow control in {WebKit's} {JavaScript} bytecode,'' in
  \emph{Principles of Security and Trust {(POST)}}, 2014, pp. 159--178.

\bibitem{BR16}
N.~Bielova and T.~Rezk, ``{A Taxonomy of Information Flow Monitors},'' in
  \emph{POST}, 2016, to appear.

\bibitem{BS06}
N.~Broberg and D.~Sands, ``{Flow Locks: Towards a Core Calculus for Dynamic
  Flow Policies},'' in \emph{European Symposium on Programming}.\hskip 1em plus
  0.5em minus 0.4em\relax Berlin, Heidelberg: Springer Berlin Heidelberg, 2006,
  pp. 180--196.

\bibitem{BDS13}
N.~Broberg, B.~van Delft, and D.~Sands, ``{Paragon for Practical Programming
  with Information-Flow Control},'' in \emph{Asian Symposium on Programming
  Languages and Systems}, 2013, pp. 217--232.

\bibitem{CP10}
D.~Cachera and D.~Pichardie, ``{A Certified Denotational Abstract
  Interpreter},'' in \emph{Interactive Theorem Proving {(ITP)}}, 2010, pp.
  9--24.

\bibitem{CKN14}
A.~Chudnov, G.~Kuan, and D.~A. Naumann, ``{Information Flow Monitoring as
  Abstract Interpretation for Relational Logic},'' in \emph{IEEE Computer
  Security Foundations Symposium}.\hskip 1em plus 0.5em minus 0.4em\relax IEEE,
  2014, pp. 48--62.

\bibitem{ChudnovN15}
A.~Chudnov and D.~A. Naumann, ``Inlined information flow monitoring for
  {JavaScript},'' in \emph{ACM SIGSAC conference on Computer and Communications
  Security}, 2015, pp. 629--643.

\bibitem{CS10}
M.~R. Clarkson and F.~B. Schneider, ``{Hyperproperties.}'' \emph{Journal of
  Computer Security}, vol.~18, no.~6, pp. 1157--1210, 2010.

\bibitem{CCP13}
A.~Cortesi, G.~Costantini, and P.~Ferrara, ``{A Survey on Product Operators in
  Abstract Interpretation.}'' \emph{Festschrift for Dave Schmidt}, vol. 129,
  pp. 325--336, 2013.

\bibitem{Cou99}
P.~Cousot, ``{The calculational design of a generic abstract interpreter},'' in
  \emph{Calculational System Design}, M.~Broy and R.~Steinbr{\"u}ggen,
  Eds.\hskip 1em plus 0.5em minus 0.4em\relax NATO ASI Series F. IOS Press,
  Amsterdam, 1999, vol. 173, pp. 421--506.

\bibitem{CC77}
P.~Cousot and R.~Cousot, ``{Abstract interpretation: a unified lattice model
  for static analysis of programs by construction or approximation of
  fixpoints},'' in \emph{ACM Symposium on Principles of Programming Languages},
  New York, New York, USA, Jan. 1977, pp. 238--252.

\bibitem{CC79}
------, ``Systematic design of program analysis frameworks,'' in \emph{ACM
  Symposium on Principles of Programming Languages}.\hskip 1em plus 0.5em minus
  0.4em\relax ACM, 1979, pp. 269--282.

\bibitem{Cal06}
P.~Cousot, R.~Cousot, J.~Feret, L.~Mauborgne, A.~Min{\'e}, D.~Monniaux, and
  X.~Rival, ``Combination of abstractions in the {ASTR{\'E}E} static
  analyzer,'' in \emph{Asian Computing Science Conference {(ASIAN)}}, 2006, pp.
  272--300.

\bibitem{CH78}
P.~Cousot and N.~Halbwachs, ``{Automatic Discovery of Linear Restraints Among
  Variables of a Program.}'' in \emph{ACM Symposium on Principles of
  Programming Languages}, 1978, pp. 84--96.

\bibitem{DaraisMH15}
D.~Darais, M.~Might, and D.~V. Horn, ``Galois transformers and modular abstract
  interpreters: reusable metatheory for program analysis,'' in
  \emph{Object-Oriented Programming, Systems, Languages, and Applications
  {(OOPSLA)}}, 2015.

\bibitem{AmorimCDDHPPPT14}
A.~A. de~Amorim, N.~Collins, A.~DeHon, D.~Demange, C.~Hritcu, D.~Pichardie,
  B.~C. Pierce, R.~Pollack, and A.~Tolmach, ``A verified information-flow
  architecture,'' in \emph{ACM Symposium on Principles of Programming
  Languages}, 2014, pp. 165--178.

\bibitem{GiacobazziM10}
R.~Giacobazzi and I.~Mastroeni, ``Adjoining classified and unclassified
  information by abstract interpretation,'' \emph{Journal of Computer
  Security}, vol.~18, no.~5, 2010.

\bibitem{Gran89}
P.~Granger, ``{Static analysis of arithmetical congruences},''
  \emph{International Journal of Computer Mathematics}, 1989.

\bibitem{Gran92}
------, ``{Improving the Results of Static Analyses Programs by Local
  Decreasing Iteration.}'' in \emph{Foundations of Software Technology and
  Theoretical Computer Science}, vol. 652, 1992, pp. 68--79.

\bibitem{HedinBS15}
D.~Hedin, L.~Bello, and A.~Sabelfeld, ``Value-sensitive hybrid information flow
  control for a javascript-like language,'' in \emph{IEEE Computer Security
  Foundations Symposium}, 2015, pp. 351--365.

\bibitem{Jal.15}
J.-H. Jourdan, V.~Laporte, S.~Blazy, X.~Leroy, and D.~Pichardie, ``{A
  Formally-Verified C Static Analyzer},'' in \emph{ACM Symposium on Principles
  of Programming Languages}.\hskip 1em plus 0.5em minus 0.4em\relax New York,
  New York, USA: ACM Press, 2015, pp. 247--259.

\bibitem{Kin96}
M.~Kindahl, ``The galois connection in interval analysis,'' Uppsala University,
  Sweden, Tech. Rep., 1994, \url{http://user.it.uu.se/~matkin/interval.ps.gz}.

\bibitem{KSF13}
M.~Kov{\'a}cs, H.~Seidl, and B.~Finkbeiner, ``{Relational abstract
  interpretation for the verification of 2-hypersafety properties},'' in
  \emph{ACM SIGSAC conference on Computer and Communications Security}.\hskip
  1em plus 0.5em minus 0.4em\relax New York, New York, USA: ACM Press, 2013,
  pp. 211--222.

\bibitem{Leg08}
G.~Le~Guernic, ``Precise dynamic verification of confidentiality,'' in
  \emph{Proceedings of the 5th International Verification Workshop in
  connection with {IJCAR}}, 2008.

\bibitem{Lal.07}
G.~Le~Guernic, A.~Banerjee, T.~P. Jensen, and D.~A. Schmidt, ``{Automata-based
  confidentiality monitoring},'' in \emph{Asian Computing Science Conference
  {(ASIAN)}}.\hskip 1em plus 0.5em minus 0.4em\relax ~Springer-Verlag, Dec.
  2006.

\bibitem{Mal.11}
P.~Mardziel, S.~Magill, M.~Hicks, and M.~Srivatsa, ``Dynamic enforcement of
  knowledge-based security policies,'' in \emph{IEEE Computer Security
  Foundations Symposium}.\hskip 1em plus 0.5em minus 0.4em\relax IEEE, 2011,
  pp. 114--128.

\bibitem{Mal.13}
------, ``{Dynamic enforcement of knowledge-based security policies using
  probabilistic abstract interpretation},'' \emph{Journal of Computer
  Security}, vol.~21, no.~4, pp. 463--532, Jan. 2013.

\bibitem{MR05}
L.~Mauborgne and X.~Rival, ``{Trace partitioning in abstract interpretation
  based static analyzers},'' in \emph{European Symposium on Programming}.\hskip
  1em plus 0.5em minus 0.4em\relax Berlin, Heidelberg: ~Springer-Verlag, Apr.
  2005, pp. 5--20.

\bibitem{MC11}
S.~Moore and S.~Chong, ``{Static Analysis for Efficient Hybrid Information-Flow
  Control},'' in \emph{IEEE Computer Security Foundations Symposium}.\hskip 1em
  plus 0.5em minus 0.4em\relax IEEE, 2011, pp. 146--160.

\bibitem{MKS15}
C.~M{\"u}ller, M.~Kov{\'a}cs, and H.~Seidl, ``{An Analysis of Universal
  Information Flow Based on Self-Composition},'' in \emph{IEEE Computer
  Security Foundations Symposium}.\hskip 1em plus 0.5em minus 0.4em\relax IEEE,
  2015, pp. 380--393.

\bibitem{Mye99}
A.~C. Myers, ``{JFlow: Practical Mostly-Static Information Flow Control.}'' in
  \emph{ACM Symposium on Principles of Programming Languages}, 1999, pp.
  228--241.

\bibitem{jif}
A.~C. Myers, N.~Nystrom, L.~Zheng, and S.~Zdancewic, \emph{{Jif}: {J}ava
  Information Flow}, May 2001, software release.
  {\url{http://www.cs.cornell.edu/jif}}.

\bibitem{NanevskiBG13}
A.~Nanevski, A.~Banerjee, and D.~Garg, ``Dependent type theory for verification
  of information flow and access control policies,'' \emph{ACM Trans. Program.
  Lang. Syst.}, vol.~35, no.~2, 2013.

\bibitem{Pic06}
D.~Pichardie, ``{Modular Proof Principles for Parameterised Concretizations},''
  in \emph{Construction and Analysis of Safe, Secure, and Interoperable Smart
  Devices Second International Workshop, {CASSIS} , Revised Selected Papers},
  G.~Barthe, B.~Gr{\'e}goire, M.~Huisman, and J.-L. Lanet, Eds.\hskip 1em plus
  0.5em minus 0.4em\relax Springer Berlin Heidelberg, 2006, pp. 138--154.

\bibitem{RS10}
A.~Russo and A.~Sabelfeld, ``{Dynamic vs. Static Flow-Sensitive Security
  Analysis},'' in \emph{IEEE Computer Security Foundations Symposium}.\hskip
  1em plus 0.5em minus 0.4em\relax IEEE, 2010, pp. 186--199.

\bibitem{SS09}
A.~Sabelfeld and D.~Sands, ``{Declassification: Dimensions and principles},''
  \emph{Journal of Computer Security}, vol.~17, no.~5, Oct. 2009.

\bibitem{IFM-inlining-CoreJS}
J.~Santos and T.~Rezk, ``An information flow monitor-inlining compiler for
  securing a core of javascript,'' in \emph{ICT Systems Security and Privacy
  Protection}, ser. IFIP Advances in Information and Communication Technology,
  2014, vol. 428.

\bibitem{Sch00}
F.~B. Schneider, ``{Enforceable security policies},'' \emph{ACM Transactions on
  Information and System Security}, vol.~3, no.~1, pp. 30--50, Feb. 2000.

\bibitem{ShroffST07}
P.~Shroff, S.~F. Smith, and M.~Thober, ``Dynamic dependency monitoring to
  secure information flow,'' in \emph{IEEE Computer Security Foundations
  Symposium}, 2007, pp. 203--217.

\bibitem{VIS96}
D.~Volpano, C.~Irvine, and G.~Smith, ``{A Sound Type System for Secure Flow
  Analysis},'' \emph{Journal of Computer Security}, vol.~4, no. 2-3, pp.
  167--187, 1996.

\bibitem{Win93}
G.~Winskel, \emph{The Formal Semantics of Programming Languages: an
  Introduction}.\hskip 1em plus 0.5em minus 0.4em\relax Cambridge, 1993.

\bibitem{Zda02}
S.~A. Zdancewic, ``Programming languages for information security,'' Ph.D.
  dissertation, Cornell University, 2002.

\bibitem{ZdancewicMyers01}
S.~Zdancewic and A.~C. Myers, ``Robust declassification,'' in \emph{IEEE
  Computer Security Foundations Workshop}, 2001, pp. 15--23.

\end{thebibliography}


  \crefalias{section}{appsec}
  \crefalias{subsection}{appsec}

\begin{appendices}
\newpage
\section{Alternative Ideal Monitor}
\label{sec:alt_monitoring_collecting}

The alternative ideal monitor relies on an alternative
collecting semantics.
In a nutshell, both collecting semantics are instrumented to keep track of a
boolean  $\boolann$, that signals if annotation commands are allowed in the
current context.
Intuitively, in a low context, the boolean $\boolann$ is set to true, meaning
that annotation commands are allowed.
Otherwise, the boolean $\boolann$ is set to false, signifying that the current
context is high, and disallowing annotation commands. 

How does the monitoring  semantics determine the security context?
It simply checks what happens for branching commands:
Are we already in a high security context? if not, are there some states that
follow a conditional branch that is different from the one taken by the major
state?
If there are some states that follow a different control path, this means that
both conditional branches ought to be treated as high 
branches.
If the conditional is in a low security context and no minor states follow a
different conditional branch, then both conditional branches are to be treated
as low branches. 

In high conditional branches, both the monitoring semantics and the
collecting semantics return an error if an annotation command is
encountered.
This way, the monitoring  semantics implicitly signals an alignment
failure, as proposed in \cite{CKN14}.

\Cref{fig:alt_monitoring_semantics,fig:alternative_static_collecting_semantics}
introduce the alternative ideal monitor and the alternative
 collecting semantics.
We conjecture that this alternative ideal monitor returns an
error iff.\ the monitor that is defined in \cite{CKN14} in terms of a tracking set
results in either an alignment failure or an assertion failure.
This conjecture remains to be proved.

\begin{figure*}[p]
  \fbox{
  \begin{mathpar}

    \seminsmalt{c}{\st}{\err}{\boolann} \triangleq \err
      
    \seminsmalt{\skipcom}{\st}{\St}{\boolann} \triangleq  \St

    \seminsmalt{id := e}{\st}{\St}{\boolann} \triangleq
    \left\{ \semins{id := e}{\stm} \mid \stm \in \St \right\}

    \seminsmalt{c_1; c_2}{\st}{\St}{\boolann} \triangleq
    \seminsmalt{c_2}{\semins{c_1}{\st}}{}{\boolann}
    \comp \seminsmalt{c_1}{\st}{\St}{\boolann}
    
    \seminsmalt{\assumecom~\Phi}{\st}{\St}{\boolann} \triangleq
    \begin{cases}
      \{ \stm \in \St \mid  \st \shortmid \stm \models \Phi \} & \text{if
      } \boolann = true \\
      \err & \text{if } \boolann = false
    \end{cases}

    \seminsmalt{\assertcom~\Phi}{\st}{\St}{\boolann} \triangleq
    \begin{cases}
      \St & \text{if } \forall \stm \in \St, \st \shortmid \stm \models \Phi
      \text{ and } \boolann = true \\
      \err & \text{otherwise}
    \end{cases}

      
    \seminsmalt{\ifcom~ b~ \thencom~ c_1~ \elsecom~ c_0}{\st}{\St}{\boolann}
    \triangleq
     \begin{array}[t]{l}
    \begin{cases}
      \seminsmalt{c_1}{\st}{}{\boolann \wedge \boolann'} \comp \guardm{\st}{b}{\St} 
      {} \concjoin {}
      \seminscalt{c_0}{}{\boolann \wedge \boolann'} \comp \guardc{\neg b}{\St}
      &
      \text{if } \semexp{b}{\st} \\
      \seminscalt{c_1}{}{\boolann \wedge \boolann'} \comp \guardc{b}{\St}
      {} \concjoin {}
      \seminsmalt{c_0}{\st}{}{\boolann \wedge \boolann'} \comp \guardm{\st}{\neg
        b}{\St} 
      & \text{otherwise} 
    \end{cases} \\
    \text{with } \boolann' \triangleq
    \begin{cases}
    (\guard{\neg b}{\St} = \emptyset) &  \text{if } \semexp{b}{\st} \\
    (\guard{ b}{\St} = \emptyset) &   \text{if } \neg \semexp{b}{\st}
    \end{cases}
     \end{array}

    
    \seminsmalt{ \whilecom~ b~\docom~ c}{\st}{\St}{\boolann} \triangleq
    \operatorname{snd} \left(
    \left(\operatorname{lfp}_{\lambda (\st,\St).(\bot,\emptyset)}^{\dotpreccurlyeq \times
      \dotconcleq} \mathcal{G}^{\boolann} \right) (\st,\St) \right)  \\
    \textit{ with }
    \mathcal{G}^{\boolann}(w) \triangleq
    \lambda (\st,\St).
    \begin{cases}
      \st,
      \seminscalt{\whilecom~b~\docom~c}{\St}{\boolann}
      & \text{if }  \neg \semexp{b}{\st} \\
      w \left( \semins{c}{\st},
      \seminsmalt{ \ifcom~ b~\thencom~ c~\elsecom~ \skipcom}{\st}{\St}{\boolann}
      \right) 
      & \text{otherwise}
    \end{cases}

  \end{mathpar}}
  \caption{Alternative ideal monitor}
  \label{fig:alt_monitoring_semantics}
\end{figure*}
\begin{figure*}[p]
  \fbox{
  \begin{mathpar}
    \seminscalt{\skipcom}{\St}{\boolann} = \St

     \seminscalt{id := e}{\St}{\boolann} = \{ \stm[ id \mapsto \semexp{e}{\stm }]  \mid \stm
     \in \St \}

     \seminscalt{c_1; c_2}{\St}{\boolann} = \seminscalt{c_2}{}{\boolann} \comp
     \seminscalt{c_1}{\St}{\boolann} 
     
     \seminscalt{\ifcom~ b~ \thencom~ c_1~ \elsecom~ c_2}{\St}{\boolann} =
     \seminscalt{c_1}{}{\boolann} \comp  \guard{b}{\St} \concjoin
     \seminscalt{c_2}{}{\boolann} \comp  \guard{\neg b}{\St}

     \seminscalt{\assumecom~\Phi}{\St}{\boolann} =
     \begin{cases}
       \St & \text{if } \boolann \\
       \err & \text{if } \neg \boolann
     \end{cases}

     \seminscalt{\assertcom~\Phi}{\St}{\boolann} =
     \begin{cases}
       \St & \text{if } \boolann \\
       \err & \text{if } \neg \boolann
     \end{cases}

     \seminscalt{\whilecom~ b~\docom~ c}{\St}{\boolann} =
    \guard{\neg b}{ \operatorname{lfp}_{\St}^{\concleq} \seminscalt{\ifcom~
        b~ \thencom~ c~ \elsecom~ \skipcom}{}{\boolann} }

    \guard{b}{\St} \triangleq \{ \stm \in \St \mid \semexp{b}{\stm} = true \}    
  \end{mathpar}}
  \caption{Alternative collecting semantics}
  \label{fig:alternative_static_collecting_semantics}
\end{figure*}

\end{appendices}

\iftechreport

\clearpage
\newpage

\begin{appendices}

     \section{Table of Symbols}

  \begin{align*}
    \States & \triangleq Var \rightharpoonup \mathbb{Z}  \text{ mappings from vars to integers} \\
    \bot & \text{ undefined outcome} \\
    \States_\bot & \triangleq \States {} \cup {} \{ \bot \}   \text{ set of outcomes} \\
    \st & \in \States  \text{ proper state} \\
    \sterr & \in \States_\bot  \text{ outcome}\\
    \err & \text{ security violation, or fault} \\
    \pstates & \text{ powerset of } \States \\
    \pstateserr & \triangleq \pstates \cup \{ \err \}  \\
    \St & \in \pstates  \text{ set of states} \\
    \Sterr & \in \pstateserr  \text{ set of states or fault} \\
    \Phi, \Psi & \in \relform \text{ relational formulas} \\
    \relform & \text{ the set of relational formulas} \\
    \Delta & \in \prelform \text{ a set of relational formulas } \\
    \prelformerr & \triangleq \prelform \cup \{\err\}  \\
    \hat{\Delta} & \in \prelformerr \text{ a set of relational formulas of fault} \\
    \semins{c}{} & \in \States_\bot \to \States_\bot  \text{ denotational
      sem. of com.} \\
    \semexp{e}{} & \in \States \to \mathbb{Z}  \text{ denotational
      sem. of exp.} \\
    \seminsc{c}{} & \in \pstateserr \to \pstateserr  \text{ collecting sem.}\\
    \seminsm{c}{\st}{} & \in  \pstateserr \to \pstateserr  \text{ ideal monitor sem.} \\
    \pured \seminsmabs{c}{}{} & \text{ purely-dynamic  monitor.}\\
    \hybmod \seminsmabs{c}{}{} & \text{ hybrid mon. with modified vars} \\
    \hybint \seminsmabs{c}{}{} & \text{ hybrid mon. with modified intervals} \\
    \pured\seminscabspured{c}{} & \text{ static analysis for purely-dynamic mon.} \\
    \hybmod\seminscabshybmod{c}{}{c'} & \text{ static analysis for hybrid mon. with modified
    vars} \\ 
    \hybint\seminscabshybint{c}{}{\st}{\st'} & \text{ static analysis for hybrid mon. with
      intervals}  \\ 
     \concleq & \text{ set inclusion lifted to } \pstateserr \\
     \concjoin & \text{ set union lifted to } \pstateserr\\
    \concmeet & \text{ set intersection lifted to } \pstateserr \\
    \absleq & \text{ ``includes'' ($\supseteq$) lifted to } \prelformerr \\
    \absjoin & \text{ set intersection lifted to } \prelformerr\\
    \absmeet & \text{ set union lifted to } \prelformerr\\
    \preccurlyeq & \text{ approximation order}\\
    \alphast & \in \pstateserr \to \prelformerr \text{ abstraction function
      param. by $\st$}\\
    \gammast & \in \prelformerr \to \pstateserr \text{ concretisation function
      param. by $\st$} \\
    \alphapair & \in \States \times \pstateserr \to \States \times \prelformerr
    \text{ generalised $\alphast$ } \\
    \gammapair & 
    \in \States \times \prelformerr \to
    \States \times \pstateserr \text{ generalised $\gammast$ }\\    
  \end{align*}

   \clearpage

  \section{Background}

  \subsection{Collecting Semantics}
  \label{sec:ap_static_collecting}
  \lemeqstaticcollecting*
  \textit{Proof.} The proof of equivalence of both collecting semantics
  is by structural induction on commands.
  We feature in this proof one simple case (assignments) as well as
  the most interesting case that is the case of while loops.
  
  1 -- Case: assignments
  \begin{align*}
    \seminsc{id:=e}{\St} & \triangleq \{ \semins{id:=e}{\st} \mid \st \in \St \mbox{
      and } \semins{id:=e}{\st}\neq\bot \} \\
    & = \{ \st[ id \mapsto \semexp{e}{\st }] \mid \st \in \St \mbox{
      and } \semins{id:=e}{\st}\neq\bot\} \\
    & = \{ \st[ id \mapsto \semexp{e}{\st }]  \mid \st
     \in \St \}
  \end{align*}

  2 -- Case: skip

  1 -- Case: loops.
  
  1.1 -- Let us first prove the following intermediate result:
  \begin{multline*}
    \seminsc{\whilecom~ b~\docom~ c}{\St} = \\ \guard{\neg b}{
  \operatorname{lfp_{\emptyset}^{\subseteq} \lambda X. \St \cup
    \seminsc{c}{}\comp \guard{b}{X} }}.
    \end{multline*}

  Indeed, let the sequence $(x_n^{\St})_{n \geq 0}$ be defined as:
  \begin{align*}
    x_{n}^{\St} & \triangleq \{ \mathcal{F}^{(n)}(\bot)(\st) \in \States \mid
    \st \in \St \} 
  \end{align*}
  Notice that for all $\st \in \St$, the sequence
  $(\mathcal{F}^{(n)}(\bot)(\st))_{n \geq 0}$ converges and is equal to the
  evaluation of 
  the while loop in the state $\st$
  $(\semins{\whilecom~ b~\docom~ c}{\st} =
  \mathcal{F}^{(\infty)}(\bot)(\st))$, by
  definition of the denotational semantics of loops.

  Let also the sequences $(y_n^\St)_{n \geq 0}$ and $(g_n^\St)_{n \geq 0}$ be
  defined as:
  \begin{align*}
    y_{n}^{\St} & \triangleq \guard{\neg b}{g_n^{\St}} \\
	  g_{n+1}^{\St} & \triangleq \St \cup \seminsc{c}{} \comp \guard{b}{g_n^{\St}} \\
    g_0^{\St} & \triangleq \emptyset
  \end{align*}

  Then, it holds that:
  \[ \forall \St \in \pset{\States}, \forall n \in \mathbb{N},
  x_n^\St = y_n^\St. \]
  Indeed, the proof proceeds by induction on $n$.
  
  - $ x_0^\St = \emptyset = y_0^\St $
  
  - Let $n \in \mathbb{N}$ such that:
  $\forall \St \in \pset{\States}, x_n^\St = y_n^\St$.
  Then:
  \begin{align*}
    x_{n+1}^\St & = \{ \mathcal{F}^{(n+1)}(\bot)(\st) \in States \mid \st \in
    \St \} \\ 
    & = \guard{\neg b}{\St} \cup \\
    & \qquad
    \{ \mathcal{F}^{(n)}(\bot)(\semins{c}{\st}) \in \States \mid
    \st \in \guard{b}{\St}   \}  \\
    & = \guard{\neg b}{\St} \cup \\
     & \qquad \{ \mathcal{F}^{(n)}(\bot)(\stm) \in \States \mid
    \stm \in \seminsc{c}{} \comp \guard{b}{\St}   \}  \\
    & =  \guard{\neg b}{\St} \cup x_n^{ \seminsc{c}{} \comp \guard{b}{\St}} \\
    & = \text{(By induction hypothesis)} \\
    & \quad \quad \guard{\neg b}{\St} \cup
    y_n^{\seminsc{c}{} \comp \guard{b}{\St}} \\
    & = \text{(By definition of $y_n^{\seminsc{c}{} \comp \guard{b}{\St}}$ )} \\
    & \quad \quad \guard{\neg b}{\St} \cup
    \guard{\neg b}{g_n^{\seminsc{c}{} \comp \guard{b}{\St}}} \\
    & = \guard{\neg b}{\St \cup g_n^{\seminsc{c}{} \comp \guard{b}{\St}}} \\
    & = 
    \text{(for all $\St$, $ g_{n}^{\St} =
	    \bigcup\limits_{0 \leq k \leq n-1} (\seminsc{c}{} \comp \guardc{b}{})^{(k)}
      (\St)$ )} \\
    & \quad \quad \guard{\neg b}{g_{n+1}^{\St}} \\
    & = y_{n+1}^{\St} \qed
  \end{align*}

  1.2 -- Let us now prove that :
  \begin{multline*} \operatorname{lfp_{\emptyset}^{\subseteq} \lambda X. \St \cup
    \seminsc{c}{}\comp \guard{b}{X} } = \\
  \operatorname{lfp_{\emptyset}^{\subseteq}} \lambda X. \St \cup
  \seminsc{\ifcom~ b~ \thencom~ c~ \elsecom~ \skipcom}{X}
  \end{multline*}

  Indeed, let the sequence $(f_n^\St)_{n \geq 0}$ be defined as:
  \begin{align*}
    f_0^\St & \triangleq \emptyset \\
    f_{n+1}^\St & \triangleq \St \cup
    \seminsc{\ifcom~ b~ \thencom~ c~ \elsecom~ \skipcom}{f_n^\St} 
  \end{align*}
  Therefore, by induction on $n \in \mathbb{N}$, it holds that $f_n = g_n$:

  - $f_0^\St = g_0^\St = \emptyset$.
  
  - let $n \in \mathbb{N}$, such that $f_n^\St = g_n^\St$. Then:
  \begin{align*}
    g_{n+1}^\St & = \St \cup \seminsc{c}{} \comp \guard{b}{g_n} \\
    &  = \text{($\guard{\neg b}{g_n} \subseteq g_n \subseteq g_{n+1}$)} \\ 
    & \quad \quad
    \St \cup \seminsc{c}{} \comp \guard{b}{g_n} \cup \guard{\neg b}{g_n} \\
    & = \St \cup \seminsc{\ifcom~ b~ \thencom~ c~ \elsecom~
      \skipcom}{g_n} \\
    & = \text{(By induction hypothesis)} \\
    & \quad \quad \St \cup  \seminsc{\ifcom~ b~ \thencom~ c~ \elsecom~
      \skipcom}{f_n^\St} \\
    & = f_{n+1}^\St \qed
  \end{align*}

  1.3 -- Finally,
  \begin{align*}
    & \seminsc{\whilecom~ b~\docom~ c}{\St}  \\
    & \quad = \guard{\neg b}{
      \operatorname{lfp_{\emptyset}^{\subseteq} \lambda X. \St \cup
        \seminsc{c}{}\comp \guard{b}{X} }} \\
    & \quad = \guardop_{\neg b} \big(
      \operatorname{lfp_{\emptyset}^{\subseteq}} \lambda X. \\
    & \quad \qquad \St \cup
      \seminsc{\ifcom~ b~ \thencom~ c~ \elsecom~ \skipcom}{X} \big)  
  \\
  & \quad = \guardop_{\neg b} \big( \\
  & \quad \qquad \operatorname{lfp_{\St}^{\subseteq}}
  \seminsc{\ifcom~ b~ \thencom~ c~ \elsecom~ \skipcom}{} \big)
  \qed
  \end{align*}

  \newpage

  \section{Monitoring Collecting Semantics}
  \label{sec:ap_monitoring_collecting}
  
  \lemmonstaticeq*
  \textit{Proof.}

  The proof is by structural induction on commands.

  1 -- Cases skip and assignments stem from the definition of the collecting
  semantics and the monitoring semantics.

  2 -- Case: sequence $c_1; c_2$

  Let $\st,\st' \in \States$, such that $\semins{c_1;c_2}{\st} = \st'$. Then in
  particular, we know that $\semins{c_1}{\st} \in \States$, meaning that it
  terminates. 
  \begin{align*}
    \seminsm{c_1;c_2}{\st}{\St} & = \seminsm{c_2}{\semins{c_1}{\st}}{} \comp
    \seminsm{c_1}{\st}{\St} \\
    & = \text{(By induction twice, since $\semins{c_1}{\st} \in \States$)} \\
    & \quad \seminsc{c_2}{} \comp \seminsc{c_1}{\St}  \\
    & = \seminsc{c_1; c_2}{\St}
  \end{align*}

  3 -- Case: conditionals
  
  Let $\st,\st' \in \States$ such that
  $\semins{\ifcom~b~\thencom~c_1~\elsecom~c_2}{\st}=\st'$
  Let us also assume that $\semexp{b}{\st} = true$.

  Then:
  \begin{align*}
    \seminsm{c}{\st}{\St} & = 
    \seminsm{c_1}{\st}{} \comp \guard{b}{\St}
      {} \cup {}
      \seminsc{c_2}{} \comp \guard{\neg b}{\St} \\
      & \text{(By induction on $c_1$)} \\
      & \quad
      \seminsc{c_1}{\St} \comp \guard{b}{\St}
       {} \cup {}
       \seminsc{c_2}{} \comp \guard{\neg b}{\St} \\
       & = \seminsc{c}{\St}
  \end{align*}

4 -- Case: loops

Let $\st,\st' \in \States$ such that $\semins{while~b~\docom~c}{\st}=\st'$.
Since the loop terminates, there exists a smallest $k \in \mathbb{N}^*$ such that
$\mathcal{F}^{(k)}(\st) = \st'$. The natural $k$ is intuitively the number of executed
iterations that must be executed before exiting the loop.

Therefore:
\begin{align*}
  & \seminsm{\whilecom~b~\docom~c}{\st}{\St} \\
  & \quad = \operatorname{snd} \left( \mathcal{G}^{(k)}(\lambda
  (\st,\St). (\bot, \emptyset)) (\st, \St) \right ) \\
  & \quad =
  \text{($\St' \triangleq
    \seminsm{(\ifcom~b~\thencom~c_1~\elsecom~\skipcom)^{(k-1)}}{\st}{\St}$)}  \\
  & \qquad \guard{\neg b}{
    \operatorname{lfp}_{\St'}^{\subseteq}
      \seminsc{ \ifcom~ b~\thencom~ c~\elsecom~  \skipcom}{}
  }
  \end{align*}

Notice that we write $(c)^{(k-1)}$ as a shorthand for the sequence of commands
$c; c; c; c; \ldots$, where $c$ is sequentially composed with itself $k-1$ times.

Additionally, by using the same proof as for conditionals, we have:
\begin{align*}
  \St' & =  \seminsc{(\ifcom~b~\thencom~c_1~\elsecom~\skipcom)^{(k-1)}}{\St} \\
  & = \seminsc{\ifcom~b~\thencom~c_1~\elsecom~\skipcom}{}^{(k-1)} \St
\end{align*}

Therefore, the fixpoint over $\St'$ can be formulated as a fixpoint over $\St$:
\begin{align*}
 & \operatorname{lfp}_{\St'}^{\subseteq}
  \seminsc{ \ifcom~ b~\thencom~ c~\elsecom~  \skipcom}{} \\
  & \quad =
   \operatorname{lfp}_{\St}^{\subseteq}
  \seminsc{ \ifcom~ b~\thencom~ c~\elsecom~  \skipcom}{} \\
\end{align*}

Finally, we deduce that:
\begin{align*}
  & \seminsm{\whilecom~b~\docom~c}{\st}{\St} \\
  & \quad =
  \guard{\neg b}{
    \operatorname{lfp}_{\St}^{\subseteq}
    \seminsc{ \ifcom~ b~\thencom~ c~\elsecom~  \skipcom}{}} \\
    &  \quad = \seminsc{\whilecom~b~\docom~c}{\St} \qed
\end{align*}

Notice that if $c$ is an assertion-free command, we have $\seminsm{c}{\st}{\St}
\subseteq \seminsc{c}{\St}$. 
The proof is exactly the same, and the only
difference proceeds by noticing that 
$\seminsm{\assumecom~\Phi}{\st}{\St}
\subseteq \seminsc{\assumecom~\Phi}{\St}$.


Let us now prove the soundness of the monitoring semantics
wrt.\ TINI.

\theomoncollectingsound*

\textit{Proof.}

Let $\st_1,\st_1' \in \States$ such that $\semins{c}{\st_1} = \st_1'$.

Assume $\seminsm{\hat{c}}{\st_1}{\States} \neq \err$.

Let $\st_2,\st_2' \in \States$  and assume $\st_1 =_{in} \st_2$, and prove
$\st_1' =_{out} \st_2'$. 

Therefore, we have:
\begin{align*}
& \seminsm{\assumecom~\agree in;~ c;~\assertcom~ \agree out}{\st_1}{\States}
  \\
  & \quad =
  \seminsm{c;~\assertcom~ \agree out}{\st_1}{} \comp
  \seminsm{\assumecom~\agree in}{\st_1}{\States} \\
  & \quad =
  \seminsm{\assertcom~ \agree out}{\st_1'}{} \comp
  \seminsm{c}{\st_1}{} \comp
  \seminsm{\assumecom~\agree in}{\st_1}{\States} \\
  & \quad \neq  \text{(by assumption)} \\
  & \qquad \err
\end{align*}

Notice that $\st_2 \in  \seminsm{\assumecom~\agree in}{\st_1}{\States}$, since
$\st_1 =_{in} \st_2$.

Therefore,
$\st_2' \in \seminsm{c}{\st_1}{} \comp
\seminsm{\assumecom~\agree in}{\st_1}{\States}$,
since the monitoring semantics is equivalent to the
collecting semantics for annotation-free commands (\Cref{lem:mon_static_eq}),
and the collecting semantics is  the lifting of the denotational
semantics over a set of states
(\Cref{lem:eq_static_collecting} and $\st_2' \in \States$). 

This means that
$\seminsm{\assertcom~ \agree out}{\st_1'}{\{ \st_2' \}} \neq \err$, by
monotonicity of the monitoring semantics.
Therefore:

\[ \st_1' =_{out} \st_2'  \]

-- Case TINI $ \implies \seminsm{\hat{c}}{\st_1}{\States} \neq \err$:

Assume for all $\st_2,\st_2' \in \States$  such that
$\semins{c}{\st_2}=\st_2'$,
then  $\st_1 =_{in} \st_2 \implies \st_1' =_{out} \st_2'$.
Prove $\seminsm{\hat{c}}{\st_1}{\States} \neq \err$.

Note that:
\[ \forall \stm \in
\seminsm{\assumecom\ \agree in}{\st_1}{\States}, \st_1 =_{in} \stm  \]

Therefore, $\forall \stm' \in   \seminsm{c}{} \comp
\seminsm{\assumecom~ \agree  in}{\st_1}{\States}$,
$\exists \stm \in \seminsm{\assumecom\ \agree  in}{\st_1}{\States}$
such that $\stm' = \semins{c}{\stm}$ and $\st =_{in} \stm$.

Thus, we deduce by assumption that
$\forall \stm' \in   \seminsm{c}{} \comp
\seminsm{\assumecom\ \agree  in}{\st_1}{\States}, \st' =_{out} \stm'$.

Consequently, it holds that:

\[ \seminsm{\assertcom\ \agree out}{\st'}{} \comp
\seminsm{c}{}{} \comp \seminsm{\assumecom\ \agree in}{\st_1}{\States}
\neq \err. \hfill \qed \]

\begin{restatable}[The monitoring semantics is
    monotone]{mylem}{lemmoncollectingmonotone}
  For all major states $\sterr \in \States_\bot$, it holds that:
  \[ \forall \Sterr,\Sterr' \in \prelformerr, \quad \Sterr \concleq \Sterr'
  \implies 
  \seminsm{c}{\sterr}{\Sterr} \concleq \seminsm{c}{\sterr}{\Sterr'} \]
\end{restatable}

\textit{Proof.}

The monitoring semantics is monotone, since the
collecting semantics is monotone, and the monitoring semantics of
both annotations is also monotone.
Notice in particular that the monotonicity of assert annotations stems from the
extension of the partial order $\subseteq$ to let $\err$ be the top element of
$\prelformerr$.

\clearpage

  \section{Abstract Domain of Relational Formulas}
  \label{sec:abs_domain_relformulas}
  
  \lemalphastgammast*

Let us recall the definitions of $\alphast$ and $\gammast$:
  \begin{align*}
  \alphast & \in  \pstateserr \to \prelformerr \\
  \alphast(\St) & \triangleqom \{ \Phi \mid \forall \stm \in \St, \st
  \shortmid \stm 
  \models \Phi \}.
  \end{align*}
  \begin{align*}
  \gammast & \in \prelformerr \to \pstateserr \\
  \gammast(\Delta) & \triangleqom
  \{ \stm \in \States \mid \forall \Phi \in \Delta, \st \shortmid \stm \models
  \Phi \}.
  \end{align*}
  
  \textit{Proof.}

Notice first that if $\St = \err$, then
  $\alphast(\err)  \absleq \Delta$
  implies
  $\Delta = \err$,
  therefore
  $ \St \concleq \gammast(\Delta) $.
  Additionally,  $\err \concleq \gammast(\Delta)$ also implies
  $\Delta = \err$, therefore
  $\alphast(\St)  \absleq \Delta$.
  
  Also, if $\Delta = \err$, then both
  $\St \concleq \gammast(\Delta)$ and
  $\alphast(\St)  \absleq \Delta$ are equivalent since they both hold.

  Let us now assume $\St \in \States$ and $\Delta \in \relform$, and prove that:
  \[  \alphast(\St)
  \supseteq \Delta \iff \St \subseteq \gammast(\Delta). \]

-- Case $\alphast(\St) \supseteq \Delta \implies \St \subseteq \gammast(\Delta)$:
 \begin{align*}
  \alphast(\St) \supseteq \Delta & \implies
  \Delta \subseteq
  \{ \Phi \mid \forall \stm \in \St, \st
  \shortmid \stm 
  \models \Phi \} \\
  & \implies \forall  \stm \in \St, \forall \Phi \in \Delta,
  \st \shortmid \stm \models \Phi \\
  & \implies \forall \stm \in \St, \stm \in \gammast(\Delta) \\
  & \implies \St \subseteq \gammast(\Delta) \qed
\end{align*}

 -- Case $\St \subseteq \gammast(\Delta) \implies \Delta \subseteq
 \alphast(\St)$: 
\begin{align*}
  \St \subseteq \gammast(\Delta) &
  \implies \forall \stm \in \St, \stm \in \gammast(\Delta) \\ 
  & \implies \forall \stm \in \St, \forall \Phi \in \Delta, \st \shortmid \stm
  \models \Phi \\
  & \implies \forall \Phi \in \Delta, \Phi \in \alphast(\St) \\
  & \implies \Delta \subseteq \alphast(\St) \qed
 \end{align*}

\lemsoundabsmon*
\textit{Proof.}

Let $\st_0,\st \in \St$ such that $\st = \semins{c}{\st_0}$.
The best abstraction of the state transformer $\seminsm{c}{\st_0}{}$ consists in
concretising an abstract state $\Delta$, applying $\seminsm{c}{\st_0}{}$ and
then abstracting again using $\alphast$:
$\alphast \comp \seminsm{c}{\st_0}{} \comp \gammares{\st_0}$.
Therefore, the monitoring abstract semantics is sound if it holds that:
\[ \alphast \comp \seminsm{c}{\st_0}{} \comp \gammares{\st_0} \dotabsleq
\seminsmabs{c}{\st_0}{}. \]
Let us now  prove the equivalence of the 3 conditions.

--
\begin{align*}
  & \alphast \comp \seminsm{c}{\st_0}{} \comp
\gammares{\st_0} \dotabsleq \seminsmabs{c}{\st_0}{} \\
& \implies \text{($(\alphast,\gammast)$ is a Galois connection)} \\
& \quad \quad  \seminsm{c}{\st_0}{} \comp \gammares{\st_0}  \dotconcleq
 \gammast \comp \seminsmabs{c}{\st_0}{} 
\end{align*}

--
\begin{align*}
  & \seminsm{c}{\st_0}{} \comp \gammares{\st_0}  \dotconcleq
  \gammast \comp \seminsmabs{c}{\st_0}{} \\
  & \implies \text{(By monotony)}\\
  &  \quad \quad
  \seminsm{c}{\st_0}{} \comp \gammares{\st_0} \comp \alphares{\st_0}
  \dotconcleq
  \gammast \comp \seminsmabs{c}{\st_0}{} \comp \alphares{\st_0} \\
  & \implies  \text{(By monotony and
    $\gammares{\st_0} \comp \alphares{\st_0}$ being extensive)} \\ 
  & \quad \quad
  \seminsm{c}{\st_0}{}
  \dotconcleq
  \gammast \comp \seminsmabs{c}{\st_0}{} \comp \alphares{\st_0} \\
  & \implies \text{(By monotony)} \\
  & \quad \quad
   \alphast \comp \seminsm{c}{\st_0}{} 
  \dotabsleq
  \alphast\comp\gammast \comp \seminsmabs{c}{\st_0}{} \comp \alphares{\st_0}\\
  & \implies \text{($ \alphast\comp\gammast$ reductive)} \\
  & \quad \quad
   \alphast \comp \seminsm{c}{\st_0}{} 
   \dotabsleq
    \seminsmabs{c}{\st_0}{} \comp \alphares{\st_0}
\end{align*}

--
\begin{align*}
  &  \alphast \comp \seminsm{c}{\st_0}{} 
   \dotabsleq
   \seminsmabs{c}{\st_0}{} \comp \alphares{\st_0} \\
   & \implies \text{(By monotony)} \\
  & \quad \quad  \alphast \comp \seminsm{c}{\st_0}{} \comp \gammares{\st_0}
   \dotabsleq
   \seminsmabs{c}{\st_0}{} \comp \alphares{\st_0} \comp \gammares{\st_0} \\
   & \implies \text{(By monotony and $\alphares{\st_0} \comp \gammares{\st_0}$
     being reductive)} \\ 
   & \quad \quad
   \alphast \comp \seminsm{c}{\st_0}{} \comp \gammares{\st_0}  \dotabsleq
    \seminsmabs{c}{\st_0}{} \qed
\end{align*}

\newpage

  \section{Monitor Derivation}
  \label{sec:ap_monitor_derivation}

  \subsection{Purely-Dynamic Monitor}
  \label{sec:ap_monitor_pured}

We first start by proving the soundness of the abstract static analysis that the
purely-dynamic monitor relies on.  
\lemsoundabsstaticpured*

Let us recall the definition of this abstract static semantics:
\begin{align*}
  \seminscabs{c}{} & \in \prelformerr \to \prelformerr \\
  \seminscabs{c}{\Delta} & \triangleq
  \begin{cases}
    \err & \text{if } \Delta = \err \\
    \relform & \text{if } \Delta = \relform \\
    \emptyset & \text{otherwise } 
  \end{cases}
\end{align*}
\textit{Proof.}

Let $\st,\st' \in \States$.

Notice first that $\seminscabs{c}{\err} = \err$, therefore
\[   \alphares{\st'} \comp \seminsc{c}{} \comp \gammast(\err)
\absleq \seminscabs{c}{\err}.  \]

Let us now assume that $\Delta \in \prelform$.

If $\Delta = \relform$, and since $\gammast(\relform) = \emptyset$ ($\relform$
contains at least an expression $e$ and its negation $\neg e$, therefore
$\relform$ is concretised to the empty set), we have:
\[ \alphares{\st'} \comp \seminsc{c}{} \comp \gammast(\relform) =
\alphares{\st'}{} \comp \seminsc{c}{\emptyset} = \alphares{\st'}(\emptyset) =
\relform  \]

Additionally, we can always approximate an element of $\prelform$ by the top
element of $\prelform$.
Therefore, it holds that:
\[ \alphares{\st'} \comp \seminsc{c}{} \comp \gammast(\Delta)
\absleq \emptyset \]
\qed

Let us now prove the soundness of the abstract semantics of the purely-dynamic
monitor we derive. 
\lemsoundabsmonitorpured*
\textit{Proof.}

Let us  first rule out the error case:
\[
\alphares{\st'} \comp \seminsc{c}{} \comp \gammast(\err) =
\err \triangleq   \seminsmabs{c}{\st}{\err} 
\]

1 -- Case: skip
\begin{align*}
\alphast \comp \seminsm{\skipcom}{\st}{} \comp \gammast(\Delta) & = \alphast
\comp 
\gammast(\Delta) \\
& \supseteq \Delta \\
& \triangleq \seminsmabs{\skipcom}{\st}{\Delta}
\end{align*}

2 -- Case: assignments

Let $\st,\st' \in \States$ such that $\st' = \semins{id := e}{\st}$. Then:

\begin{align*}
 & \alphares{\st'} \comp \seminsm{id := e}{\st}{} \comp \gammast(\Delta)  \\
  & \qquad = \alphares{\st'} \comp
  \seminsm{id := e}{\st}{} \comp \{ \stm  \mid \forall \Phi \in \Delta, \st
    \shortmid \stm \models \Phi \}  \\
    & \qquad =  \alphares{\st'} \comp
    \{ \stm[id \mapsto \semexp{e}{\stm} ]  \mid \forall \Phi \in \Delta, \st
    \shortmid \stm \models \Phi \} 
\end{align*}

2.1 -- Note that $\forall \Phi \in \Delta$, such that $id \not \in \fv{\Phi}$,
it holds that~\cite[Lemma 2 in Section II.B]{CKN14}:
\[\st \shortmid \stm \models \Phi \implies
\st' \shortmid \stm[id \mapsto \semexp{e}{\stm}] \models \Phi.\]

Therefore, $\forall \Phi \in \Delta \text{ such that } id \not \in \fv{\Phi}$,
it holds that:
\[ \Phi \in \alphares{\st'} \comp
\{ \stm[id \mapsto \semexp{e}{\stm} ]  \mid \forall \Phi \in \Delta, \st
\shortmid \stm \models \Phi \}. \]

2.2 -- Otherwise, it holds that:
\begin{align*}
 & \left(\stm \in
  \gammast(\Delta) \wedge (\Delta \entail \agree e)\right)
  \\ & \quad \implies 
\st[id  \mapsto \semexp{e}{\st}] \shortmid \tau[id
  \mapsto \semexp{e}{\stm}] \models \agree id
\\ & \quad \implies \text{(since $\st' = \st[id  \mapsto \semexp{e}{\st}]$)}
\\ & \qquad \quad
\st' \shortmid \tau[id
  \mapsto \semexp{e}{\stm}] \models \agree id
\end{align*}

\textit{It is worthwhile to note that this proof explicitly uses the fact that
  $\st'$ is the result of evaluation of the assignment {\ttfamily id:=e} on
  $\st$.
This means that if we were to derive an abstract static semantics tracking
relational formulas, we would not be able to deduce $\agree e$, even if $\Delta
\entail \agree e$. This is because we have no way of relating the abstraction
of relational formulas
wrt.\ $\st$ to an abstraction of relational formulas wrt.\ $\st'$, without
additional information.}

Therefore, if  $\Delta \entail \agree e$, then it holds that:
\[ \agree id \in
\alphares{\st'} \comp
\{ \stm[id \mapsto \semexp{e}{\stm} ]  \mid \forall \Phi \in \Delta, \st
\shortmid \stm \models \Phi \}.\]

2.3 -- Finally:

\begin{align*}
   & \alphares{\st'} \comp \seminsm{id := e}{\st}{} \comp \gammast(\Delta)  \\
  & \qquad \supseteq
  \{ \Phi \in \Delta \mid id \not \in \fv{\Phi} \} \cup
  \begin{cases}
    {\agree id} & \text{if } \Delta \entail \agree e \\
    \emptyset & \text{otherwise}
  \end{cases}
  \\
  & \qquad \triangleq \seminsmabs{id := e}{\st}{\Delta}
\end{align*}

3 -- Case: conditionals

Let $\st,\st' \in \States$ such that
$\st' = \semins{\ifcom~ b~ \thencom~ c_1~ \elsecom~ c_2}{\st}$.
Let us consider the case that $\semexp{b}{\st} = true$. Then:
\begin{align*}
  & \alphares{\st'} \comp
  \seminsm{\ifcom~ b~ \thencom~ c_1~ \elsecom~ c_2}{\st}{}
  \comp \gammast(\Delta)  \\ 
  & \quad = \alphares{\st'}\big(
  \seminsm{c_1}{\st}{} \comp \guardop_{b} \comp \gammast(\Delta) {} \cup {} \\
  & \qquad\qquad 
  \seminsc{c_2}{} \comp \guardop_{\neg b} \comp \gammast(\Delta) \big)\\
  & \quad =  \alphares{\st'} \comp
  \seminsm{c_1}{\st}{} \comp \guardop_{b} \comp \gammast(\Delta) {} \cap {} \\
  & \qquad\qquad \alphares{\st'} \comp
  \seminsc{c_2}{} \comp \guardop_{\neg b} \comp \gammast(\Delta)
\end{align*}

We will now treat both branches separately before merging them.

3.1 -- Then-branch:
\begin{align*}
& \alphares{\st'} \comp
  \seminsm{c_1}{\st}{} \comp \guardop_{b} \comp \gammast(\Delta) \\
  & \quad \supseteq \text{(by monotonicity and $\gammast \comp \alphast$ being 
    extensive)} \\
  & \qquad \alphares{\st'} \comp
  \seminsm{c_1}{\st}{} \comp \gammast \comp \alphast \comp \guardop_{b} \comp
  \gammast(\Delta) \\
  & \quad \supseteq \text{(by induction hypothesis)} \\
   & \qquad \seminsmabs{c_1}{\st}{} \comp \alphast \comp \guardop_{b} \comp
  \gammast(\Delta) 
\end{align*}

Note that $\guardop_b \comp \gammast(\Delta) \subseteq \gammast(\Delta)$,
therefore:
\[\alphast \comp \guardop_b \comp \gammast(\Delta) \supseteq \alphast
\comp \gammast(\Delta) \supseteq \Delta \]
Additionally, since $\semexp{b}{\st} = true$, and also:
\[\forall \stm \in \guardop_b \comp \gammast(\Delta), \semexp{b}{\stm} = true\]
then it holds that: $\forall \stm \in \guardop_b \comp \gammast(\Delta),
\st \shortmid \stm \models \bagree b$.
Notice that we explicitly use the assumption that the major state evaluates to
true. 
Therefore: \[ \bagree b \in \alphast \comp \guardop_{b} \comp
\gammast(\Delta).\]

To sum up, we obtain an approximation of $\guardop_b$:

\[
\alphast \comp \guardop_b \comp \gammast(\Delta) \supseteq \Delta \cup \{
  \bagree b \}
\]

Therefore, in the then-branch we have:
\[
 \alphares{\st'} \comp \seminsm{c_1}{\st}{} \comp \guardop_{b} \comp
 \gammast(\Delta) \supseteq
 \seminsmabs{c_1}{\st}{(\Delta \cup \{ \bagree b \})}.
 \] 

3.2 -- Else-branch:
\begin{align*}
& \alphares{\st'} \comp
  \seminsc{c_2}{} \comp \guardop_{\neg b} \comp \gammast(\Delta) \\
  & \quad  \supseteq \text{(by monotonicity and $\gammast \comp \alphast$ being 
    extensive)} \\
  & \qquad \alphares{\st'} \comp \seminsc{c_2}{} \comp \gammast \comp \alphast
  \comp \guardop_{\neg b} \comp \gammast(\Delta)
\end{align*}

Since the major state $\st$ is assumed to evaluate to true, it holds that:
\[ \alphast \comp \guardop_{\neg b} \comp \gammast(\Delta) \supseteq
\begin{cases}
  \mathcal{L} & \text{if } \Delta \entail \agree b \\
  \Delta & \text{otherwise}
\end{cases}
\]

Therefore, in the else branch we have:

\begin{align*}
& \alphares{\st'} \comp
  \seminsc{c_2}{} \comp \guardop_{\neg b} \comp \gammast(\Delta) \\
  & \quad  \supseteq \alphares{\st'} \comp \seminsc{c_2}{} \comp \gammast \comp
  \alphast 
  \comp \guardop_{\neg b} \comp \gammast(\Delta) \\
  & \quad \supseteq \seminscabs{c_2}{} \comp \lambda \Delta. \begin{cases}
  \mathcal{L} & \text{if } \Delta \entail \agree b \\
  \Delta & \text{otherwise}
  \end{cases} \\
  & \quad =
  \begin{cases}
    \mathcal{L} & \text{if } \Delta \entail \agree b \\
    \emptyset & \text{otherwise}
  \end{cases}
\end{align*}

3.3 -- Finally, we merge both approximations of the then-branch and the
else-branch:

\begin{align*}
  & \alphares{\st'} \comp
  \seminsm{\ifcom~ b~ \thencom~ c_1~ \elsecom~ c_2}{\st}{}
  \comp \gammast(\Delta)  \\
  & \quad \supseteq  \alphares{\st'} \comp
  \seminsm{c_1}{\st}{} \comp \guardop_{b} \comp \gammast(\Delta) {} \cap {} \\
  & \qquad\qquad \alphares{\st'} \comp
  \seminsc{c_2}{} \comp \guardop_{\neg b} \comp \gammast(\Delta) \\
  & \quad \supseteq \seminsmabs{c_1}{\st}{(\Delta \cup \{ \bagree b \})}
  {} \cap {}
  \begin{cases}
    \mathcal{L} & \text{if } \Delta \entail \agree b \\
    \emptyset & \text{otherwise } 
  \end{cases} \\
  & \quad =
  \begin{cases}
    \seminsmabs{c_1}{\st}{(\Delta \cup \{\bagree b \})} & \text{if } \Delta
    \entail \agree b \\
    \emptyset & \text{otherwise}
  \end{cases} \\
  & \quad \triangleq \text{(when $\semexp{b}{\st} = true$)} \\
  & \qquad
  \seminsmabs{\ifcom~ b~ \thencom~ c_1~ \elsecom~c_2}{\st}{\Delta}
\end{align*}

The case where $\semexp{b}{\st} = false$ is symmetric.

4 -- Case: sequences.

Let $\st_1 = \semins{c_1}{\st}$ and $\st_2 = \semins{c_2}{\st_1}$. Then:
\begin{align*}
  & \alphares{\st_2} \comp \seminsm{c_1;c_2}{\st}{} \comp \gammast(\Delta)
  \\
  & \quad = \alphares{\st_2} \comp \seminsm{c_2}{\st_1}{} \comp
  \seminsm{c_1}{\st}{} \comp \gammast(\Delta) \\
  & \quad \supseteq \text{($ \gammares{\st_1} \comp
    \alphares{\st_1}$ is extensive)}\\ 
  & \qquad \alphares{\st_2} \comp \seminsm{c_2}{\st_1}{}
  \comp \gammares{\st_1} \comp \alphares{\st_1} \comp
  \seminsm{c_1}{\st}{} \comp \gammast(\Delta) \\
  & \quad \supseteq \text{(By induction hypothesis)} \\
  & \qquad \seminsmabs{c_2}{\st_1}{} \comp \seminsmabs{c_1}{\st}{\Delta} \\
  & \quad \triangleq \seminsmabs{c_1;c_2}{\st}{\Delta}
\end{align*}

5 -- Case: assumptions
\begin{align*}
  & \alphares{\st} \comp \seminsm{\assumecom~\Phi}{\st}{} \comp
  \gammast(\Delta) \\
  & \quad = \alphares{\st} \comp \{ \stm \in \gammast(\Delta) \mid \st \shortmid \stm
  \models \Phi \} \\
  & \quad \supseteq \Delta \cup \{ \Phi \} \\
  & \quad  \triangleq \seminsmabs{\assumecom~\Phi}{\st}{\Delta}
\end{align*}

6 -- Case: assertions
\begin{align*}
  & \alphares{\st} \comp \seminsm{\assertcom~\Phi}{\st}{} \comp
  \gammast(\Delta) \\
  & \quad = \alphast \comp 
  \begin{cases}
    \gammast(\Delta) & \text{ if } \forall \stm \in \gammast(\Delta), \st
    \shortmid \stm \models \Phi \\ 
    \err & \text{otherwise}
  \end{cases} \\
  & \quad \supseteq
  \begin{cases}
    \Delta \cup \{ \Phi \} & \text{if } \Delta \entail \Phi \\
    \err & \text{otherwise}
  \end{cases}
\end{align*}

7 -- Case: loops

\[  \seminsm{\whilecom~ (e)~\docom~ c}{\st}{\St} \triangleq
\operatorname{snd} \left(
(\operatorname{lfp}_{\lambda (\st,\St).(\bot,\emptyset)}^{\preccurlyeq \times
  \dotsubseteq} \mathcal{G})  (\st,\St)\right)
\]

with:

\begin{multline*} 
\mathcal{G}(w) \triangleq
    \lambda (\st,\St). \\
    \begin{cases}
      \st,  \seminsc{\whilecom~b~\docom~c}{\St}
      & \text{if }  \neg \semexp{b}{\st} \\
      w \left( \semins{c}{\st},
      \seminsm{ \ifcom~ b~\thencom~ c~\elsecom~ \skipcom}{\st}{\St} \right)
      & \text{otherwise}
    \end{cases}
\end{multline*}

and:

\begin{align*}
  \concdom & \triangleq
\States \times \pstateserr {} \cup  {}  \{(\bot,\emptyset), (\top,\err)\} \\
\absdom & \triangleq
\States \times \prelformerr {} \cup {} \{ (\bot,\relform), (\top,\err)\}
 \end{align*}

\begin{align*}
  \alphapair & \in  \concdom \to \absdom \\
  \alphapair(\st,\St) &  \triangleq
  \begin{cases}
    \bot, \relform & \text{if } \st,\St = \bot,\emptyset \\
    \top, \err & \text{if } \st,\St = \top, \err \\
    \st, \alphast(\St) & \text{otherwise }
  \end{cases}
\end{align*}

\begin{align*}
  \gammapair & \in \absdom \to \concdom \\
  \gammapair(\st,\Delta) & \triangleq
  \begin{cases}
    \bot, \emptyset & \text{if } \st,\Delta = \bot,\relform \\
    \top, \err & \text{if } \st,\Delta = \top,\err \\
    \st, \gammast(\Delta) & \text{otherwise}
  \end{cases}
\end{align*}

1 -- First, we prove that $(\alphapair,\gammapair)$ is a Galois connection:
\[ 
  (\concdom; \preccurlyeq \times \concleq) 
\galois{\alphapair}{\gammapair} 
(\absdom;  \preccurlyeq \times \absleq)
\]

Let $(\st,\St) \in \concdom$, and $(\st',\Delta) \in \absdom$.

  1.1 -- Assume
  $\alphapair(\st, \St) \preccurlyeq \times \absleq (\st',\Delta)$.
  
  Prove $(\st,\St) \preccurlyeq \times \concleq \gammapair(\st',\Delta)$.

  If $(\st,\St) = (\bot, \emptyset)$, then it holds that
  $(\st,\St) \preccurlyeq \times \concleq\gammapair(\st',\Delta)$.

  Otherwise, if $(\st,\St) = (\top,\err)$, then $(\st',\Delta) = (\top,\err)$,
  and it holds that
  $(\st,\St) \preccurlyeq \times \concleq \gammapair(\st',\Delta)$. 

  Otherwise, if $(\st,\St) \in \States \times \pstateserr$, then either
  $\st = \st'$, or $\st' = \top$. If $\st' = top$, then $\Delta = \err$ and it
  holds that $(\st,\St) \preccurlyeq \times \concleq \gammapair(\st',\Delta)$.
  If $\st = \st'$, then
  $\alphast(\St) \supseteq \Delta \implies \St \concleq \gammast(\Delta)$
  since $(\alphast,\gammast)$ is a Galois connection.
  Therefore, it also holds that 
  $(\st,\St) \preccurlyeq \times \concleq \gammapair(\st',\Delta)$.

   1.2 -- Assume 
   $(\st,\St)  \preccurlyeq \times \concleq \gammapair(\st',\Delta)$.

   Prove $\alphapair(\st, \St) \preccurlyeq \times \absleq (\st',\Delta)$.

   If $(\st',\Delta) = (\bot,\relform)$, then $(\st,\St) = (\bot,\emptyset)$, thus it holds that
   $\alphapair(\st, \St) \preccurlyeq \times \absleq (\st',\Delta)$.

   Otherwise, if $(\st',\Delta) = (\top,\err)$, then it holds that
   $\alphapair(\st, \St) \preccurlyeq \times \absleq (\st',\Delta)$.

   Otherwise, if $(\st',\Delta) \in \States \times \prelformerr$, then
   $\st \preccurlyeq \st'$ implies that either $\st = \bot$, or $\st = \st'$.
   If $\st = \bot$, then $\St = \emptyset$ and it holds that
   $\alphapair(\st, \St) \preccurlyeq \times \absleq (\st',\Delta)$.
   If $\st = \st'$, then
   $\St \concleq \gammast(\Delta) \implies \alphast(\St) \absleq \Delta$
   since $(\alphast,\gammast)$ is a Galois connection.
   Therefore, it holds that: 
   $\alphapair(\st, \St) \preccurlyeq \times \absleq (\st',\Delta)$.

   2 -- Approximating the fixpoint
   $ \operatorname{lfp}_{\lambda (\st,\St).(\bot,\emptyset)}^{\preccurlyeq
     \times \dotconcleq} \mathcal{G} \comp \gammapair(\st,\Delta)$.

   2.1 -- Applying the fixpoint transfer theorem:

   \begin{align*}
     & \alphapair \comp (\operatorname{lfp}_{\lambda
       (\st,\St).(\bot,\emptyset)}^{\dotpreccurlyeq \times \dotconcleq}
     \mathcal{G}) \comp \gammapair(\st,\Delta) \\
     & \quad
       \preccurlyeq \times \absleq \text{(assuming $\mathcal{G}^\sharp$ is a
         sound approximating of $\mathcal{G}$)} \\
       & \quad \quad
      (\operatorname{lfp}_{\lambda (\st,\Delta).(\bot,\relform)}^{\dotpreccurlyeq
        \times \dotabsleq} 
    \mathcal{G}^\sharp)  (\st,\Delta)
   \end{align*}
   
   2.2 -- Deriving a sound approximation of $\mathcal{G}^\sharp$:

   \begin{align*}
     & \alphapair \comp \mathcal{G}(w) \comp \gammapair(\st,\Delta) \\
     & = 
     \alphapair \comp 
     \begin{cases}
       \st, \seminsc{\whilecom~b~\docom~c}{\gammast(\Delta)}
       \quad \text{ if }  \neg \semexp{b}{\st} \\
       w \big( \semins{c}{\st}, \\
       \quad 
      \seminsm{ \ifcom~ b~\thencom~ c~\elsecom~ \skipcom}{\st}{} \comp
      \gammast(\Delta) \big) 
      \text{ oth.}
     \end{cases}
     \\
     & =  \begin{cases}
       \st, \alphast \comp \seminsc{\whilecom~b~\docom~c}{} \comp
       \gammast(\Delta)
       \quad \text{ if }  \neg \semexp{b}{\st} \\
       \alphapair \comp w \big( \semins{c}{\st}, \\
       \quad 
       \seminsm{ \ifcom~ b~\thencom~ c~\elsecom~ \skipcom}{\st}{} \comp
       \gammast(\Delta) \big)
      \text{ oth.}
      \end{cases}
      \\
      & \preccurlyeq \times \absleq
       \begin{cases}
      \st, \alphast \comp \seminsc{\whilecom~b~\docom~c}{} \comp \gammast(\Delta)
       \quad \text{ if }  \neg \semexp{b}{\st} \\
       \alphapair \comp w \comp \gammapair \comp  \alphapair \comp \big(
       \semins{c}{\st}, \\
       \quad
      \seminsm{ \ifcom~ b~\thencom~ c~\elsecom~ \skipcom}{\st}{} \comp \gammast(\Delta) \big)
      \text{ oth.}
       \end{cases} \\
       & \preccurlyeq \times \absleq
       \begin{cases}
      \st, \alphast \comp \seminsc{\whilecom~b~\docom~c}{} \comp \gammast(\Delta)
       \quad \text{ if }  \neg \semexp{b}{\st} \\
       \alphapair \comp w \comp \gammapair \comp 
        \big( \semins{c}{\st}, \\
        \quad \alphares{\semins{c}{\st}} \comp
        \seminsm{ \ifcom~ b~\thencom~c~\elsecom~
          \skipcom}{\st}{} \comp \\
        \qquad \qquad \gammast(\Delta)
        \big) 
      \qquad \qquad \qquad \text{ oth.}
       \end{cases} \\
       &
        \preccurlyeq \times \absleq
       \begin{cases}
       \st, \alphast \comp \seminsc{\whilecom~b~\docom~c}{} \comp \gammast(\Delta)
       \quad \text{ if }  \neg \semexp{b}{\st} \\
      w^\sharp \comp 
        \big( \semins{c}{\st}, \\
        \quad \alphares{\semins{c}{\st}} \comp
        \seminsm{ \ifcom~ b~\thencom~c~\elsecom~
          \skipcom}{\st}{} \comp \\
        \qquad \qquad \gammast(\Delta)
        \big) 
      \qquad \qquad \qquad \text{ oth.}
       \end{cases}
       \\
       & \preccurlyeq \times \absleq
       \begin{cases}
       \st, \alphast \comp \seminsc{\whilecom~b~\docom~c}{} \comp \gammast(\Delta)
       \quad \text{ if }  \neg \semexp{b}{\st} \\
      w^\sharp \comp 
        \big( \semins{c}{\st}, \\
        \quad 
        \seminsmabs{ \ifcom~ b~\thencom~c~\elsecom~
          \skipcom}{\st}{\Delta}
        \big) 
      \text{ oth.}
       \end{cases} \\
       & \text{(Notice that the derivation above is still generic.)} \\
       & \text{(Specialising it with the abstract static semantics now.)}
       \\
        & \preccurlyeq \times \absleq
       \begin{cases}
         \st, \err
       \qquad \qquad \qquad \text{ if }  \neg \semexp{b}{\st} \wedge \Delta = \err \\
       \st, \Delta \cup \{ \bagree \neg b \}
       \quad \quad \text{ if }  \neg \semexp{b}{\st} \wedge (\Delta \entail \agree b)
       \\
       \st, \{ \bagree \neg b \}
       \qquad \qquad \text{ otherwise if }  \neg \semexp{b}{\st} \\
      w^\sharp \comp 
        \big( \semins{c}{\st}, \\
        \, \,
        \seminsmabs{ \ifcom~ b~\thencom~c~\elsecom~
          \skipcom}{\st}{\Delta}
        \big) 
      \text{ otherwise}
       \end{cases}
   \end{align*}

   3 -- Finally,

   \begin{align*}
     & \seminsmabs{\whilecom~b~\docom~c}{\st}{\Delta} \triangleq
     \\
     \operatorname{snd} 
     (\operatorname{lfp}_{\lambda (\st,\Delta).(\bot,\relform)}^{\dotpreccurlyeq
        \times \dotabsleq} 
    \mathcal{G}^\sharp)  (\st,\Delta)
   \end{align*}

   with:
   \begin{multline*}
   \mathcal{G}^\sharp \triangleq 
   \lambda w^\sharp. 
   \lambda (\st,\Delta) . \\
      \begin{cases}
         \st, \err
       \qquad \qquad \qquad \text{ if }  \neg \semexp{b}{\st} \wedge \Delta = \err \\
       \st, \Delta \cup \{ \bagree \neg b \}
       \quad \quad \text{ if }  \neg \semexp{b}{\st} \wedge (\Delta \entail \agree b)
       \\
       \st, \{ \bagree \neg b \}
       \qquad \qquad \text{ otherwise if }  \neg \semexp{b}{\st} \\
      w^\sharp \comp 
        \big( \semins{c}{\st}, \\
        \, \,
        \seminsmabs{ \ifcom~ b~\thencom~c~\elsecom~
          \skipcom}{\st}{\Delta}
        \big) 
      \text{ otherwise}
       \end{cases}
   \end{multline*}
  
\newpage
\subsection{Hybrid Monitor with the Modified Variables}

We start by first proving an intermediate result, introduced in
\Cref{lem:framerule}.
We denote by $\st_{\rest V}$ the restriction of the state
$\st$ to the set $V$ of variables.
\Cref{lem:framerule}
\begin{restatable}[]{mylem}{lemframerule}
  \label{lem:framerule}
  $\forall \st,\stm, \stm' \in \States$, $\forall \Phi \in \mathcal{L}$,
  $\forall V \subseteq Var$,  if:
  \begin{enumerate}
  \item $\st \shortmid \stm \models \Phi$
  \item $\stm_{\rest V} = \stm'_{\rest V}$
  \item $ \fv{\Phi} \subseteq V$
  \end{enumerate}
  then,
  \[ \st \shortmid \stm' \models \Phi \]
\end{restatable}
\textit{Proof.}
The proof of this lemma is straightforward, by structural induction on
relational formulas as well as a structural induction on expressions, by
remarking that $\forall e \in Exp, (\fv{e} \subseteq V \implies
\semexp{e}{\stm} = \semexp{e}{\stm'})$. 

\lemsoundabsstatichybrid*
with $\hybmod\seminscabshybmod{c}{}{c'}$ defined as:
\begin{align*}
  \hybmod\seminscabshybmod{c}{}{c'} & \in \prelformerr \to \prelformerr \\
  \hybmod\seminscabshybmod{c}{\Delta}{c'} & \triangleq
  \begin{cases}
    \err & \text{if } \Delta = \err \\
    \mathcal{L} & \text{if } \Delta = \mathcal{L} \\
    \big\{ \Phi \in \Delta \mid \forall id \in \fv{\Phi},  & \\
     \quad  id \not \in \Mod(c) \cup
    \Mod(c') \big\} & \text{otherwise }  
  \end{cases}
\end{align*}
\textit{Proof.}

Let a command {\ttfamily c'} and  $\st,\st' \in \States$ such that 
$\st' = \semins{c'}{\st}$.

Then:
\[ \alphares{\st'} \comp \seminsc{c}{} \comp \gammast(\mathcal{L}) =
\alphares{\st'}{} \comp \seminsc{c}{\emptyset} = \alphares{\st'}(\emptyset) =
\mathcal{L}  \]

Also,
\[ \alphares{\st'} \comp \seminsc{c}{} \comp \gammast(\err) =
\alphares{\st'}{} \comp \seminsc{c}{\err} = \alphares{\st'}(\err) =
\err  \]

Additionally, assuming $\Delta \neq \relform$ and $\Delta \neq \err$:
\begin{align*}
  & \alphares{\st'} \comp \seminsc{c}{} \comp \gammast(\Delta) \\
  & \quad \text{(By monotonicity of $\alphares{\st'}$, and $\gammast \comp
    \alphast$ being extensive)} \\
  & \quad \quad 
   \alphares{\st'} \comp \gammast \comp \alphast \comp \seminsc{c}{} \comp
   \gammast(\Delta) \\ 
  & \quad \absleq \text{(By applying \Cref{lem:framerule}, since for all
    $\stm \in \gammast(\Delta)$, } \\
  & \qquad \text{ for all $\stm'$ such that $\stm' =
    \semins{c}{\stm},\stm' =_{Var\backslash \Mod(c)} \stm$)} 
  \\
  & \qquad \quad 
  \alphares{\st'} \comp \gammast
  \left( \{ \Phi \in \Delta \mid \fv{\Phi} \cap \Mod(c) = \emptyset \} \right) \\
  & \quad \absleq \text{(By applying \Cref{lem:framerule}, since for all
    $\st, \st'$ such that} \\
  & \qquad \text{ $\st' = \semins{c'}{\st}$, $\st' =_{Var\backslash \Mod(c')}
    \st$, and by defining $\Delta'$ as} \\
  & \qquad \text{$\Delta' \triangleq 
    \{ \Phi \in \Delta \mid \fv{\Phi} \cap \Mod(c) = \emptyset \} $)}
  \\
  & \qquad \quad \{ \Phi \in \Delta' \mid \fv{\Phi} \cap \Mod(c') = \emptyset \}
  \\ 
  & \quad =
  \{ \Phi \in \Delta \mid \fv{\Phi} \cap (\Mod(c) \cup \Mod(c')) = \emptyset \}
  \\
  & \quad \triangleq \hybmod\seminscabshybmod{c}{\Delta}{c'} \qed
\end{align*}

\lemsoundabsmonitorhybrid*
\textit{Proof.}

Let $\St \in States$ and $\Delta \in \mathcal{L}$. Let us derive a hybrid
monitoring semantics relying on a static analysis over-approximating the set of
variables that may be modified by a command {\ttfamily c}.
We will consider only the case of branching instructions, since the derivation
of the abstract monitoring semantics of the other commands is similar to the one
in \Cref{lem:soundabsmonitorpured}.

1 -- Case : conditionals

Let $\st,\st' \in \States$ such that
$\st' = \semins{\ifcom~ b~ \thencom~ c_1~ \elsecom~ c_2}{\st}$.
Let us also assume that $\semexp{b}{\st} = true$, which means that 
$\st' = \semins{c_1}{\st}$.
Then, similarly to \Cref{lem:soundabsmonitorhybrid} we have:

\begin{align*}
  & \alphares{\st'} \comp
  \seminsm{\ifcom~ b~ \thencom~ c_1~ \elsecom~ c_2}{\st}{}
  \comp \gammast(\Delta)  \\ 
  & \quad = \alphares{\st'}\big(
  \seminsm{c_1}{\st}{} \comp \guardop_{b} \comp \gammast(\Delta) {} \cup {} \\
  & \qquad\qquad 
  \seminsc{c_2}{} \comp \guardop_{\neg b} \comp \gammast(\Delta) \big)\\
  & \quad =  \alphares{\st'} \comp
  \seminsm{c_1}{\st}{} \comp \guardop_{b} \comp \gammast(\Delta) {} \cap {} \\
  & \qquad\qquad \alphares{\st'} \comp
  \seminsc{c_2}{} \comp \guardop_{\neg b} \comp \gammast(\Delta)
\end{align*}
as well as:
\[
 \alphares{\st'} \comp \seminsm{c_1}{\st}{} \comp \guardop_{b} \comp
 \gammast(\Delta) \absleq
 \seminsmabs{c_1}{\st}{(\Delta \cup \{ \bagree b \})}.
 \]

 As for the non-executed branch, we have a more precise abstract static
 semantics by \Cref{lem:soundabsstatichybrid}: 
\begin{align*}
& \alphares{\st'} \comp
\seminsc{c_2}{} \comp \guardop_{\neg b} \comp \gammast(\Delta) \\
& \absleq
  \begin{cases}
    \mathcal{L} & \text{if } \Delta \entail \agree b \\
    \big\{ \Phi \in \Delta \mid \forall id \in \fv{\Phi},  & \\
     \quad  id \not \in \Mod(c_1) \cup
    \Mod(c_2) \big\} & \text{otherwise }  
  \end{cases}
\end{align*}

Finally,
\begin{align*}
  & \alphares{\st'} \comp
  \seminsm{\ifcom~ b~ \thencom~ c_1~ \elsecom~ c_2}{\st}{}
  \comp \gammast(\Delta)  \\
 & \quad \absleq  \alphares{\st'} \comp
  \seminsm{c_1}{\st}{} \comp \guardop_{b} \comp \gammast(\Delta) {} \cap {} \\
  & \qquad\qquad \alphares{\st'} \comp
  \seminsc{c_2}{} \comp \guardop_{\neg b} \comp \gammast(\Delta) \\
  & \quad \absleq \seminsmabs{c_1}{\st}{(\Delta \cup \{ \bagree b \})}
          {} \cap {} \\
  & \quad \qquad \qquad
   \begin{cases}
    \mathcal{L} & \text{if } \Delta \entail \agree b \\
    \big\{ \Phi \in \Delta \mid \forall id \in \fv{\Phi},  & \\
     \quad  id \not \in \Mod(c_1) \cup
    \Mod(c_2) \big\} & \text{otherwise }  
   \end{cases} \\
   & =
   \begin{cases}
      \seminsmabs{c_1}{\st}{(\Delta \cup \{ \bagree b \})} & \text{if } \Delta
      \entail \agree b  \\
      \seminsmabs{c_1}{\st}{(\Delta \cup \{ \bagree b \})}
                 {} \cap {}
       & \\
      \{ \Phi \in \Delta \mid \fv{\Phi} \cap ( \Mod(c_1) \cup
      \Mod(c_2)) = \emptyset \}
      & \text{otherwise} 
   \end{cases}
\end{align*}

2 -- Case: loops

We rely on  the generic derivation of loops in \Cref{lem:soundabsmonitorpured},
and specialize it with the static analysis relying on the modified variables.

Therefore, we continue the derivation from the generic approximation obtained in
\Cref{lem:soundabsmonitorpured}:

   \begin{align*}
     & \alphapair \comp \mathcal{G}(w) \comp \gammapair(\st,\Delta) \\
            & \preccurlyeq \times \absleq
       \begin{cases}
       \st, \alphast \comp \seminsc{\whilecom~b~\docom~c}{} \comp \gammast(\Delta)
       \quad \text{ if }  \neg \semexp{b}{\st} \\
      w^\sharp \comp 
        \big( \semins{c}{\st}, \\
        \quad 
        \seminsmabs{ \ifcom~ b~\thencom~c~\elsecom~
          \skipcom}{\st}{\Delta}
        \big) 
      \text{ oth.}
       \end{cases} \\
       & \preccurlyeq \times \absleq \\
       &
       \begin{cases}
         \st, \err
       \qquad \qquad \qquad \text{ if }  \neg \semexp{b}{\st} \wedge \Delta = \err \\
       \st, \Delta \cup \{ \bagree \neg b \}
        \quad \quad \text{ if }  \neg \semexp{b}{\st} \wedge (\Delta \entail \agree b)
       \\
       \st, 
       \big( \{ \Phi \in \Delta \mid \fv{\Phi} \cap \Mod(c)=  \emptyset \}
           {} \cap {} \\
       \qquad \Delta  \big) {} \cup {}  \{ \bagree \neg b \} 
      \quad  \text{ otherwise if }  \neg \semexp{b}{\st} \\
      w^\sharp \comp 
        \big( \semins{c}{\st}, \\
        \, \,
        \seminsmabs{ \ifcom~ b~\thencom~c~\elsecom~
          \skipcom}{\st}{\Delta}
        \big) 
      \text{ otherw.}
       \end{cases}
   \end{align*}

   Therefore:

     \[
     \seminsmabs{\whilecom~b~\docom~c}{\st}{\Delta} \triangleq
     \operatorname{snd} \left(
     (\operatorname{lfp}_{\lambda (\st,\Delta).(\bot,\relform)}^{\dotpreccurlyeq
        \times \dotabsleq} 
    \mathcal{G}^\sharp)  (\st,\Delta) \right) 
  \]

  with:

   \begin{multline*}
   \mathcal{G}^\sharp \triangleq 
   \lambda w^\sharp. 
   \lambda (\st,\Delta) . \\
       \begin{cases}
         \st, \err
       \qquad \qquad \qquad \text{ if }  \neg \semexp{b}{\st} \wedge \Delta = \err \\
       \st, \Delta \cup \{ \bagree \neg b \}
        \quad \quad \text{ if }  \neg \semexp{b}{\st} \wedge (\Delta \entail \agree b)
       \\
       \st, 
       \big( \{ \Phi \in \Delta \mid \fv{\Phi} \cap \Mod(c)=  \emptyset \}
           {} \cap {} \\
       \qquad \Delta  \big) {} \cup {}  \{ \bagree \neg b \} 
      \quad  \text{ otherwise if }  \neg \semexp{b}{\st} \\
      w^\sharp \comp 
        \big( \semins{c}{\st}, \\
        \, \,
        \seminsmabs{ \ifcom~ b~\thencom~c~\elsecom~
          \skipcom}{\st}{\Delta}
        \big) 
      \text{ otherw.}
       \end{cases}
   \end{multline*}
   
\clearpage

\subsection{Hybrid Monitor with Intervals}
\label{sec:ap_hybrid_intervals}

The abstract semantics of an interval analysis, inspired by \cite{Kin96},
 is presented in \Cref{fig:abstract_interval_analysis}.

\begin{figure}[hbp]
  \fbox{
    \begin{mathpar}
      \guardint{b}( \err ) \triangleq \err

      \\
      
    \blwint{[a,b]} \triangleq [-\infty, b]

    \abvint{[a,b]} \triangleq [a, +\infty]
    
    \guardint{ e_1 \leq e_2 }( \imath ) \triangleq
    \appint{e_1}{\blwint{\imath(e_2)}}(\imath)
    \sqcap^\sharp
    \appint{e_2}{\abvint{\imath(e_1)}}(\imath)

    \guardint{ e_1 < e_2}(\imath) \triangleq
    \appint{e_1}{\blwint{\imath(e_2) -1 }}(\imath)
    \sqcap^\sharp
    \appint{e_2}{\abvint{\imath(e_1) +1 }}(\imath)

    \guardint{ e_1 = e_2}(\imath) \triangleq
    \appint{e_1}{\imath(e_2) }(\imath)
    \sqcap^\sharp
    \appint{e_2}{\imath(e_1)}(\imath)

    \guardint{ \neg( e_1 = e_2) }(\imath) \triangleq \guardint{ (e_1 < e_2)
      \wedge (e_2 > e_1) }(\imath) \triangleq
    \guardint{ b_1 \wedge b_2}(\imath) \triangleq
    \guardint{b_1}(\imath) \sqcap^\sharp \guardint{b_2}(\imath)

      \guardint{ b_1 \vee b_2}(\imath) \triangleq
    \guardint{b_1}(\imath) \sqcup^\sharp \guardint{b_2}(\imath)
    
  \end{mathpar}}
  
  \fbox{
  \begin{mathpar}
    \appint{x}{i}(\imath) \triangleq \imath[x \mapsto \imath(x) \sqcap^\sharp i]
    
    \appint{e_1 + e_2}{i}(\imath) \triangleq
    \appint{e_1}{i - \imath(e_2)}(\imath) \sqcap^\sharp \appint{e_2}{i -
      \imath(e_1)}(\imath)

     \appint{e_1 - e_2}{i}(\imath) \triangleq
    \appint{e_1}{i + \imath(e_2)}(\imath) \sqcap^\sharp \appint{e_2}{
      \imath(e_1) -i}(\imath)
    
    \appint{n}{i}(\imath) \triangleq \top^{\intervals}
  \end{mathpar}}

  \fbox{
  \begin{mathpar}
    \imath(n) \triangleq [n ,n ]


    \imath(e_1 + e_2) \triangleq \imath(e_1) + \imath(e_2)

    \imath(e_1 - e_2) \triangleq \imath(e_1) - \imath(e_2)

  \end{mathpar}}
   \fbox{
    \begin{mathpar}
  \seminscabsint{c}{\err} \triangleq \err
      
  \seminscabsint{\skipcom}{\imath} \triangleq \imath
  
  \seminscabsint{c_1; c_2}{\imath} \triangleq
  \seminscabsint{c_2}{} \comp \seminscabsint{c_1}{\imath}  

  \seminscabsint{id := e}{\imath} \triangleq \imath[id \mapsto \imath(e)]
  
  \seminscabsint{\assumecom~\Phi}{\imath} \triangleq \imath

  \seminscabsint{\assertcom~\Phi}{\imath} \triangleq \imath

  \seminscabsint{\ifcom~ b~ \thencom~ c_1~ \elsecom~ c_2}{\imath}
  \triangleq
  \seminscabsint{c_1}{} \comp \guardint{b}(\imath) \sqcup^\sharp
   \seminscabsint{c_2}{} \comp \guardint{\neg b}(\imath) 

   \seminscabsint{\whilecom~ (e)~ c}{\imath} \triangleq
   \guardint{\neg b} \left(
   \operatorname{lfp}_{\imath}^{\sqsubseteq^\sharp}
   \seminscabsint{\ifcom~ b~ \thencom~ c_1~ \elsecom~ \skipcom}{}
   \right)

\end{mathpar}}
  \caption{Abstract semantics of an interval analysis}
  \label{fig:abstract_interval_analysis}
\end{figure}

\lemgrangerprod*
\textit{Proof.}

Let $\Delta \in \prelformerr$. Then:
\begin{align*}
 &  \alphaint \comp \gammast ( \Delta ) \\
 & \quad = \alphaint \left( \gammaint(\err) \cap  \gammast ( \Delta )\right) \\
  & \quad = \text{(By definition of $\gammastint$)} \\
   &\quad \qquad \alphaint \comp \gammastint(\err,\Delta) \\
  & \quad= \text{(By the soundness condition of a Granger's pair)} \\
  &\quad \qquad \alphaint \comp \gammastint\left(
  \redtointst(\err,\Delta),\Delta\right) \\ 
  & \quad = \text{(By definition of $\alphastint$)} \\
  & \quad \qquad \operatorname{fst} \alphastint \comp \gammastint\left(
  \redtointst(\err,\Delta),\Delta\right) \\
  & \quad \leqintervals \text{(Since $\alphastint \comp \gammastint$ is
    reductive)} \\ 
  & \quad \qquad \operatorname{fst} \left(
  \redtointst(\err,\Delta),\Delta\right) \\
  & \quad=   \redtointst(\err,\Delta)
\end{align*}

Similarly, let $\imath \in \statesint$. Then:
\begin{align*}
  \alphast \comp \gammaint(\imath) & = \alphast \left(
  \gammaint(\imath) \cap \gammast(\err)   \right) \\
  & = \operatorname{snd} \alphastint \comp \gammastint \left(
  \imath, \err \right) \\
  & = \operatorname{snd}\alphastint \comp \gammastint \left(
  \imath, \redtoprelformst(\imath,\err) \right) \\
  & \absleq \redtoprelformst(\imath,\err) \qed
\end{align*}

\lemsoundredprod*

\textit{Proof.}

The error cases are straightforward:
$\redtointst(\imath,\err) \triangleq \imath$ and
$\redtoprelformst(\err,\Delta) \triangleq \Delta$.

Let us restrict ourselves to $\imath \in \statesint$ and
$\Delta \in \prelform$.

1 -- $\redtointst$ is sound.

1.1 -- Case : $\redtointst(\imath,\Delta)$.

We will prove that:
\[ \gammastint\left(  \mathop{\sqcap^{\sharp}}\limits_{\Phi \in \Delta}
    \redtointst\left(\imath,\{ \Phi \}\right), \Delta \right) =
    \gammastint(\imath, \Delta) \]

Indeed,    

\begin{align*}
   &  \gammastint(\imath,\Delta) \\
   & =\gammaint(\imath) \cap \gammast(\Delta) \\
   & = \text{($\Delta$ is interpreted conjunctively by $\gammast$)} \\
   & \qquad
   \gammaint(\imath)  \mathop{\cap}\limits_{\Phi \in \Delta}
   \gammast\left(\{ \Phi\}\right) \\
   & = \mathop{\cap}\limits_{\Phi \in \Delta}
   \gammaint(\imath)  \mathop{\cap}\limits_{\Phi \in \Delta}
   \gammast\left(\{ \Phi\}\right) \\
   & = 
   \mathop{\cap}\limits_{\Phi \in \Delta}
   \gammastint\left(\imath, \{ \Phi \} \right) \\
   & = \text{(By cases 1.2 and 1.3, proven below)} \\
   & \qquad
  \mathop{\cap}\limits_{\Phi \in \Delta} \gammastint\left(
   \redtointst\left(\imath,\{ \Phi \}\right), \{ \Phi \}
   \right) \\
   & =
    \mathop{\cap}\limits_{\Phi \in \Delta} \gammaint\left(
  \redtointst\left(\imath,\{ \Phi \}\right) \right)
  \mathop{\cap}\limits_{\Phi \in \Delta} \gammast(\{ \Phi \}) \\
  & = \text{($\Delta$ is interpreted conjunctively by $\gammast$)} \\
  & \qquad
  \mathop{\cap}\limits_{\Phi \in \Delta} \gammaint\left(
  \redtointst\left(\imath,\{ \Phi \}\right) \right)
  \cap \gammast(\Delta) \\
  &  = \text{($\gammaint$ is multiplicative)} \\
  & \qquad
  \gammaint \left( \mathop{\sqcap^{\sharp}}\limits_{\Phi \in \Delta}
  \redtointst\left(\imath,\{ \Phi \}\right)  \right)
  \cap \gammast(\Delta) \\
  & = \gammastint\left(  \mathop{\sqcap^{\sharp}}\limits_{\Phi \in \Delta}
  \redtointst\left(\imath,\{ \Phi \}\right), \Delta \right) \\
& \triangleq \gammastint\left( \redtoint(\imath,\Delta) , \Delta \right) \\
& \qquad \text{with }\redtoint(\imath,\Delta) \triangleq 
\mathop{\sqcap^{\sharp}}\limits_{\Phi \in \Delta} 
\redtointst\left(\imath,\{ \Phi \}\right) 
\end{align*}

1.2 -- Case :  $\redtointst(\imath,\{ \agree e \})$

\begin{align*}
  & \gammastint \left( \imath, \{ \agree e \} \right) \\
  & \quad  = \gammaint(\imath) \cap
  \gammast(\{ \agree e \}) \\
  & \quad = \gammaint(\imath) \cap \{ \stm \in \States \mid \st \shortmid \stm \models
  \agree e\} \\
  & \quad = \gammaint(\imath) \cap
  \{ \stm \in \States \mid \semexp{e}{\stm} = \semexp{e}{\st}\} \\
  & \quad =  \gammaint(\imath) \cap
  \guardc{e = \semexp{e}{\st}}(\States) \\
  & \quad = \gammaint(\imath) \cap
  \guardc{e = \semexp{e}{\st}} \comp \gammaint(\imath) \\
  & \quad = \text{(By soundness of
    $\guardint{e = \semexp{e}{\st}}$)} \\
 & \qquad  \gammaint(\imath) \cap
  \gammaint \comp
  \guardint{e = \semexp{e}{\st}}(\imath) \cap
  \gammast(\{\agree e\}) \\
  & \quad = \gammaint\left( \imath \capintervals
  \guardint{e = \semexp{e}{\st}}(\imath) \right)
  \cap \gammast(\{ \agree e \}) \\
  & \quad = \gammastint\left(\imath \capintervals
  \guardint{e = \semexp{e}{\st}}(\imath),
  \{ \agree e\}  \right)
\end{align*}

1.3 -- Case: $\redtointst(\imath, \{ \bagree b \})$

Similarly to case 1.2, we have:

\begin{align*}
  & \gammastint \left( \imath, \{ \bagree e \} \right) \\
  & = \gammastint \left(
  \imath \capintervals \guardint{b}(\imath),
  \{ \bagree b \}  \right) \qed
\end{align*}

\lemsoundabsmonitorinterval*
\textit{Proof.}

Let $\St \in States$ and $\Delta \in \mathcal{L}$. Let us derive a hybrid
monitoring semantics relying on an interval static analysis.
We will consider only the case of branching instructions, since the derivation
of the abstract monitoring semantics of the other commands is similar to the one
in \Cref{lem:soundabsmonitorpured}.

1 -- Case : conditionals

Let $\st,\st' \in \States$ such that
$\st' = \semins{\ifcom~ b~ \thencom~ c_1~ \elsecom~ c_2}{\st}$.
Let us also assume that $\semexp{b}{\st} = true$.
Then, similarly to \Cref{lem:soundabsmonitorhybrid} we have:
\begin{align*}
  & \alphares{\st'} \comp
  \seminsm{\ifcom~ b~ \thencom~ c_1~ \elsecom~ c_2}{\st}{}
  \comp \gammast(\Delta)  \\ 
& \quad =  \alphares{\st'} \comp
  \seminsm{c_1}{\st}{} \comp \guardop_{b} \comp \gammast(\Delta) {} \cap {} \\
  & \qquad\qquad \alphares{\st'} \comp
  \seminsc{c_2}{} \comp \guardop_{\neg b} \comp \gammast(\Delta)
\end{align*}

as well as:
\[
 \alphares{\st'} \comp \seminsm{c_1}{\st}{} \comp \guardop_{b} \comp
 \gammast(\Delta) \absleq
 \seminsmabs{c_1}{\st}{(\Delta \cup \{ \bagree b \})}.
 \]

  As for the non-executed branch, we have a more precise abstract static
 semantics by \Cref{lem:soundabsstaticinterval}: 
\begin{align*}
& \alphares{\st'} \comp
\seminsc{c_2}{} \comp \guardop_{\neg b} \comp \gammast(\Delta) \\
& \absleq
    \begin{cases}
      \err \qquad \qquad \text{if } \Delta = \err \\
      \mathcal{L} \qquad \qquad \text{if } \Delta \entail \agree b \\
       \lambda \imath. \redtoprelformres{\st'}(\imath,\emptyset)  
       \comp \seminscabsint{c_2}{}
       \comp  \guardint{ \neg b} 
      \comp \redtointst(\top^{\intervals}, \Delta)
       \text{oth.}
    \end{cases}
\end{align*}

Finally,

\begin{align*}
  & \alphares{\st'} \comp
  \seminsm{\ifcom~ b~ \thencom~ c_1~ \elsecom~ c_2}{\st}{}
  \comp \gammast(\Delta)  \\
 & \quad \absleq \alphares{\st'} \comp
  \seminsm{c_1}{\st}{} \comp \guardop_{b} \comp \gammast(\Delta) {} \cap {} \\
  & \qquad\qquad \alphares{\st'} \comp
  \seminsc{c_2}{} \comp \guardop_{\neg b} \comp \gammast(\Delta) \\
   & \quad \absleq \seminsmabs{c_1}{\st}{(\Delta \cup \{ \bagree b \})}
        {} \cap {} \\
         & \quad 
      \begin{cases}
      \mathcal{L} & \text{if } \Delta \entail \agree b \\
       \lambda \imath.\redtoprelformres{\st'}(\imath,\emptyset)  
       \comp \seminscabsint{c_2}{} \comp & \\
       \qquad 
         \guardint{ \neg b} 
      \comp \redtointst(\top^{\intervals}, \Delta)
      & \text{otherwise}
      \end{cases}
      \\
  & \quad \absleq
      \begin{cases}
        \seminsmabs{c_1}{\st}{(\Delta \cup \{ \bagree b \})} & \text{if } \Delta
        \entail \agree b \\
      \seminsmabs{c_1}{\st}{(\Delta \cup \{ \bagree b \})}
                 {} \cap {} &  \\             
      \quad \lambda \imath.\redtoprelformres{\st'}(\imath,\emptyset)  
      \comp \seminscabsint{c_2}{} \comp & \\
      \qquad  \guardint{ \neg b} 
      \comp \redtointst(\top^{\intervals}, \Delta)
      & \text{otherwise}
        \end{cases}
\end{align*}

2 -- Case : loops

We rely on  the generic derivation of loops in \Cref{lem:soundabsmonitorpured},
and specialize it with the static analysis relying on the modified variables.

Therefore, we continue the derivation from the generic approximation obtained in
\Cref{lem:soundabsmonitorpured}:

 \begin{align*}
   & \alphapair \comp \mathcal{G}(w) \comp \gammapair(\st,\Delta) \\
   & \preccurlyeq \times \absleq
   \begin{cases}
         \st, \alphast \comp \seminsc{\whilecom~b~\docom~c}{} \comp \gammast(\Delta)
         \quad \text{ if }  \neg \semexp{b}{\st} \\
         w^\sharp \comp 
        \big( \semins{c}{\st}, \\
        \quad 
        \seminsmabs{ \ifcom~ b~\thencom~c~\elsecom~
          \skipcom}{\st}{\Delta}
        \big) 
        \text{ oth.}
   \end{cases} \\
   & \preccurlyeq \times \absleq \\
   &
   \begin{cases}
         \st, \err
         \qquad \qquad \qquad \text{ if }  \neg \semexp{b}{\st} \wedge \Delta = \err \\
         \st, \Delta \cup \{ \bagree \neg b \}
         \quad \quad \text{ if }  \neg \semexp{b}{\st} \wedge (\Delta \entail \agree b)
         \\
         \st,
         \big( \lambda \imath.\redtoprelformres{\st}(\imath,\emptyset)  
         \comp \seminscabsint{\whilecom~b~\docom~c}{} 
         \comp
         \\
         \qquad \redtointst(\top^{\intervals}, \Delta)
        {} \cap {} \Delta \big) \cup \{ \bagree \neg b \}
          \text{ otherwise if }  \neg \semexp{b}{\st} \\
             w^\sharp \comp 
             \big( \semins{c}{\st}, \\
             \, \,
             \seminsmabs{ \ifcom~ b~\thencom~c~\elsecom~
          \skipcom}{\st}{\Delta}
             \big) 
             \text{ otherw.}
   \end{cases}
 \end{align*}

   Therefore:

     \[
      \seminsmabs{\whilecom~b~\docom~c}{\st}{\Delta} \triangleq
     \operatorname{snd} \left(
     (\operatorname{lfp}_{\lambda (\st,\Delta).(\bot,\relform)}^{\dotpreccurlyeq
        \times \dotabsleq} 
    \mathcal{G}^\sharp)  (\st,\Delta) \right) 
    \]
    
    with:
    
   \begin{multline*}
   \mathcal{G}^\sharp \triangleq 
   \lambda w^\sharp. 
   \lambda (\st,\Delta) . \\
    \begin{cases}
         \st, \err
         \qquad \qquad \qquad \text{ if }  \neg \semexp{b}{\st} \wedge \Delta = \err \\
         \st, \Delta \cup \{ \bagree \neg b \}
         \quad \quad \text{ if }  \neg \semexp{b}{\st} \wedge (\Delta \entail \agree b)
         \\
         \st,
         \big( \lambda \imath.\redtoprelformres{\st}(\imath,\emptyset)  
         \comp \seminscabsint{\whilecom~b~\docom~c}{} 
         \comp
         \\
         \qquad \redtointst(\top^{\intervals}, \Delta)
        {} \cap {} \Delta \big) \cup \{ \bagree \neg b \}
          \text{ otherwise if }  \neg \semexp{b}{\st} \\
             w^\sharp \comp 
             \big( \semins{c}{\st}, \\
             \, \,
             \seminsmabs{ \ifcom~ b~\thencom~c~\elsecom~
          \skipcom}{\st}{\Delta}
             \big) 
             \text{ otherw.}
    \end{cases}
    \end{multline*}
\end{appendices}

\clearpage
\fi

\end{document}